\documentclass[11pt,hyper,a4paper]{article}
\usepackage{jheppub}
\usepackage{amsmath,amssymb,amsfonts,amscd,bm, amsthm, slashed, mathrsfs, bm}
\usepackage{graphicx, wrapfig}
\usepackage{multirow}
\usepackage{verbatim}
\usepackage[title]{appendix}
\usepackage{fancybox}
\usepackage[utf8]{inputenc}

\usepackage{url}
\usepackage{float}

\usepackage[normalem]{ulem}
\usepackage{graphicx}

\usepackage{color}
\usepackage[usenames,dvipsnames,svgnames,table]{xcolor}
\definecolor{darkblue}{cmyk}{0.9,0.9,0,0}
\hypersetup{colorlinks=false, linkcolor=darkblue,linkcolor=darkblue, citecolor=darkblue}

\usepackage{multirow}
\usepackage{verbatim}

\usepackage{bbm}
\usepackage{comment}
\usepackage{enumitem}
\usepackage{float}
\usepackage{empheq}
\usepackage[usenames,dvipsnames,svgnames,table]{xcolor}
\usepackage{enumerate}

\title{BPS Invariants for Seifert Manifolds}

\author{Hee-Joong Chung}

\affiliation{Yau Mathematical Sciences Center, Tsinghua University, Haidian District, Beijing 100084, China}

\abstract{We calculate the homological blocks for Seifert manifolds from the exact expression for the $G=SU(N)$ Witten-Reshetikhin-Turaev invariants of Seifert manifolds obtained by Lawrence, Rozansky, and Mari\~no.
For the $G=SU(2)$ case, it is possible to express them in terms of the false theta functions and their derivatives.
For $G=SU(N)$, we calculate them as a series expansion and also discuss some properties of the contributions from the abelian flat connections to the Witten-Reshetikhin-Turaev invariants for general $N$.
We also provide an expected form of the $S$-matrix for general cases and the structure of the Witten-Reshetikhin-Turaev invariants in terms of the homological blocks.
}

\begin{document}

\maketitle


\section{Introduction}
\label{sec:intro}

The Chern-Simons partition function on the knot complements in the 3-sphere $S^3$ with boundary conditions or on $S^3$ with Wilson loops supported on knots provide knot polynomial invariants, \textit{e.g.} the Jones or the HOMFLY polynomial with colors or with the refinement.
The Jones polynomial, which is a polynomial with integer powers and integer coefficients, can be understood as the graded Euler characteristic of the Khovanov homology, and this homology provides the categorification of the Jones polynomial.
Those integer coefficients can be understood as the dimension of the vector space.
Such categorification of knot polynomials have been studied in many literatures, but it has not been much studied for closed 3-manifolds. 
For example, a mathematical definition or a construction that categorifies the Chern-Simons (CS) partition function or the Witten-Reshetikhin-Turaev (WRT) invariant \cite{Witten-Jones, Reshetikhin-Turaev} for closed 3-manifolds is not available yet.

The WRT invariants for closed 3-manifolds have been calculated for a number of 3-manifolds including Seifert manifolds.
From what is originally expressed, it was not obvious to see whether such WRT invariants can be expressed in terms of the $q$-series with integer powers and integer coefficients, which are the properties suitable for the categorification.

A systematic (physical) approach to the subject firstly appeared in \cite{Gukov-Putrov-Vafa,Gukov-Pei-Putrov-Vafa} and it was conjectured that the WRT invariant for closed 3-manifold can be expressed in terms of the so called homological block, which is a $q$-series with integer powers and integer coefficients so that it may admit the categorification.\footnote{It has been found in \cite{Lawrence-Zagier} that the WRT invariant on the Poincar\'e homology sphere $\Sigma(2,3,5)$ can be expressed as a linear combination of false theta functions with the modular parameter $\tau$ approaching to $1/K$ from the upper half plane of the complex $\tau$-plane where $K$ is the quantum-corrected Chern-Simons level.
As the Poincar\'e homology sphere $\Sigma(2,3,5)$ is an integer homology sphere, in this case it can be seen rather easily from the perspective of \cite{Gukov-Putrov-Vafa,Gukov-Pei-Putrov-Vafa} that the WRT invariant is expressed as a $q$-series with integer powers and integer coefficients.
After \cite{Lawrence-Zagier}, a number of examples including the case of rational homology spheres have been calculated in \cite{Hikami-Bsphere, Hikami-lattice1, Hikami-lattice2, Hikami-spherical, Hikami-decomp}.
However, the integrality of the WRT invariant was not obvious in those examples on rational homology spheres.
}
The conjecture for the case of $G=U(N)$ was also discussed in \cite{Gukov-Pei-Putrov-Vafa}.	\\

Here, we briefly summarize the conjecture of \cite{Gukov-Putrov-Vafa, Gukov-Pei-Putrov-Vafa}.
The conjecture states that the partition function of the Chern-Simons theory with $G=SU(2)$ can be decomposed into homological blocks $\widehat{Z}_b(q)$, 
\begin{align}
Z_{SU(2)}(M_3) 
=\sum_{\substack{a, b \, \in \\ \text{Tor} \, H_1(M_3, \mathbb{Z})/\mathbb{Z}_2}} e^{\pi i K lk(a,a)} S_{ab} \widehat{Z}_b(q) \, \bigg|_{q \searrow e^{ \frac{2\pi i}{K}}  }	\label{su-2-decomp}
\end{align} 
with the $S$-matrix
\begin{align}
S_{ab} = \frac{e^{2\pi i lk(a,b)} + e^{-2\pi i lk(a,b)}}{\left| \mathcal{W}_a \right| \left| \text{Tor} \, H_1(M_3, \mathbb{Z}) \right|^{1/2}}	\label{S-matrix-GPPV}
\end{align}
where $\widehat{Z}_b(q)$, which is defined on $|q|<1$, takes a form of $q^{\Delta_b} \mathbb{Z}[[q]]$ with a rational number $\Delta_b$.
$K \in \mathbb{Z}$ is the quantum-corrected level and $q \searrow e^{ \frac{2\pi i}{K}}$ means that $\tau$ of $q=e^{2\pi i \tau}$ approaches to $\frac{1}{K}$ from the upper half plane of the complex $\tau$-plane.
$\mathcal{W}_a$ denotes the stabilizer subgroup $\text{Stab}_{\mathbb{Z}_2}(a)$ for $a$ in the Weyl group $\mathbb{Z}_2$ of $SU(2)$.
The label $a$ denotes the abelian flat connections or equivalently the reducible flat connections when $G=SU(2)$.
Here, $lk(\cdot, \cdot)$ denotes the linking form on $\text{Tor} \, H_1(M_3,\mathbb{Z})$.
More specifically, given two elements $a, b \in \text{Tor} \, H_1(M_3,\mathbb{Z})$, there is a 2-chain $B'$ such that $s \, b= \partial B'$ for some $s \in \mathbb{Z}_{\neq 0}$.
Then the linking form is defined as
\begin{align}
lk(a,b) = \frac{\# (a \cap B')}{s}	\quad 	\text{mod} \ \mathbb{Z}
\end{align}
where $\#$ denotes the intersection number.
Also, the linking form $lk(a,a)$ can be interpreted as the Chern-Simons invariant of the abelian flat connection $a$.
In addition, the WRT invariant $Z_{SU(2)}(M_3)$ can be written as 
\begin{align}
Z_{SU(2)}(M_3) = \sum_{ a \in \text{Tor} \, H_1(M_3, \mathbb{Z})/\mathbb{Z}_2} e^{\pi i K lk(a,a)} Z_a  \, \bigg|_{q \searrow e^{ \frac{2\pi i}{K}}  }	\,	.
\end{align}
Here, $Z_a$ means the contribution from the abelian flat connection to the WRT invariant, which can be obtained from the Borel resummation of the perturbative expansion with respect to the abelian flat connection $a$.
Several examples including Lens space, some rational homology Seifert manifolds, $\mathcal{O}(1) \rightarrow \Sigma_g$, and plumbed 3-manifolds were worked out and they supported the conjecture in \cite{Gukov-Putrov-Vafa, Gukov-Pei-Putrov-Vafa}.	\\

Interestingly, the homological blocks are labelled by the abelian flat connections, and this can be understood in the context of resurgent analysis \cite{Gukov-Marino-Putrov}.
According to the resurgent analysis on the $G=SU(2)$ case, the exact partition function is expressed in terms of the Borel resummation of the perturbative expansion around abelian flat connections, while the contributions from non-abelian flat connections are encoded in transseries expansions of the Borel resummation of the perturbative expansion with respect to abelian flat connections.

The homological blocks can be understood in the context of the M-theory configuration,	
\begin{align}
\begin{tabular}{r c c c c c c}
\text{space-time :}		&	&	$\mathbb{R}$ 	&$\times$ 	&$TN$ 	&$\times$ 	&$T^* M_3$	\\
\text{M5 branes :}	&	&	$\mathbb{R}$ 	&$\times$ 	&$D^2$ 	&$\times$ 	&$M_3$
\end{tabular}
\label{M-conf}
\end{align}
where $D^2$ is a disc, $TN$ is the Taub-NUT space, and $D^2 \subset TN$.
There are two $U(1)$ symmetries in this system, which are the rotational symmetry $U(1)_{q}$ on $D^2$ and the $U(1)_R$ symmetry.
These two symmetries provide two gradings $\mathbb{Z} \times \mathbb{Z}$ in the homological invariants that lead to homological blocks.
When $M_3$ is a Seifert manifold, which is the 3-manifold that is considered in this paper, there is an additional symmetry $U(1)_\beta$ that arises due to the existence of the semi-free $U(1)$ action on the Seifert manifold.
This will lead to another extra grading $\mathbb{Z}$ in the homological invariants.

Via the 3d-3d correspondence\footnote{Some aspects of the 3d-3d correspondence for Seifert manifolds have been discussed in \cite{CDGS,Gukov-Pei,GGP-4manifold,Alday:2017yxk}.}, the homological invariants for general $M_3$ can be realized as the Hilbert space $\mathcal{H}_{\mathcal{T}[M_3,G]}$ of BPS states of the 3d $\mathcal{N}=2$ theory $\mathcal{T}[M_3, G]$ on $D^2 \times \mathbb{R}$ where $\mathbb{R}$ is regarded as a time direction.
Since there is a boundary $\partial D^2 = S^1$, the Hilbert space $\mathcal{H}_{\mathcal{T}[M_3,G]}$ is specified by the boundary condition and this is labelled by $b \in \text{Tor} \, H_1 (M_3,\mathbb{Z})/\mathbb{Z}_2$.
Thus, the Hilbert space $\mathcal{H}_{\mathcal{T}[M_3,G]}$ can be decomposed into 
\begin{align}
\mathcal{H}_{\mathcal{T}[M_3,G]} = \bigoplus_{b \in \text{Tor} H_1 (M_3,\mathbb{Z})/\mathbb{Z}_2} \mathcal{H}_b	\qquad	\text{with }		
\mathcal{H}_b	=	\bigoplus_{\substack{i \in \mathbb{Z}+\Delta_b \\ j \in \mathbb{Z}}} \mathcal{H}_{b}^{i, j}
\end{align}
where $i$ and $j$ denote the grading for $U(1)_q$ and $U(1)_R$.
Then the homolgical block $\widehat{Z}_b$ is given by
\begin{align}
\widehat{Z}_b(q) = \text{Tr}_{\mathcal{H}_b} q^{i} (-1)^{j}
\end{align}
and this is the partition function or the half index of $\mathcal{T}[M_3,G]$ on $D^2 \times_q S^1$ with the boundary condition $b$.
If $M_3$ is a Seifert manifold, there is an additional grading due to $U(1)_\beta$, 
\begin{align}
\mathcal{H}_b	=	\bigoplus_{\substack{i \in \mathbb{Z}+\Delta_b \\ j, l \in \mathbb{Z}}} \mathcal{H}_{b}^{i, j; l}
\end{align}
and the homological block is given by
\begin{align}
\widehat{Z}_b (q,t) = \text{Tr}_{\mathcal{H}_b} q^{i} (-1)^{j} t^{l}	\label{homblock}
\end{align}
where $t$ denotes the fugacity of $U(1)_\beta$.
In this paper, we only consider the case $t=1$ of \eqref{homblock}.

The $S$-transform in \eqref{su-2-decomp} can be understood as the $S$-duality of the Type IIB string theory in the context of several dualities from the M-theory configuration \eqref{M-conf} now with $S^1$ instead of $\mathbb{R}$.
The Hilbert space $\mathcal{H}_b$ encodes the spectrum of massless BPS particles of $\mathcal{T}[M_3,G]$, which are realized as M2 branes ending on M5 branes.
If considering the $G=SU(2)$ case, at the boundary, M2 branes are wrapped on torsion 1-cycles $(b', -b')$ on $M_3$ where $[b'] = b \in \text{Tor} H_1(M_3,\mathbb{Z})/\mathbb{Z}_2$ and $b$ is interpreted as the charge of the spectrum.
Meanwhile, taking the Type IIA limit of \eqref{M-conf} on $S^1$ of $D^2$ and then T-dualizing along $S^1$ of \eqref{M-conf}, the resulting configuration becomes the D3-D5 system in the Type IIB string theory.
Taking the $S$-duality of the Type IIB string, it becomes the D3-NS5 system and the boundary condition of the system at the infinity of $\mathbb{R}_+$ is given by the connected component of the moduli space of $SL(2,\mathbb{C})$ flat connections \cite{Witten-M5knots}.
Meanwhile, since $Z_{SU(2)}(M_3)$ is written only in terms of the contributions $Z_a$ from the abelian flat connections $a$ in \eqref{su-2-decomp}, the boundary conditions labelled only by the abelian flat connections $a$ of $SU(2)$ are taken into account.\footnote{The reason why only abelian flat connections are taken into account, not all flat connections, is not known yet, though the resurgent analysis provide some explanation on it.} 
Then, the subscript $a$ of the $S$-matrix $S_{ab}$ corresponds to the connected component of the moduli space of abelian flat connections $\text{Hom}(\text{Tor} H_1(M_3,\mathbb{Z}), U(1))/\mathbb{Z}_2$, while the subscript $b$ of $S_{ab}$ denotes the M2 brane charges in the configuration \eqref{M-conf}.
We also note that the linking form provides a pairing or isomorphism between $b \in \text{Tor} H_1(M_3,\mathbb{Z})/\mathbb{Z}_2$ and $a \in (\text{Tor} H_1(M_3,\mathbb{Z}))^*/\mathbb{Z}_2:=\text{Hom}(\text{Tor} H_1(M_3,\mathbb{Z}), U(1))/\mathbb{Z}_2$,
\begin{align}
\text{Tor} H_1(M_3,\mathbb{Z})/\mathbb{Z}_2		\stackrel{lk}{\cong}	
(\text{Tor} H_1(M_3,\mathbb{Z}))^*/\mathbb{Z}_2	\cong
\pi_0 \mathcal{M}^{\text{ab}}_{\text{flat}}(M_3,SU(2))	\,	.
\end{align}
\\
\vspace{-5mm}

In this paper, we consider the $G=SU(N)$ WRT invariant for general Seifert manifolds $X(P_1/Q_1, \ldots, P_F/Q_F)$ where $P_j$ and $Q_j$ are coprime for each $j=1, \ldots, F$ and $P_j$'s are pairwise coprime.
We provide a formula \eqref{sun-a-wrt2} that calculates $Z_a$'s from which we can calculate the homological blocks exactly for $G=SU(2)$ and as a $q$-series expansion for $G=SU(N)$ from the exact expression given by Lawrence and Rozansky \cite{Lawrence-Rozansky} and Mari\~no \cite{Marino2004}.
We see in section \ref{ssec:sun-prop} that examples calculated in this paper fit into the expected structures, \eqref{sun-wrt0} and \eqref{sun-wrt1} with \eqref{sun-smat}, of the WRT invariant for Seifert manifolds in terms of homological blocks.
In addition, we find two new symmetries that the contributions from abelian flat connections to the WRT invariant are the same under the action of the center of $G$ on abelian flat connections and under complex conjugation of abelian flat connections, as discussed in section \ref{ssec:sun-prop}.
We also discuss the integral expression of Lawrence, Rozansky, and Mari\~{n}o in the context of the resurgent analysis. 
For example, when $G=SU(2)$ and $|\text{Tor} H_1(M_3, \mathbb{Z})|=1$, we see that the integral expression for the contribution of abelian flat connection $Z_0$ to the WRT invariant can be understood as the Borel resummation of the Borel transform of the perturbative expansion of $Z_0$.

The organization of this paper is as follows.
In section \ref{sec:su2f3}, we calculate homological blocks from the integral expression of the $G=SU(2)$ WRT invariant of Seifert manifolds with three singular fibers obtained by Lawrence and Rozansky, which serves as a basic example in this paper.
The calculation covers any values of $|\text{Tor} H_1(M_3, \mathbb{Z})|$, which fills a gap in the literature.
Considering the case that $|\text{Tor} H_1(M_3, \mathbb{Z})|$ is an even number, the formula for the $S$-transform \eqref{S-matrix-GPPV} is generalized. 
From the calculation, we see that the homological blocks are expressed in terms of the false theta function that is known in literature.
We also discuss the calculation performed in this paper in the context of resurgent analysis.
Then, we move on to the case of the higher rank $G=SU(N)$ in section \ref{sec:higher-rank}.
We discuss some properties of the formula that we obtained and provide some number of examples.
From those examples, we provide a general form \eqref{sun-smat} of the $S$-matrix for $G=SU(N)$ and also the structures, \eqref{sun-wrt0} and \eqref{sun-wrt1}, of the WRT invariant in terms of homological blocks.	
In Appendix \ref{sec:su2f4}, we perform similar calculations for the case of four singular fibers with $G=SU(2)$.
In this case, the homological block is expressed in terms of the false theta function of a different type that is discussed in \cite{Zagier-identity}.
We also calculate homological blocks for Seifert manifolds with more singular fibers and with the higher genus base surface when $G=SU(2)$ in Appendix \ref{sec:su2fg}.
In these cases, the homological blocks are expressed in terms of false theta functions discussed in section \ref{sec:su2f3} and Appendix \ref{sec:su2f4}, and their derivatives with respect to $q$.	\\

\noindent \textit{Note added:} While preparing the manuscript, we found that there are some partial overlaps with \cite{Cheng:2018vpl} on the $G=SU(2)$ case.


\section{$G=SU(2)$ and three singular fibers}
\label{sec:su2f3}
We consider the WRT invariant for $G=SU(2)$ and Seifert manifolds $X(P_1/Q_1, P_2/Q_2, P_3/Q_3)$ with three singular fibers with the genus of the base surface being zero where $P_j$'s are coprime to $Q_j$'s for each $j$ and also $P_j$'s are pairwise coprime.\footnote{We chose a condition that $P_j$'s are pairwise coprime because this condition was used in derivation of the integral expression of the WRT invariant in \cite{Lawrence-Rozansky, Marino2004}, which is a starting point of this paper.} 
The integer-valued $b$ of the Seifert invariants, which is (minus) the Euler number of the $S^1$ bundle of the Seifert fibration, is set to zero.\footnote{For the case that $b$ is not equal to zero, we can set it to zero by using the symmetries $b \rightarrow b \pm 1$ and $Q_j \rightarrow Q_j \mp P_j$ for a given $j$.}
The order of the torsion of the first homology group $| \text{Tor} \, H_1(M, \mathbb{Z}) |$ is given by
\begin{align}
H := P \sum_{i=1}^3 \frac{Q_i}{P_i} = \pm |\text{Tor} \, H_1(M, \mathbb{Z}) |
\end{align}
where $P=P_1 P_2 P_3$.
Due to the conditions that $P_j$'s and $Q_j$'s are coprime for each $j$ and $P_j$'s are pairwise coprime, $P$ and $H$ are coprime, which is also the case for arbitrary number of singular fibers.
We denote the quantum-corrected level as $K=k+2$ and 
\begin{align}
q := e^{\frac{2\pi i}{K}}	\,	.
\end{align}

According to \cite{Lawrence-Rozansky}, contributions from abelian flat connections including the trivial flat connection of $G=SU(2)$ to the WRT invariant on Seifert manifolds described above can be written as
\begin{align}
Z^{\text{ab}}_{SU(2)}(M_3) = \frac{B}{2\pi i} q^{-\phi_3/4} \sum_{t=0}^{H-1} \int_{\Gamma_t} dy 
\ e^{-\frac{K}{2 \pi i} \frac{H}{P} y^2 - 2 K t y} \
\frac{\prod_{j=1}^{3} e^{\frac{y}{P_j}} - e^{-\frac{y}{P_j}} }{e^{y} - e^{-y}}
\label{int-formula}
\end{align}
where
\begin{align}
B	&=	-\frac{\text{sign}P}{4 \sqrt{|P|}} e^{\frac{3}{4}\pi i \, \text{sign} \, \left( \frac{H}{P} \right)}	\,	,	\\
\phi_F	&=	3 \, \text{sign}\left( \frac{H}{P} \right) + \sum_{j=1}^{F} \left( 12 s(Q_j, P_j) - \frac{Q_j}{P_j} \right)		\,	,
\end{align}
and $s(Q,P)$ is the Dedekind sum
\begin{align}
s(Q,P) = \frac{1}{4P} \sum_{l=1}^{P-1} \cot \Big( \frac{\pi l}{P} \Big) \cot \Big( \frac{\pi Q l}{P} \Big)	\,	
\end{align}
for $P>0$, which has a property $s(-Q,P)=-s(Q,P)$.
The integration cycle $\Gamma_t$ is chosen in such a way that for each $t$ the integrand is convergent on both ends of infinity, \textit{e.g.} when $K \in \mathbb{Z}_{+}$ and $\frac{P}{H}$ is positive, $\Gamma_0$ is a line from $-(1+i)\infty$ to $(1+i)\infty$ through the origin and $\Gamma_t$ is parallel to $\Gamma_0$ and passes through $y=-2\pi i \frac{P}{H}t$, which is a stationary phase point of the integrand.
When $\frac{P}{H}$ is negative, the contour is given by a clockwise rotation of $\Gamma_0$ by $\frac{\pi}{2}$ and similarly for $\Gamma_t$.
Also, we note that the reducible flat connection in the $G=SU(2)$ case is the abelian flat connection, so we use both interchangeably for $G=SU(2)$.
Here, $t$ labels the abelian flat connections where $t=0$ corresponds to the trivial flat connection.

In \eqref{int-formula} and in the rest of paper, we use the physics normalization
\begin{align}
Z_{SU(2)}(S^1 \times S^2) = 1	\,	,	\qquad	Z_{SU(2)}(S^3) = \sqrt{\frac{2}{K}} \sin \Big(\frac{\pi}{K}\Big)	\,	.
\end{align}
Also we put an additional $1/2$ to the expression in \cite{Lawrence-Rozansky} to have the same overall coefficient with the result of \cite{Marino2004}.
We would like to express \eqref{int-formula} in terms of convergent $q$-series for the analytically continued theory.


\subsection{Calculation of the partition function}
\label{ssec:calculation}

The last factor in the integral formula \eqref{int-formula} can be expanded as
\begin{align}
\frac{\prod_{j=1}^{3} e^{\frac{y}{P_j}} - e^{-\frac{y}{P_j}} }{e^{y} - e^{-y}}
=\sum_{n=0}^{\infty} \chi_{2P}(n) e^{-\frac{n}{P} y}		\label{3f-exp}
\end{align}
when $\sum_{j=1}^{3} \frac{1}{P_j} <1$ and
\begin{align}
\frac{\prod_{j=1}^{3} e^{\frac{y}{P_j}} - e^{-\frac{y}{P_j}} }{e^{y} - e^{-y}}
= e^{y (\sum_{j=1}^{3}\frac{1}{P_j}-1) } + e^{-y (\sum_{j=1}^{3}\frac{1}{P_j}-1) }
+\sum_{n=0}^{\infty} \chi_{2P}(n) e^{-\frac{n}{P} y}		\label{3f-exp-add}
\end{align}
when $\sum_{j=1}^{3} \frac{1}{P_j} > 1$.
Here, $\chi_{2P}(n)$ is a periodic function with period $2P$
\begin{align}
\chi_{2P}(n) =
\begin{cases}
\mp \epsilon_1 \epsilon_2 \epsilon_3		&	\quad	\text{if } n \equiv P\left( 1 \pm \sum_{j=1}^3 \frac{\epsilon_j}{P_j} \right)	\quad	\text{mod } 2P	\\
0								&	\quad	\text{otherwise}
\end{cases}	\label{chi-def}
\end{align}
and $\epsilon_j=\pm1$, $j=1,2,3$.
This depends on the choice of $P_j$, but for simplicity of notation we denote it as $\chi_{2P}(n)$.
Also, we assumed $\text{Re} \, y > 0$ and $P>0$.
We will discuss the case of other possible ranges in section \ref{ssec:other-ranges}.
Then for $\sum_{j=1}^{3} \frac{1}{P_j} <1$, \eqref{int-formula} can be written as
\begin{align}
Z^{\text{ab}}_{SU(2)}(M_3) 
&= \frac{B}{2\pi i} q^{-\phi_3/4} \sum_{t=0}^{H-1} e^{2 \pi i K \frac{P}{H} t^2} \int_{\Gamma_t} dy 
\ e^{-\frac{K}{2\pi i} \frac{H}{P} \big(y + 2 \pi i \frac{P}{H}t \big)^2} \
\sum_{n=0}^{\infty} \chi_{2P}(n) e^{-\frac{n}{P} y}	\\
&= \frac{B}{2\pi i} q^{-\phi_3/4} \sum_{t=0}^{H-1} e^{2 \pi i K \frac{P}{H} t^2} \int_{\Gamma_0}  dy 
\ e^{-\frac{K}{2 \pi i} \frac{H}{P} y^2} \
\sum_{n=0}^{\infty} \chi_{2P}(n) e^{-\frac{n}{P} y} e^{2 \pi i \frac{t}{H} n }	\,	.	\label{3f-integral}
\end{align}
Here we didn't move the integration cycle from $\Gamma_t$ to $\Gamma_0$ but just changed the integration variable from $y+2\pi i \frac{P}{H}t$ to $y$ for each $t$.
We analytically continue $K$ and take $\text{Im} \, K < 0$, \textit{i.e.} $|q|<1$.
We also take $H>0$.
Then, instead of $\Gamma_0$, we choose the integration contour $\gamma$ as a line parallel to the imaginary axis of the $y$-plane that passes through $\text{Re} \, y >0$.
Calculating the integral, we obtain the partition function of the analytically continued $SU(2)$ theory for abelian flat connections,
\begin{align}
Z^{\text{ab}}_{SU(2)}(M_3)
&= \frac{B}{2 i} q^{-\phi_3/4} \bigg( \frac{2i}{K} \frac{P}{H} \bigg)^{1/2} \sum_{t=0}^{H-1} e^{2 \pi i K \frac{P}{H} t^2} \
\sum_{n=0}^{\infty} \chi_{2P}(n) e^{2 \pi i \frac{t}{H} n } q^{\frac{n^2}{4HP}}	\,	.	\label{3f-a-wrt}
\end{align}
When $\sum_{j=1}^{3} \frac{1}{P_j} > 1$, there is an additional term from $\big(e^{y(\sum_{j=1}^{3}\frac{1}{P_j}-1) } + e^{-y(\sum_{j=1}^{3}\frac{1}{P_j}-1) } \big)$ in \eqref{3f-exp-add} and this contributes to the WRT invariant as
\begin{align}
\frac{B}{2 i} q^{-\phi_3/4} \bigg( \frac{2i}{K} \frac{P}{H} \bigg)^{1/2} \sum_{t=0}^{H-1} e^{2 \pi i K \frac{P}{H} t^2} 
\big( e^{-2\pi i (\sum_{j=1}^{3}\frac{1}{P_j}-1) \frac{P}{H}t } + e^{2\pi i (\sum_{j=1}^{3}\frac{1}{P_j}-1) \frac{P}{H}t } \big) 
q^{\frac{P}{4H} \big( \sum_{j=1}^{3}\frac{1}{P_j}-1 \big)^2}	\,	.	\label{3f-add-int}
\end{align}
We note that though $Z^{\text{ab}}_{SU(2)}(M_3)$ in \eqref{3f-a-wrt} seemingly contains contributions only from abelian flat connections, it is expected to be the full partition function $Z_{SU(2)}(M_3)$ of the analytically continued theory, which contains contributions from non-abelian flat connections as transseries.
This will be discussed in section \ref{ssec:resurgent}.
So in the following discussion we will denote \eqref{3f-a-wrt} by $Z_{SU(2)}(M_3)$ instead of $Z_{SU(2)}^{\text{ab}}(M_3)$.


\subsection{The case $H=1$}
\label{ssec:su2h1}
When $H=1$, there is only a trivial flat connection and the partition function is
\begin{align}
Z_{SU(2)}(M_3)
= \frac{B}{2i} q^{-\phi_3/4} \Big( \frac{2i}{K} P \Big)^{1/2} 
\times 
\begin{cases}
\sum_{n=0}^{\infty} \chi_{2P}(n) q^{\frac{n^2}{4 P}}	&	\text{when }	\sum_{j=1}^{3} \frac{1}{P_j} < 1		\\
2q^{\frac{P}{4} \big( \sum_{j=1}^{3}\frac{1}{P_j}-1 \big)^2} + \sum_{n=0}^{\infty} \chi_{2P}(n) q^{\frac{n^2}{4 P}}	&	\text{when }	\sum_{j=1}^{3} \frac{1}{P_j} > 1		
\end{cases}	\,	.
\end{align}
It is possible to decompose $\chi_{2P}(n)$ into another periodic function $\psi^{(l)}_{2P}(n)$, 
\begin{align}
\psi^{(l)}_{2P}(n) =
\begin{cases}
\pm 1	&	\quad	\text{if } n \equiv \pm l	\quad	\text{mod } 2P	\\
0		&	\quad	\text{otherwise}		\,	,
\end{cases}	\label{psi}
\end{align}
which satisfies $\psi^{(l)}_{2P}(n)=-\psi^{(2P-l)}_{2P}(n)$.
Then we have
\begin{align}
\chi_{2P}(n) = \, & \psi_{2P}^{(P(1-(1/{P_1} + 1/{P_2} + 1/{P_3})))}(n) 	\nonumber	\\
&+ \psi_{2P}^{(P(1-(1/{P_1} - 1/{P_2} - 1/{P_3})))}(n) + \psi_{2P}^{(P(1-(-1/{P_1} + 1/{P_2} - 1/{P_3})))}(n) + \psi_{2P}^{(P(1-(-1/{P_1} - 1/{P_2} + 1/{P_3})))}(n)	\nonumber	\\
=:& \sum_{s=0}^{3} \psi_{2P}^{(R_s)}(n)	\label{3f-chi-decomp}
\end{align}
where, for simplicity, we denote $P(1-(1/{P_1} + 1/{P_2} + 1/{P_3}))$, $P(1-(1/{P_1} - 1/{P_2} - 1/{P_3}))$, $P(1-(-1/{P_1} + 1/{P_2} - 1/{P_3}))$, and $P(1-(-1/{P_1} - 1/{P_2} + 1/{P_3}))$ by $R_0, R_1, R_2$, and $R_3$, respectively.
Then, when $\sum_{j=1}^{3} \frac{1}{P_j} < 1$, the partition function can be written as
\begin{align}
Z_{SU(2)}(M_3) = \frac{B}{2i} q^{-\phi_3/4} \Big( \frac{2i}{K} P \Big)^{1/2} \sum_{s=0}^{3} \widetilde{\Psi}^{(R_s)}_{P} (q)
\end{align}
where 
\begin{align}
\widetilde{\Psi}^{(l)}_P(q) := \sum_{n=0}^{\infty} \psi_{2P}^{(l)}(n) \, q^{\frac{n^2}{4P}}	\,	.
\end{align}
It is known that $\widetilde{\Psi}^{(l)}_P(q)$ is a false theta function, which is the Eichler integral of the modular form $\Psi^{(l)}_P(q) := \sum_{n=0}^{\infty} n \psi_{2P}^{(l)}(n) q^{\frac{n^2}{4P}}$ of half-integer weight $3/2$ \cite{Lawrence-Zagier}.

The only coprime $P_j$'s that satisfy $\sum_{j=1}^{3} \frac{1}{P_j} > 1$ is $(P_1,P_2,P_3)=(2,3,5)$ for the case of three singular fibers.
Thus, in this case, the partition function is given by
\begin{align}
Z_{SU(2)}(M_3) = \frac{B}{2i} q^{-\phi_3/4} \Big( \frac{60i}{K} \Big)^{1/2}  \big( 2q^{\frac{1}{120}} -\widetilde{\Psi}^{(1)}_{30}(q)-\widetilde{\Psi}^{(11)}_{30}(q)-\widetilde{\Psi}^{(19)}_{30}(q)-\widetilde{\Psi}^{(29)}_{30}(q) \big)	\,	.	\label{sigma235}
\end{align}
The $q$-series in parenthesis takes a form of $q^{\frac{1}{120}}\mathbb{Z}[[q]]$.
This agrees with the result of \cite{Lawrence-Zagier} on the Poincar\'e homology sphere $\Sigma(2,3,5)$.
As it is an integer homology sphere, there is one homological block from the trivial flat connection, which is \eqref{sigma235}.


\subsection{The case $H \geq 2$}
\label{ssec:su2h2}
When $H \geq 2$, we have rational homology Seifert manifolds and there are contributions from other abelian flat connections in addition to the trivial flat connection.
The periodic function $\psi^{(l)}_{2P}(n)$ can be decomposed in terms of $\psi^{(l)}_{2HP}(n)$'s as
\begin{align}
\begin{split}
\psi^{(l)}_{2P}(n) &= \psi^{(l)}_{2HP}(n) - \psi^{(2P-l)}_{2HP}(n) + \cdots  - \psi^{((H-1)P-l)}_{2HP}(n) + \psi^{((H-1)P+l)}_{2HP}(n) 	\\
&= \sum_{h=0}^{\frac{H-1}{2}} \psi^{(2hP+l)}_{2HP}(n) - \sum_{h=0}^{\frac{H-1}{2}-1} \psi^{(2(h+1)P-l)}_{2HP}(n)	\,		
\label{psi-decomp-odd}
\end{split}
\end{align}
when $H$ is odd and
\begin{align}
\begin{split}
\psi^{(l)}_{2P}(n) &= \psi^{(l)}_{2HP}(n) - \psi^{(2P-l)}_{2HP}(n) + \cdots  +\psi^{((H-2)P+l)}_{2HP}(n) - \psi^{(HP-l)}_{2HP}(n)	\\
&= \sum_{h=0}^{\frac{H}{2}-1} \psi^{(2hP+l)}_{2HP}(n) - \sum_{h=0}^{\frac{H}{2}-1} \psi^{(2(h+1)P-l)}_{2HP}(n)	\,		
\label{psi-decomp-even}
\end{split}
\end{align}
when $H$ is even.
By introducing the floor and the ceiling function
\begin{align}
\lfloor x \rfloor 	&= \text{max} \{ m \in \mathbb{Z} \, | \, m \leq x \}	\,	,	\\
\lceil x \rceil 	&= \text{min} \{ m \in \mathbb{Z} \, | \, m \geq x \}	\,	,
\end{align}
\eqref{psi-decomp-odd} and \eqref{psi-decomp-even} can be written as
\begin{align}
\psi_{2P}^{(l)}(n) &= \sum_{h=0}^{\big\lceil\frac{H}{2}-1\big\rceil} \psi^{(2hP+l)}_{2HP}(n) - \sum_{h=0}^{\big\lfloor\frac{H}{2}-1\big\rfloor} \psi^{(2(h+1)P-l)}_{2HP}(n)	\,	.
\label{psi-decomp}	
\end{align}
Therefore from \eqref{3f-chi-decomp} and \eqref{psi-decomp}, 
\begin{align}
\chi_{2P}(n) = \sum_{s=0}^{3} \Big( \sum_{h=0}^{\big\lceil\frac{H}{2}-1\big\rceil} \psi^{(2hP+R_s)}_{2HP}(n) - \sum_{h=0}^{\big\lfloor\frac{H}{2}-1\big\rfloor} \psi^{(2(h+1)P-R_s)}_{2HP}(n) \Big)	\,	.
\label{chi-decomp}
\end{align}
Then, when $\sum_{j=1}^{3} \frac{1}{P_j} < 1$, the partition function is
\begin{align}
\begin{split}
Z_{SU(2)}(M_3) 
=& \frac{B}{2 i} q^{-\phi_3/4} \bigg( \frac{2i}{K} \frac{P}{H} \bigg)^{1/2} 
\Bigg[ \sum_{n=0}^{\infty} \bigg( \sum_{s=0}^{3} \Big( \sum_{h=0}^{\big\lceil\frac{H}{2}-1\big\rceil} \psi^{(2hP+R_s)}_{2HP}(n) - \sum_{h=0}^{\big\lfloor\frac{H}{2}-1\big\rfloor} \psi^{(2(h+1)P-R_s)}_{2HP}(n) \Big) \bigg) q^{\frac{n^2}{4HP}}	\,		\\
&\hspace{0mm}   + \sum_{t=1}^{H-1} e^{2 \pi i K \frac{P}{H} t^2} \
\sum_{n=0}^{\infty} e^{2 \pi i \frac{t}{H} n } \bigg( \sum_{s=0}^{3} \Big( \sum_{h=0}^{\big\lceil\frac{H}{2}-1\big\rceil} \psi^{(2hP+R_s)}_{2HP}(n) - \sum_{h=0}^{\big\lfloor\frac{H}{2}-1\big\rfloor} \psi^{(2(h+1)P-R_s)}_{2HP}(n) \Big) \bigg) q^{\frac{n^2}{4HP}}	\Bigg]	\,	
\end{split}	\label{3f-result}
\end{align}
where we put the contribution from the trivial flat connection $(t=0)$ separately. 	\\

In \eqref{3f-result}, we see that the $e^{2\pi i \frac{t}{H}n}$ factor can be taken out of the summation over $n$ when $K \in \mathbb{Z}$.
Indeed, consider 
\begin{align}
\sum_{t=1}^{H-1} e^{2\pi i K \frac{P}{H}t^2} \sum_{n=0}^{\infty} e^{2\pi i \frac{t}{H}n} \psi_{2HP}^{(l)}(n) q^{\frac{1}{4HP} n^2}	\label{3f-preresult}
\end{align}
in \eqref{3f-result}.
It is nonzero when $n=2HPm+l$ and $2HPm'-l$ with $m, m' \in \mathbb{Z}_{\geq 0}$.
Since $e^{2\pi i \frac{t}{H}(2HPm+l)} = e^{2\pi i \frac{t}{H}l}$, the $n=2HPm+l$ part of \eqref{3f-preresult} is given by
\begin{align}
\sum_{t=1}^{H-1} e^{2\pi i K \frac{P}{H}t^2} \sum_{\substack{n=2HPm+l \\ m \in \mathbb{Z}_{\geq 0}}}^{\infty} e^{2\pi i \frac{t}{H}l} \psi_{2HP}^{(l)}(n) q^{\frac{1}{4HP} n^2}	
= \sum_{t=1}^{H-1} e^{2\pi i K \frac{P}{H}t^2} e^{2\pi i \frac{t}{H}l} \sum_{\substack{n=2HPm+l \\ m \in \mathbb{Z}_{\geq 0}}}^{\infty} \psi_{2HP}^{(l)}(n) q^{\frac{1}{4HP} n^2}	\label{3f-pre-pos}	\,	.
\end{align}
While, for the $n=2HPm-l$ part of \eqref{3f-preresult}, we have
\begin{align}
\sum_{t=1}^{H-1} e^{2\pi i K \frac{P}{H}t^2} \sum_{\substack{n=2HPm-l \\ m \in \mathbb{Z}_{\geq 0}}}^{\infty} e^{-2\pi i \frac{t}{H}l} \psi_{2HP}^{(l)}(n) q^{\frac{1}{4HP} n^2}	
= \sum_{t=1}^{H-1} e^{2\pi i K \frac{P}{H}t^2} e^{-2\pi i \frac{t}{H}l} \sum_{\substack{n=2HPm-l \\ m \in \mathbb{Z}_{\geq 0}}}^{\infty} \psi_{2HP}^{(l)}(n) q^{\frac{1}{4HP} n^2}	\label{3f-pre-neg}	\,	.
\end{align}
With $t'=H-t$, \eqref{3f-pre-neg} can be written as
\begin{align}
\sum_{t'=1}^{H-1} e^{2\pi i K ( \frac{P}{H}t'^2 -2Pt' +HP )} e^{2\pi i \frac{t'}{H}l} \sum_{\substack{n=2HPm-l \\ m \in \mathbb{Z}_{\geq 0}}}^{\infty} \psi_{2HP}^{(l)}(n) q^{\frac{1}{4HP}n^2}	\label{3f-pre-neg2}	\,	.
\end{align}
If we want to obtain the WRT invariant, the limit $q \searrow e^{\frac{2\pi i}{K}}$ with $K \in \mathbb{Z}$ is taken, so $e^{2\pi i K ( \frac{P}{H}t'^2 -2Pt' +HP )}$ becomes $e^{2\pi i K \frac{P}{H}t'^2}$.
Then, we see that \eqref{3f-pre-pos} and \eqref{3f-pre-neg2} share the same $e^{2\pi i \frac{t}{H}l}$, so \eqref{3f-preresult} can be written as
\begin{align}
\sum_{t=1}^{H-1} e^{2\pi i K \frac{P}{H}t^2} e^{2\pi i \frac{t}{H}l} \sum_{n=0}^{\infty} \psi_{2HP}^{(l)}(n) q^{\frac{1}{4HP}n^2}	\,	.
\end{align}
Thus, from \eqref{3f-result}, the WRT invariant is given by
\begin{align}
\begin{split}
Z_{SU(2)}(M_3) 
=& \frac{B}{2i} q^{-\phi_3/4} \bigg( \frac{2i}{K} \frac{P}{H} \bigg)^{1/2} 
\Bigg[ \sum_{n=0}^{\infty} \bigg( \sum_{s=0}^{3} \Big( \sum_{h=0}^{\big\lceil\frac{H}{2}-1\big\rceil} \psi^{(2hP+R_s)}_{2HP}(n) - \sum_{h=0}^{\big\lfloor\frac{H}{2}-1\big\rfloor} \psi^{(2(h+1)P-R_s)}_{2HP}(n) \Big) \bigg) q^{\frac{1}{4 P H} n^2}	\,		\\
&\hspace{20mm} + \sum_{t=1}^{H-1} e^{2 \pi i K \frac{P}{H} t^2} \
\sum_{s=0}^{3} 
\bigg( \sum_{h=0}^{\big\lceil\frac{H}{2}-1\big\rceil} e^{2 \pi i \frac{t}{H} (2hP+R_s) }  \sum_{n=0}^{\infty} \psi^{(2hP+R_s)}_{2HP}(n) q^{\frac{1}{4 P H} n^2}	\\
&\hspace{40mm} -  \sum_{h=0}^{\big\lfloor\frac{H}{2}-1\big\rfloor} e^{2 \pi i \frac{t}{H} (2(h+1)P-R_s) }  \sum_{n=0}^{\infty} \psi^{(2(h+1)P-R_s)}_{2HP}(n) q^{\frac{1}{4 P H} n^2}	\bigg)	
\Bigg]	\,	\Bigg|_{q \searrow e^{\frac{2\pi i}{K}}}	\,	.	
\end{split}	\label{3f-wrt-result}
\end{align}
In terms of the false theta function $\widetilde{\Psi}_{HP}^{(a)}(q)$, when $H$ is odd, \eqref{3f-wrt-result} becomes
\begin{align}
\begin{split}
Z_{SU(2)}(M_3) 
=& \frac{B}{2i} q^{-\phi_3/4} \bigg( \frac{2i}{K} \frac{P}{H} \bigg)^{1/2} 
\Bigg[ \sum_{s=0}^{3} \Big( \sum_{h=0}^{\frac{H-1}{2}} \widetilde{\Psi}^{(2hP+R_s)}_{HP}(q) - \sum_{h=0}^{\frac{H-1}{2}-1} \widetilde{\Psi}^{(2(h+1)P-R_s)}_{HP}(q) \Big) 		\\
&   + \sum_{s=0}^{3} \sum_{t=1}^{\frac{H-1}{2}} e^{2 \pi i K \frac{P}{H} t^2} 
\bigg( \sum_{h=0}^{\frac{H-1}{2}} (e^{-2\pi i \frac{t}{H} (2hP+R_s)} + e^{2\pi i \frac{t}{H} (2hP+R_s)}) \widetilde{\Psi}^{(2hP+R_s)}_{HP}(q) 	\\ 
&\hspace{25mm}	- \sum_{h=0}^{\frac{H-1}{2}-1} (e^{-2\pi i \frac{t}{H} (2(h+1)P-R_s)} + e^{2\pi i \frac{t}{H} (2(h+1)P-R_s)}) \widetilde{\Psi}^{(2(h+1)P-R_s)}_{HP}(q) \bigg) 	\Bigg]	\,	\Bigg|_{q \searrow e^{\frac{2\pi i}{K}}}	\,	.
\end{split}
\label{3f-odd}
\end{align}
When $H$ is even,
\begin{align}
\begin{split}
Z_{SU(2)}(M_3) 
=& \frac{B}{2i} q^{-\phi_3/4} \bigg( \frac{2i}{K} \frac{P}{H} \bigg)^{1/2} 
\Bigg[ \sum_{s=0}^{3} \sum_{h=0}^{\frac{H}{2}-1} (\widetilde{\Psi}^{(2hP+R_s)}_{HP} - \widetilde{\Psi}^{(2(h+1)P-R_s)}_{HP} )		\\
& \hspace{-10mm}  + \sum_{s=0}^{3} \sum_{h=0}^{\frac{H}{2}-1} \bigg( \sum_{t=1}^{\frac{H-2}{2}} e^{2 \pi i K \frac{P}{H} t^2} \Big( (e^{-2\pi i \frac{t}{H} (2hP+R_s)} + e^{2\pi i \frac{t}{H} (2hP+R_s)}) \widetilde{\Psi}^{(2hP+R_s)}_{HP} 	\\ 
&\hspace{37mm}	-(e^{-2\pi i \frac{t}{H} (2(h+1)P-R_s)} + e^{2\pi i \frac{t}{H} (2(h+1)P-R_s)}) \widetilde{\Psi}^{(2(h+1)P-R_s)}_{HP} \Big) 	\bigg)	\\
&\hspace{15mm}	+e^{\frac{\pi i}{2} K PH} \Big( e^{\pi i (2hP+R_s)} \widetilde{\Psi}^{(2hP+R_s)}_{HP} - e^{\pi i (2(h+1)P-R_s)} \widetilde{\Psi}^{(2(h+1)P-R_s)}_{HP} \Big)	
\Bigg]	\,	\Bigg|_{q \searrow e^{\frac{2\pi i}{K}}}	\,	.
\end{split}
\label{3f-even}
\end{align}
Here, we write the expression as a sum over distinct Weyl orbits where Weyl reflection is given by $t \leftrightarrow -t \equiv H-t \ \text{mod}\, H$.
Meanwhile, for $\sum_{j=1}^{3} \frac{1}{P_j} > 1$, there is an additional term \eqref{3f-add-int}, which is
\begin{align}
\begin{split}
&\frac{B}{2i} q^{-\phi_3/4} \bigg( \frac{2i}{K} \frac{P}{H} \bigg)^{1/2}
\Bigg[
2  q^{\frac{P}{4H} \Big( \sum_{j=1}^{3}\frac{1}{P_j}-1 \Big)^2}	\\
&\hspace{20mm} + \sum_{t=1}^{\frac{H-1}{2}} e^{2 \pi i K \frac{P}{H} t^2} 	
\Big( e^{-2\pi i (\sum_{j=1}^{3}\frac{1}{P_j}-1) \frac{P}{H} t } + e^{2\pi i (\sum_{j=1}^{3}\frac{1}{P_j}-1) \frac{P}{H} t } \Big) 
q^{\frac{P}{4H} \Big( \sum_{j=1}^{3}\frac{1}{P_j}-1 \Big)^2}	
\Bigg]	\,	\Bigg|_{q \searrow e^{\frac{2\pi i}{K}}}
\end{split}
\end{align}
when $H$ is odd and
\begin{align}
\begin{split}
&\frac{B}{2i} q^{-\phi_3/4} \bigg( \frac{2i}{K} \frac{P}{H} \bigg)^{1/2}
\Bigg[
2  q^{\frac{P}{4H} \Big( \sum_{j=1}^{3}\frac{1}{P_j}-1 \Big)^2}	\\
&\hspace{20mm} + \sum_{t=1}^{\frac{H-2}{2}} e^{2 \pi i K \frac{P}{H} t^2} 	
\Big( e^{-2\pi i (\sum_{j=1}^{3}\frac{1}{P_j}-1) \frac{P}{H} t } + e^{2\pi i (\sum_{j=1}^{3}\frac{1}{P_j}-1) \frac{P}{H} t } \Big) 
q^{\frac{P}{4H} \Big( \sum_{j=1}^{3}\frac{1}{P_j}-1 \Big)^2}	\\
&\hspace{20mm} + e^{\frac{\pi i}{2} K PH} 	
\Big( e^{- \pi i P (\sum_{j=1}^{3}\frac{1}{P_j}-1) } + e^{\pi i P (\sum_{j=1}^{3}\frac{1}{P_j}-1) } \Big) 
q^{\frac{P}{4H} \Big( \sum_{j=1}^{3}\frac{1}{P_j}-1 \Big)^2}	
\Bigg]	\,	\Bigg|_{q \searrow e^{\frac{2\pi i}{K}}}
\end{split}
\end{align}
when $H$ is even.


\subsection{Properties of the formula}
\label{ssec:su2-prop}

Before providing some examples, we study properties of \eqref{3f-a-wrt} or \eqref{3f-wrt-result}.

The variable or the label $t$ in \eqref{3f-result} is regarded as $t$ of the diagonal matrix $\text{diag} \, (t, -t) \in (\mathbb{Z}_H)^2$, which gives the holonomy $\text{diag} \, (e^{2 \pi i \frac{P}{H}t}, e^{-2 \pi i \frac{P}{H}t})$ where we note that $H$ and $P$ are coprime.
So the Weyl group action on $(t,-t)$ gives $(-t,t)$. 
The Weyl orbit of $(t,-t)$ corresponds to the abelian flat connection where $t=0$, in particular, corresponds to the trivial flat connection \cite{Lawrence-Rozansky}, \textit{c.f.} section \ref{sssec:flat-connections}.
Given a $t$, we see that the summand in \eqref{3f-odd} and \eqref{3f-even} is invariant under $t \leftrightarrow -t$.
We note that $t \leftrightarrow -t$ provide the complex conjugate of $(t,-t)$ at the level of holonomy.
Thus, it can be said that the contributions from the abelian flat connection corresponding to the Weyl orbit of $(t,-t)$ and from the conjugate abelian flat connection corresponding to the Weyl orbit of $(-t,t)$ are the same though in the case of $SU(2)$ these two abelian flat connections are equivalent so that contributions from them are obviously the same from the beginning.		\\

In addition, the abelian flat connections that are related by the action of the center of $SU(2)$ give the same contribution.
The center of $SU(2)$ is given by $e^{2 \pi i \frac{c}{2}} I_2$, $c \in \mathbb{Z}_2$, which we denote by $(c,-c)$ at the level of $(\mathbb{Z}_2)^2/\mathbb{Z}_2$.
As we see below, there are cases that elements in $(\mathbb{Z}_H)^2/\mathbb{Z}_2$ are related by the nontrivial center, $(1,-1) \equiv (1,1) \ \text{mod } (\mathbb{Z}_2)^2$ and it can only be possible when $H$ is even.
That is because in order to relate them the center should also be expressed as $(m,-m) \in (\mathbb{Z}_H)^2$, which gives the holonomy $\text{diag} \, (e^{2\pi i \frac{P}{H}m},e^{-2\pi i \frac{P}{H}m})$, and the nontrivial center is given by $(\frac{H}{2}, -\frac{H}{2}) \in (\mathbb{Z}_H)^2$.
Therefore, if $H$ is odd then $m=\frac{H}{2}$ is not an integer, so $H$ should be even.

When $H$ is even, upon $t \rightarrow t+ \frac{H}{2}$, $e^{2\pi i \frac{l}{H}t} + e^{-2\pi i \frac{l}{H}t}$ gets an additional factor $e^{\pi i l}$.
From the conditions that $P_j$'s and $Q_j$'s are coprime for each $j$ and $P_j$'s are coprime to each other, $P_j$ should be all odd when $H$ is even, so $l$'s are always even for the case of three singular fibers $F=3$.
$l$ is also even for other number of singular fibers and the higher genus case when $H$ is even, considering \eqref{sinh-gen} and \eqref{cosh-gen} in section \ref{sec:su2fg}.
Hence, $e^{\pi i l}=1$, so the abelian flat connections that are related by the action of the center have the same $e^{2\pi i \frac{l}{H}t} + e^{-2\pi i \frac{l}{H}t}$, so contributions from them are the same up to the $e^{2\pi i K \frac{P}{H} t^2}$ factor.

For the $e^{2\pi i K \frac{P}{H} t^2}$ factor, upon $t \rightarrow t+ \frac{H}{2}$, we have an additional factor $e^{\pi i K \frac{PH}{2}}$ when $K \in \mathbb{Z}$.
Therefore, when $H$ is a multiple of 2 but not of 4, there is an additional factor $e^{\pi i K}$ for the abelian flat connection $(t+\frac{H}{2}, -t-\frac{H}{2})$ compared to the case of $(t,-t)$.
There is no such factor when $H$ is a multiple of 4.
Thus, the contributions from abelian flat connections that are related by the action of the center can have a different factor by $e^{\pi i K}$ when $H$ is a multiple of 2 but not of 4.	\\

We denote the Weyl orbit of $(t,-t)$ in $(\mathbb{Z}_H)^2/\mathbb{Z}_2$ as $W_{t}$.
When $H$ is even, Weyl orbits $W_t$ and $W_{t+\frac{H}{2}}$ that are related by the action of the center give the same contribution to the WRT invariant up to an overall factor $e^{\pi i K}$, so we group $W_t$ and $W_{t+\frac{H}{2}}$ by orbits under the action of the center, which we denote by $C_a$ where $a$ is a label for the abelian flat connection.
The range of $a$ is $a=0,1, \ldots, \frac{H-2}{4}$ when $H$ is a multiple of 2 but not of 4, and $a=0,1, \ldots, \frac{H}{4}$ when $H$ is a multiple of 4.
We also denote elements in the Weyl orbit $W_t$ by $\tilde{t}$ and a representative of any of $W_t$ in $C_b$ by $\tilde{b}$.

With the setup above, the $S$-matrix can be written as
\begin{align}
S_{ab}= \frac{1}{ \sqrt{\text{gcd}(2,H)} }\sum_{W_t \in C_a} \frac{ \sum_{\tilde{t} \in W_t} e^{2\pi i  lk (\tilde{t},\tilde{b})}}{ |\text{Tor} \, H_1(M_3,\mathbb{Z})|^{\frac{1}{2}}}	\label{su2-smat0}
\end{align}
with
\begin{align}
lk(t,t') = \frac{P}{H} \sum_{j=1}^2 t_j t'_j = \frac{2P}{H} t_1 t'_1	\label{su2-lk}
\end{align}
where $t=(t_1,-t_1)$ and $t'=(t'_1,-t'_1)$.\footnote{Or the $S$-matrix can also be written as 
\begin{align}
S_{ab}= \frac{1}{\sqrt{\text{gcd} (2, H) }}  \sum_{W_t \in C_a} \frac{e^{2\pi i lk (\tilde{t},\tilde{b})} + e^{-2\pi i lk (\tilde{t},\tilde{b})}}{|\text{Stab}_{\mathbb{Z}_2}(a)| |\text{Tor} \, H_1(M_3,\mathbb{Z})|^{\frac{1}{2}}}	\label{su2-smat1}
\end{align}
where $\text{Stab}_{\mathbb{Z}_2}(a)$ is $\mathbb{Z}_2$ if $a\equiv-a$ mod $\mathbb{Z}_H$ or is $1$ otherwise.
}
We often use the notation $lk(a,b) := lk(\tilde{a}, \tilde{b})$.
The overall normalization of $S_{ab}$ was chosen so that it satisfies the condition $S^2=I$, which is natural when considering that the $S$-matrix corresponds to the $S$-duality of Type IIB string theory as discussed in the introduction.
When $H$ is odd, there is only a single Weyl orbit $W_t$ in $C_a$, which we denote by $W_a$, as there is no non-trivial center that relates $(t,-t)$'s.
So in that case, we simply have
\begin{align}
S_{ab}= \frac{ \sum_{\tilde{t} \in W_a} e^{2\pi i lk (\tilde{t},\tilde{b})}}{ |\text{Tor} \, H_1(M_3,\mathbb{Z})|^{\frac{1}{2}}}		\,	,	\label{su2-smat2}
\end{align}
which agrees with the $S$-matrix in \cite{Gukov-Putrov-Vafa, Gukov-Pei-Putrov-Vafa} for odd $H$.

Then, from the examples considered, the WRT invariant is given by
\begin{align}
Z_{SU(2)}(M_3) = \frac{B}{2i} q^{-\phi/4} (-2K)^{\frac{1}{2} b_1(M_3)} \bigg( \frac{2i}{K} \frac{P}{H} \bigg)^{1/2} \sqrt{\text{gcd} (2,H) H} \sum_{a,b} e^{\pi i K lk(a,a)} S_{ab} \widehat{Z}_b(q)	\,	\Big|_{q \searrow e^{\frac{2\pi i}{K}}}	\label{su2-wrt0}
\end{align}
when $H$ is odd or a multiple of 4.
Here, $b_1(M_3)$ is the first Betti number, which is $2g$ in our setup of Seifert manifolds when the genus of the base surface is $g$.
The linking form $lk(a,a)$ of $a$ is the Chern-Simons invariant for the abelian flat connection $a$, so we denote it as $CS_a = \frac{1}{2} lk(a,a)$.
We also note that the factor $e^{2\pi i K \frac{P}{H}t^2}$ in \eqref{3f-odd} and \eqref{3f-even} can be written as $e^{\pi i K lk(t,t)}$.

When $H$ is a multiple of 2 but not of 4, the WRT invariant can be written as
\begin{align}
Z_{SU(2)}(M_3) = \frac{B}{2i} (-2K)^{\frac{1}{2}b_1(M_3)} q^{-\phi/4} \bigg( \frac{2i}{K} \frac{P}{H} \bigg)^{1/2} \sqrt{ \frac{H}{2}} \sum_{\dot{a},\dot{b}} e^{\pi i K lk(\dot{a},\dot{a})} (Y \otimes S_{ab})_{\dot{a}\dot{b}} \widehat{Z}_{\dot{b}}(q)	\,	\Big|_{q \searrow e^{\frac{2\pi i}{K}}}	
\label{su2-wrt1}
\end{align}
where $\dot{a}, \dot{b}= \dot{0}, \ldots, \dot{\frac{H}{2}}$, $Y= \begin{pmatrix} 1 & 1 \\ 1 & 1 \end{pmatrix}$ so that $(Y \otimes S_{ab})_{\dot{a}\dot{b}}=\begin{pmatrix} S_{ab} & S_{ab} \\ S_{ab} & S_{ab} \end{pmatrix}$, and $e^{\pi i K lk(\dot{a},\dot{a})}= (e^{\pi i K lk(\dot{0},\dot{0})},$ $\ldots, e^{\pi i K lk(\dot{\frac{H-2}{4}},\dot{\frac{H-2}{4})}},$ $e^{\pi i K(lk(\dot{0},\dot{0})+1)}, \ldots, e^{\pi i K (lk(\dot{\frac{H-2}{4}},\dot{\frac{H-2}{4}})+1)})$.
Since $Z_a = Z_{\frac{H}{2}-a}$ for $a=0,1, \ldots, \frac{H-2}{4}$, in the notation of \eqref{su2-wrt1} we have $Z_{\dot{a}}(q) = Z_{\dot{a+\frac{H+2}{4}}}(q)$, $\dot{a}=\dot{0}, \ldots, \dot{\frac{H-2}{4}}$ with $Z_{\dot{a}}(q) = (Y \otimes S_{ab})_{\dot{a}\dot{b}} \widehat{Z}_{\dot{b}}(q)$ where $Z_{\dot{a}}(q) = Z_{a}(q)$ for $a=0,1, \ldots, \frac{H-2}{4}$ and $Z_{\dot{a}}(q) = Z_{\frac{3H+2}{4}-a}(q)$ for $a= \frac{H}{2}, \frac{H}{2}-1, \ldots, \frac{H+2}{4}$.
Similarly, we denote $\widehat{Z}_{\dot{b}}(q) = \widehat{Z}_{b}(q)$ for $b=0,1, \ldots, \frac{H-2}{4}$ and $\widehat{Z}_{\dot{b}}(q) = \widehat{Z}_{\frac{3H+2}{4}-b}(q)$ for $b= \frac{H}{2}, \frac{H}{2}-1, \ldots, \frac{H+2}{4}$.
One can also obtain $(Y \otimes S_{ab})_{\dot{a}\dot{b}}$ in \eqref{su2-wrt1} directly from \eqref{su2-smat2} and \eqref{su2-lk} where indices $a$ and $b$ in \eqref{su2-smat2} run over $a,b=0,1, \ldots, \frac{H-2}{4}, \frac{H}{2}, \frac{H}{2}-1, \ldots, \frac{H+2}{4}$.


\subsection{Some examples with homological blocks}
\label{ssec:su2-ex}
From the expression \eqref{3f-odd} and \eqref{3f-even} above, we provide some concrete examples with homological blocks.

\subsubsection*{\textbullet \ $H=2$}
We can have $H=2$, for example, from a choice $(P_1,P_2,P_3) = (3,5,7)$ and $(Q_1,Q_2,Q_3)=(1,2,-5)$.
Then the WRT invariant is given by
\begin{align}
\begin{split}
Z_{SU(2)}(M_3) 
=& \frac{B}{2i} q^{-\phi_3/4} \bigg( \frac{105i}{K}  \bigg)^{1/2} 
\Bigg[ \sum_{s=0}^{3} (\widetilde{\Psi}^{(R_s)}_{210}(q) - \widetilde{\Psi}^{(210-R_s)}_{210}(q) )		\\
& \hspace{50mm}  + e^{\pi i K} \sum_{s=0}^{3} \Big( \widetilde{\Psi}^{(R_s)}_{210}(q) - \widetilde{\Psi}^{(210-R_s)}_{210}(q) \Big) 
\Bigg]	\Bigg|_{q \searrow e^{\frac{2\pi i}{K}}}	
\end{split}	\label{3f-2h}
\end{align}
where $R_0=34, R_1=106, R_2=134$, and $R_3=146$.
Contributions come from the trivial flat connection $(0,0)$ and the central flat connection $(1,-1)$ in $(\mathbb{Z}_2)^2/\mathbb{Z}_2$.
Thus, in this case, we have $W_0 = \{ (0,0) \}$ and $W_{1}= \{ (1,-1) \}$, and $C_0 = \{ W_0, W_1\}$.
From \eqref{3f-2h}, their contributions $Z_0(q)$ and $Z_1(q)$ are the same,
\begin{align}
\hspace{-5mm}Z_0(q) = Z_1(q) &= \sum_{s=0}^{3} (\widetilde{\Psi}^{(R_s)}_{210}(q) - \widetilde{\Psi}^{(210-R_s)}_{210}(q) )	\\
\begin{split}
&=\widetilde{\Psi }_{210}^{(34)}+\widetilde{\Psi }_{210}^{(106)}+\widetilde{\Psi }_{210}^{(134)}+\widetilde{\Psi }_{210}^{(146)}	
-\widetilde{\Psi }_{210}^{(64)}-\widetilde{\Psi }_{210}^{(76)}-\widetilde{\Psi }_{210}^{(104)}-\widetilde{\Psi }_{210}^{(176)}
\end{split}
\end{align}
where we denote $\widetilde{\Psi}^{(a)}_{HP}(q)$ by $\widetilde{\Psi}^{(a)}_{HP}$ for simplicity.
Then \eqref{3f-2h} is written as
\begin{align}
Z_{SU(2)}(M_3) 
= \frac{B}{2i} q^{-\phi_3/4} \bigg( \frac{105i}{K}  \bigg)^{1/2} (1+ e^{\pi i K}) Z_0(q)	\,	\bigg|_{q \searrow e^{\frac{2\pi i}{K}}}		\,	.
\end{align}
Homological blocks are
\begin{align}
\widehat{Z}_0(q) &= \widetilde{\Psi }_{210}^{(34)}+\widetilde{\Psi }_{210}^{(106)}+\widetilde{\Psi }_{210}^{(134)}+\widetilde{\Psi }_{210}^{(146)}	\,	,	\\
\widehat{Z}_1(q) &= -\widetilde{\Psi }_{210}^{(64)}-\widetilde{\Psi }_{210}^{(76)}-\widetilde{\Psi }_{210}^{(104)}-\widetilde{\Psi }_{210}^{(176)}
\end{align}
where $\widehat{Z}_0 = q^{\frac{79}{210}} \mathbb{Z}[[q]]$ and $\widehat{Z}_1 = q^{\frac{92}{105}} \mathbb{Z}[[q]]$.
So in terms of homological blocks, the WRT invariant can be expressed as 
\begin{align}
Z_{SU(2)}(M) = \frac{B}{2i} q^{-\phi_3/4} \bigg( \frac{105i}{K}  \bigg)^{1/2} (1+ e^{\pi i K}) (\widehat{Z}_0 + \widehat{Z}_1)	\,	\bigg|_{q \searrow e^{\frac{2\pi i}{K}}}	\,	.
\end{align}
Or we put it in the form of
\begin{align}
Z_{SU(2)}(M) = \frac{B}{2i} q^{-\phi_3/4} \bigg( \frac{105i}{K}  \bigg)^{1/2} \sum_{a,b=0}^{1} e^{2 \pi i K CS_a} Y_{ab} \widehat{Z}_b	\,	\bigg|_{q \searrow e^{\frac{2\pi i}{K}}}	\,	,
\end{align}
where $Y_{ab} = \begin{pmatrix} 1 & 1 \\ 1 & 1 \end{pmatrix}$ and $(CS_0,CS_1) = (0,\frac{1}{2})$.
This ``$S$"-matrix $Y_{ab}$ can also be obtained from \eqref{su2-smat2} and \eqref{su2-lk} with $W_0$ and $W_1$.
However, it doesn't satisfy $Y^2=1$ and actually there is no $S$-matrix giving $Z_a = \sum_{b} S_{ab} \widehat{Z}_b$ that satisfy $S^2=1$ at the same time.
This case $H=2$, $\text{Tor}H_1(M_3,\mathbb{Z}) = \mathbb{Z}_2$, was not a part of conjecture in \cite{Gukov-Putrov-Vafa, Gukov-Pei-Putrov-Vafa} where it was assumed that $\mathbb{Z}_2$ doesn't appear in $\text{Tor}H_1(M_3,\mathbb{Z})$.\footnote{Recently, some examples with even $H$ were also considered in \cite{Cheng:2018vpl} with consideration on the action of the center of $SU(2)$.}

\subsubsection*{\textbullet \ $H=3$}
$H=3$ can be obtained, for example, by choosing $(P_1,P_2,P_3)=(2,5,7)$ and $(Q_1,Q_2,Q_3)=(1,2,-6)$.
Then, the WRT invariant is given by
\begin{align}
\begin{split}
Z_{SU(2)}(M_3) 
=& \frac{B}{2i} q^{-\phi_3/4} \bigg( \frac{140i}{3K} \bigg)^{1/2} 
\Bigg[ \sum_{s=0}^{3} \Big( \sum_{h=0}^{1} \widetilde{\Psi}^{(140h+R_s)}_{210}(q) - \widetilde{\Psi}^{(140-R_s)}_{210}(q) 	\Big)	\\
&   + e^{\frac{2 \pi i}{3} K}  \sum_{s=0}^{3} 
\bigg( \sum_{h=0}^{1} (e^{-\frac{2\pi i}{3} (140h+R_s)} + e^{\frac{2\pi i}{3} (140h+R_s)}) \widetilde{\Psi}^{(140h+R_s)}_{210}(q) 	\\ 
&\hspace{30mm}	-(e^{- \frac{2 \pi i}{3} (140-R_s)} + e^{\frac{2 \pi i}{3}(140-R_s)}) \widetilde{\Psi}^{(140-R_s)}_{210}(q) \bigg) 	\Bigg]	\,	\Bigg|_{q \searrow e^{\frac{2\pi i}{K}}}	
\end{split}
\end{align}
where $R_0=11, R_1=59, R_2=101$, and $R_3=109$.
There are two abelian flat connections; the trivial flat connection $W_0=\{(0,0)\}$ and the abelian flat connection $W_1=\{(1,-1),(2,-2) \equiv (-1,1)\}$ in $(\mathbb{Z}_3)^2/\mathbb{Z}_2$.
From the above expression, $Z_a$'s are, respectively, 
\begin{align}
\begin{split}
Z_0(q) =& \,
\widetilde{\Psi }_{210}^{(11)}-\widetilde{\Psi }_{210}^{(31)}-\widetilde{\Psi }_{210}^{(39)}+\widetilde{\Psi }_{210}^{(59)}-\widetilde{\Psi }_{210}^{(81)}+\widetilde{\Psi}_{210}^{(101)}	\\
&+\widetilde{\Psi }_{210}^{(109)}-\widetilde{\Psi }_{210}^{(129)}+\widetilde{\Psi }_{210}^{(151)}+\widetilde{\Psi }_{210}^{(199)}+\widetilde{\Psi}_{210}^{(241)}+\widetilde{\Psi }_{210}^{(249)}	\,	,
\end{split}	\\
\begin{split}
Z_1(q) =&
-\widetilde{\Psi }_{210}^{(11)}+\widetilde{\Psi }_{210}^{(31)}-2 \widetilde{\Psi }_{210}^{(39)}-\widetilde{\Psi }_{210}^{(59)}-2 \widetilde{\Psi }_{210}^{(81)}-\widetilde{\Psi}_{210}^{(101)}	\\
&-\widetilde{\Psi }_{210}^{(109)}-2 \widetilde{\Psi }_{210}^{(129)}-\widetilde{\Psi }_{210}^{(151)}-\widetilde{\Psi }_{210}^{(199)}-\widetilde{\Psi}_{210}^{(241)}+2 \widetilde{\Psi }_{210}^{(249)}	\,	.
\end{split}
\end{align}
There are two homological blocks, $\widehat{Z}_0$ and $\widehat{Z}_1$,
\begin{align}
\widehat{Z}_0 &= 
-\widetilde{\Psi }_{210}^{(39)}-\widetilde{\Psi }_{210}^{(81)}-\widetilde{\Psi }_{210}^{(129)}+\widetilde{\Psi }_{210}^{(249)}	\,	,	\\
\begin{split}
\widehat{Z}_1 &= 
\widetilde{\Psi }_{210}^{(11)}-\widetilde{\Psi }_{210}^{(31)}+\widetilde{\Psi }_{210}^{(59)}+\widetilde{\Psi }_{210}^{(101)}+\widetilde{\Psi }_{210}^{(109)}+\widetilde{\Psi}_{210}^{(151)}+\widetilde{\Psi }_{210}^{(199)}+\widetilde{\Psi }_{210}^{(241)}
\end{split}
\end{align}
where $\widehat{Z}_0 = q^{\frac{227}{280}} \mathbb{Z}[[q]]$ and $\widehat{Z}_1 = q^{\frac{121}{840}} \mathbb{Z}[[q]]$.
From this, we obtain the $S$-matrix 
\begin{align}
S_{ab} =\frac{1}{\sqrt{3}} 
\begin{pmatrix}
1	&	1	\\
2	&	-1
\end{pmatrix}	\,	.	\label{s-mat-su2-h3}
\end{align}
Or the $S$-matrix can also be calculated from \eqref{su2-smat0} and \eqref{su2-lk} and it agrees with \eqref{s-mat-su2-h3}.
Therefore, the WRT invariant in this case can be written as
\begin{align}
\begin{split}
Z_{SU(2)_K}(M) 
=& \frac{B}{i} q^{-\phi_3/4} \bigg( \frac{70i}{K} \bigg)^{1/2} 
\sum_{a,b=0}^{1} e^{2 \pi i K CS_a }S_{ab} \widehat{Z}_{b}	\,	\bigg|_{q \searrow e^{\frac{2\pi i}{K}}}	
\end{split}
\end{align}
with $(CS_0,CS_1)=(0,\frac{1}{3})$.

\subsubsection*{\textbullet \ $H=4$}
We can have $H=4$ by taking, for example, $(P_1,P_2,P_3)=(3,5,7)$ and $(Q_1,Q_2,Q_3)=(1,-4,-3)$.
In this case, there are three abelian flat connections, $W_0=\{(0,0)\}$, $W_1=\{(1,-1),(3,-3)\}$, and $W_2=\{(2,-2)\}$ in $(\mathbb{Z}_4)^2/\mathbb{Z}_2$.

The WRT invariant is given by
\begin{align}
\begin{split}
Z_{SU(2)}(M) =
& \frac{B}{2i} q^{-\phi_3/4} \bigg( \frac{105i}{2K} \bigg)^{1/2} 
\big( Z_0 + e^{\frac{\pi i}{2} K} Z_1 + Z_2 \big)	\,	\bigg|_{q \searrow e^{\frac{2\pi i}{K}}}		\,	.
\label{3f-h4-result0}
\end{split}
\end{align}
Here we denoted the contributions from $W_0$, $W_1$, and $W_2$ by $Z_0$, $Z_1$, and $Z_2$, respectively, which are
\begin{align}
Z_0=&
\sum_{s=0}^{3} \sum_{h=0}^{1} (\widetilde{\Psi}^{(210h+R_s)}_{420}(q) - \widetilde{\Psi}^{(210(h+1)-R_s)}_{420}(q) )	\,	,	\\
Z_1=&
\sum_{s=0}^{3} \sum_{h=0}^{1} 
\Big( 2 \cos \big( \frac{\pi}{2} (210h+R_s) \big) \, \widetilde{\Psi}^{(210h+R_s)}_{420}(q) 	
-2\cos \big( \frac{\pi}{2} (210(h+1)-R_s) \big) \, \widetilde{\Psi}^{(210(h+1)-R_s)}_{420}(q) \Big)	\,	, 	\\
Z_2=& \sum_{s=0}^{3} \sum_{h=0}^{1} \Big( \widetilde{\Psi}^{(210h+R_s)}_{420}(q) - \widetilde{\Psi}^{(210(h+1)-R_s)}_{420}(q) \Big) = Z_0
\label{3f-ex-4}
\end{align}
where $R_0=34, R_1=106, R_2=134$, and $R_3=146$.
More specifically,
\begin{align}
\begin{split}
Z_0 =& \, Z_2  \\
=& \, \widetilde{\Psi }_{420}^{(34)}-\widetilde{\Psi }_{420}^{(64)}-\widetilde{\Psi }_{420}^{(76)}-\widetilde{\Psi }_{420}^{(104)}+\widetilde{\Psi }_{420}^{(106)}+\widetilde{\Psi}_{420}^{(134)}+\widetilde{\Psi }_{420}^{(146)}-\widetilde{\Psi }_{420}^{(176)}	\\
&+\widetilde{\Psi }_{420}^{(244)}-\widetilde{\Psi }_{420}^{(274)}-\widetilde{\Psi}_{420}^{(286)}-\widetilde{\Psi }_{420}^{(314)}+\widetilde{\Psi }_{420}^{(316)}+\widetilde{\Psi }_{420}^{(344)}+\widetilde{\Psi }_{420}^{(356)}-\widetilde{\Psi}_{420}^{(386)}	\,	,
\end{split}	\\
\begin{split}
\frac{1}{2}Z_1 =& 
-\widetilde{\Psi }_{420}^{(34)}-\widetilde{\Psi }_{420}^{(64)}-\widetilde{\Psi }_{420}^{(76)}-\widetilde{\Psi }_{420}^{(104)}-\widetilde{\Psi }_{420}^{(106)}-\widetilde{\Psi}_{420}^{(134)}-\widetilde{\Psi }_{420}^{(146)}-\widetilde{\Psi }_{420}^{(176)}	\\
&+\widetilde{\Psi }_{420}^{(244)}+\widetilde{\Psi }_{420}^{(274)}+\widetilde{\Psi}_{420}^{(286)}+\widetilde{\Psi }_{420}^{(314)}+\widetilde{\Psi }_{420}^{(316)}+\widetilde{\Psi }_{420}^{(344)}+\widetilde{\Psi }_{420}^{(356)}+\widetilde{\Psi}_{420}^{(386)}	\,	.
\end{split}
\end{align}
$W_0$ and $W_2$ are in the same orbit $C_0 = \{ W_0, W_2 \}$ by the center, and we see that $Z_0$ and $Z_2$ are the same as discussed in section \ref{ssec:su2-prop}.

From $Z_a$'s, the homological blocks are given by
\begin{align}
\begin{split}
\widehat{Z}_0 =
-\widetilde{\Psi }_{420}^{(64)}-\widetilde{\Psi }_{420}^{(76)}-\widetilde{\Psi }_{420}^{(104)}-\widetilde{\Psi }_{420}^{(176)}+\widetilde{\Psi }_{420}^{(244)}+\widetilde{\Psi
   }_{420}^{(316)}+\widetilde{\Psi }_{420}^{(344)}+\widetilde{\Psi }_{420}^{(356)}	\,	,
\end{split}	\\
\begin{split}
\widehat{Z}_1 =
\widetilde{\Psi }_{420}^{(34)}+\widetilde{\Psi }_{420}^{(106)}+\widetilde{\Psi }_{420}^{(134)}+\widetilde{\Psi }_{420}^{(146)}-\widetilde{\Psi }_{420}^{(274)}-\widetilde{\Psi
   }_{420}^{(286)}-\widetilde{\Psi }_{420}^{(314)}-\widetilde{\Psi }_{420}^{(386)}
\end{split}
\end{align}
where $\widehat{Z}_0 = q^{\frac{46}{105}} \mathbb{Z}[[q]]$ and $\widehat{Z}_1 = q^{\frac{289}{420}} \mathbb{Z}[[q]]$.
So we have $Z_0 = \widehat{Z}_0 + \widehat{Z}_1$ and $\frac{1}{2}Z_1 = \widehat{Z}_0 - \widehat{Z}_1$.
Thus, the $S$-matrix is given by
\begin{align}S_{ab} = 
\frac{1}{\sqrt{2}}\begin{pmatrix}
1	&	1	\\	1	&	-1
\end{pmatrix}	\,	.	\label{s-mat-su2-h4}
\end{align}
This can also be calculated from \eqref{su2-smat0} and \eqref{su2-lk} with $C_0 = \{ W_0, W_2 \}$ and $C_1 = \{ W_1 \}$.
Thus, \eqref{3f-h4-result0} can be written by
\begin{align}
\begin{split}
Z_{SU(2)}(M_3) 
=& \frac{B}{i} q^{-\phi_3/4} \bigg( \frac{105 i}{K} \bigg)^{1/2} \sum_{a,b=0}^{1} e^{2 \pi i K CS_a} S_{ab} \widehat{Z}_{b}	\,	\bigg|_{q \searrow e^{\frac{2\pi i}{K}}}	
\end{split}
\end{align}
where $(CS_0,CS_1)=(0,\frac{1}{4})$.

\subsubsection*{\textbullet \ $H=5$}
We take, for example, $(P_1,P_2,P_3)=(2,3,7)$ and $(Q_1,Q_2,Q_3)=(1,1,-5)$ to have $H=5$.
There are three abelian flat connections $W_0=\{(0,0)\}$, $W_1=\{(1,-1),(4,-4)\}$, and $W_2=\{(2,-2),(3,-3)\}$ and we denote their contributions by $Z_0$, $Z_1$, and $Z_2$, respectively.
Then the WRT invariant is
\begin{align}
Z_{SU(2)}(M_3) 
=& \frac{B}{2i} q^{-\phi_3/4} \bigg( \frac{84i}{5K} \bigg)^{1/2} 
\big( Z_0 + e^{\frac{4 \pi i}{5} K} Z_1 +e^{\frac{6 \pi i}{5} K} Z_2 \big)	\,	\bigg|_{q \searrow e^{\frac{2\pi i}{K}}}	
\label{3f-h5-result0}
\end{align}
with 
\begin{align}
\begin{split}
Z_0 &= \sum_{s=0}^{3} \Big( \sum_{h=0}^{2} \widetilde{\Psi}^{(84h+R_s)}_{210}(q) - \sum_{h=0}^{1} \widetilde{\Psi}^{(84(h+1)-R_s)}_{210}(q) \Big) 	\,	,
\end{split}	\\
\begin{split}
\hspace{-10mm}Z_1 &= \sum_{s=0}^{3} \Big( \sum_{h=0}^{2} (e^{-\frac{2\pi i}{5} (84h+R_s)} + e^{\frac{2\pi i}{5} (84h+R_s)}) \widetilde{\Psi}^{(84h+R_s)}_{210}(q) 	\\ 
&\hspace{20mm}	- \sum_{h=0}^{1} (e^{-\frac{2\pi i}{5}(84(h+1)-R_s)} + e^{\frac{2\pi i}{5} (84(h+1)-R_s)}) \widetilde{\Psi}^{(84(h+1)-R_s)}_{210}(q) \Big)	\,	,	\\
\end{split} 	\\
\begin{split}
Z_2 &= \sum_{s=0}^{3} \Big( \sum_{h=0}^{2} (e^{-\frac{4\pi i}{5} (84h+R_s)} + e^{\frac{4\pi i}{5} (84h+R_s)}) \widetilde{\Psi}^{(84h+R_s)}_{210}(q) 	\\ 
&\hspace{20mm}	- \sum_{h=0}^{1} (e^{-\frac{4\pi i}{5} (84(h+1)-R_s)} + e^{\frac{4\pi i}{5} (84(h+1)-R_s)}) \widetilde{\Psi}^{(84(h+1)-R_s)}_{210}(q) \Big)
\end{split} 
\end{align}
where $R_0=1, R_1=41, R_2=55$, and $R_3=71$.
More explicitly,
\begin{align}
\begin{split}
Z_0(q) =& \,
\widetilde{\Psi }_{210}^{(1)}-\widetilde{\Psi }_{210}^{(13)}-\widetilde{\Psi }_{210}^{(29)}+\widetilde{\Psi }_{210}^{(41)}-\widetilde{\Psi }_{210}^{(43)}+\widetilde{\Psi}_{210}^{(55)}+\widetilde{\Psi }_{210}^{(71)}-\widetilde{\Psi }_{210}^{(83)}+\widetilde{\Psi }_{210}^{(85)}	\\
&-\widetilde{\Psi }_{210}^{(97)}-\widetilde{\Psi}_{210}^{(113)}+\widetilde{\Psi }_{210}^{(125)}-\widetilde{\Psi }_{210}^{(127)}+\widetilde{\Psi }_{210}^{(139)}+\widetilde{\Psi }_{210}^{(155)}-\widetilde{\Psi}_{210}^{(167)}+\widetilde{\Psi }_{210}^{(169)}	\\
&+\widetilde{\Psi }_{210}^{(209)}+\widetilde{\Psi }_{210}^{(223)}+\widetilde{\Psi }_{210}^{(239)}	\,	,
\end{split}	\\
\begin{split}
Z_1(q) =& \,
B \widetilde{\Psi }_{210}^{(1)}-A \widetilde{\Psi }_{210}^{(13)}-B \widetilde{\Psi }_{210}^{(29)}+B \widetilde{\Psi }_{210}^{(41)}-A \widetilde{\Psi }_{210}^{(43)}+2
   \widetilde{\Psi }_{210}^{(55)}+B \widetilde{\Psi }_{210}^{(71)}-A \widetilde{\Psi }_{210}^{(83)}	\\
&+2 \widetilde{\Psi }_{210}^{(85)}-A \widetilde{\Psi }_{210}^{(97)}  -A
   \widetilde{\Psi }_{210}^{(113)}+2 \widetilde{\Psi }_{210}^{(125)}-A \widetilde{\Psi }_{210}^{(127)}+B \widetilde{\Psi }_{210}^{(139)}+2 \widetilde{\Psi }_{210}^{(155)}	\\
&-A
   \widetilde{\Psi }_{210}^{(167)}+B \widetilde{\Psi }_{210}^{(169)}+B \widetilde{\Psi }_{210}^{(209)}+A \widetilde{\Psi }_{210}^{(223)}+B \widetilde{\Psi }_{210}^{(239)}	\,	,
\end{split}	\\
\begin{split}
Z_2(q) =&
A \widetilde{\Psi }_{210}^{(1)}-B \widetilde{\Psi }_{210}^{(13)}-A \widetilde{\Psi }_{210}^{(29)}+A \widetilde{\Psi }_{210}^{(41)}-B \widetilde{\Psi }_{210}^{(43)}+2\widetilde{\Psi }_{210}^{(55)}+A \widetilde{\Psi }_{210}^{(71)}-B \widetilde{\Psi }_{210}^{(83)}	\\
&+2 \widetilde{\Psi }_{210}^{(85)}-B \widetilde{\Psi }_{210}^{(97)}-B\widetilde{\Psi }_{210}^{(113)}+2 \widetilde{\Psi }_{210}^{(125)}-B \widetilde{\Psi }_{210}^{(127)}+A \widetilde{\Psi }_{210}^{(139)}+2 \widetilde{\Psi }_{210}^{(155)}	\\
&-B\widetilde{\Psi }_{210}^{(167)}+A \widetilde{\Psi }_{210}^{(169)}+A \widetilde{\Psi }_{210}^{(209)}+B \widetilde{\Psi }_{210}^{(223)}+A \widetilde{\Psi }_{210}^{(239)}
\end{split}
\end{align}
where $A:=\frac{1}{2}(-\sqrt{5}-1)$ and $B:=\frac{1}{2}(\sqrt{5}-1)$ for convenience.
The homological blocks are given by
\begin{align}
\begin{split}
\widehat{Z}_0 = \widetilde{\Psi }_{210}^{(55)}+\widetilde{\Psi }_{210}^{(85)}+\widetilde{\Psi }_{210}^{(125)}+\widetilde{\Psi }_{210}^{(155)}	\,	,
\end{split}	\\
\begin{split}
\widehat{Z}_1 = \widetilde{\Psi }_{210}^{(1)}-\widetilde{\Psi }_{210}^{(29)}+\widetilde{\Psi }_{210}^{(41)}+\widetilde{\Psi }_{210}^{(71)}+\widetilde{\Psi }_{210}^{(139)}+\widetilde{\Psi}_{210}^{(169)}+\widetilde{\Psi }_{210}^{(209)}+\widetilde{\Psi }_{210}^{(239)}	\,	,
\end{split}	\\
\begin{split}
\widehat{Z}_2 = -\widetilde{\Psi }_{210}^{(13)}-\widetilde{\Psi }_{210}^{(43)}-\widetilde{\Psi }_{210}^{(83)}-\widetilde{\Psi }_{210}^{(97)}-\widetilde{\Psi }_{210}^{(113)}-\widetilde{\Psi}_{210}^{(127)}-\widetilde{\Psi }_{210}^{(167)}+\widetilde{\Psi }_{210}^{(223)}
\end{split}
\end{align}
where $\widehat{Z}_0(q) = q^{\frac{505}{840}} \mathbb{Z}[[q]]$, $\widehat{Z}_1(q) = q^{\frac{1}{840}} \mathbb{Z}[[q]]$, and $\widehat{Z}_2(q) = q^{\frac{169}{840}} \mathbb{Z}[[q]]$.
Therefore, the $S$-matrix is given by
\begin{align}
S_{ab} = \frac{1}{\sqrt{5}}
\begin{pmatrix}
1	&	1					&	1	\\
2	&	\frac{1}{2}(\sqrt{5}-1)		&	\frac{1}{2}(-\sqrt{5}-1)	\\
2	&	\frac{1}{2}(-\sqrt{5}-1)	&	\frac{1}{2}(\sqrt{5}-1)	
\end{pmatrix}
=
\begin{pmatrix}
1	&	1	&	1	\\
2	&	B	&	A	\\
2	&	A	&	B
\end{pmatrix}	\,	.
\end{align}
This can also be calculated from \eqref{su2-smat0} and \eqref{su2-lk} with $C_a = \{ W_a \}$, $a=0,1,2$.
Thus, \eqref{3f-h5-result0} can be written as
\begin{align}
\begin{split}
Z_{SU(2)}(M_3) 
=& \frac{B}{i} q^{-\phi_3/4} \bigg( \frac{42i}{K} \bigg)^{1/2} 
\sum_{a,b=0}^{2} e^{2\pi i K CS_a} S_{ab} \widehat{Z}_{b}(q)	\,	\bigg|_{q \searrow e^{\frac{2\pi i}{K}}}	
\end{split}
\end{align}
where $(CS_0,CS_1,CS_2)=(0,\frac{2}{5},\frac{3}{5})$.

\subsubsection*{\textbullet \ $H=6$}
We take, for example, $(P_1,P_2,P_3)=(5,7,11)$ and $(Q_1,Q_2,Q_3)=(-2,8,-8)$ to have $H=6$.
There are four abelian flat connections, $W_0=\{(0,0)\}$, $W_1=\{(1,-1),(5,-5)\}$, $W_2=\{(2,-2),(4,-4)\}$, and $W_3=\{(3,-3)\}$. 
We denote their contributions by $Z_0$, $Z_1$, $Z_2$, and $Z_3$, respectively.
Then, the WRT invariant is given by
\begin{align}
Z_{SU(2)}(M_3) 
=& \frac{B}{2i} q^{-\phi_3/4} \bigg( \frac{385i}{3K} \bigg)^{1/2} 
\Big( Z_0 + e^{2 \pi i K \frac{1}{6}} Z_1 + e^{ \pi i K} Z_2 + e^{2 \pi i K \frac{4}{6}} Z_3 
\Big)	\,	\bigg|_{q \searrow e^{\frac{2\pi i}{K}}}	
\label{3f-h6-result0}
\end{align}
where 
\begin{align}
\begin{split}
Z_0 &= \sum_{s=0}^{3} \sum_{h=0}^{2} (\widetilde{\Psi}^{(770h+R_s)}_{2310}(q) - \widetilde{\Psi}^{(770(h+1)-R_s)}_{2310}(q) )	\,	,
\end{split}	\\
\begin{split}
Z_1 &= \sum_{s=0}^{3} \sum_{h=0}^{2} \Big( (e^{-2\pi i \frac{1}{6} (770h+R_s)} + e^{2\pi i \frac{1}{6} (770h+R_s)}) \widetilde{\Psi}^{(770h+R_s)}_{2310}(q)	\\
&\hspace{25mm}-(e^{-2\pi i \frac{1}{6} (770(h+1)-R_s)} + e^{2\pi i \frac{1}{6} (770(h+1)-R_s)}) \widetilde{\Psi}^{(770(h+1)-R_s)}_{2310}(q) \Big)	\,	,
\end{split}	\\
\begin{split}
Z_2 &= \sum_{s=0}^{3} \sum_{h=0}^{2} \Big( (e^{-2\pi i \frac{2}{6} (770h+R_s)} + e^{2\pi i \frac{2}{6} (770h+R_s)}) \widetilde{\Psi}^{(770h+R_s)}_{2310}(q)	\\
&\hspace{25mm}-(e^{-2\pi i \frac{2}{6} (770(h+1)-R_s)} + e^{2\pi i \frac{2}{6} (770(h+1)-R_s)}) \widetilde{\Psi}^{(770(h+1)-R_s)}_{2310}(q) \Big)	\,	,
\end{split}	\\
\begin{split}
Z_3 &= \sum_{s=0}^{3} \sum_{h=0}^{2} \Big( e^{\pi i (770h+R_s)} \widetilde{\Psi}^{(770h+R_s)}_{2310}(q) - e^{\pi i (770(h+1)-R_s)} \widetilde{\Psi}^{(770(h+1)-R_s)}_{2310}(q) \Big)
\end{split}
\end{align}
with $R_0 = 218$, $R_1 = 398$, $R_2 = 442$, and $R_3 = 482$.
From them, we have
\begin{align}
\begin{split}
Z_0 =Z_3
=&\widetilde{\Psi }_{2310}^{(218)}-\widetilde{\Psi }_{2310}^{(288)}-\widetilde{\Psi }_{2310}^{(328)}-\widetilde{\Psi }_{2310}^{(372)}+\widetilde{\Psi }_{2310}^{(398)}+\widetilde{\Psi}_{2310}^{(442)}+\widetilde{\Psi }_{2310}^{(482)}-\widetilde{\Psi }_{2310}^{(552)}+\widetilde{\Psi }_{2310}^{(988)}	\\
&-\widetilde{\Psi }_{2310}^{(1058)}-\widetilde{\Psi}_{2310}^{(1098)}-\widetilde{\Psi }_{2310}^{(1142)}+\widetilde{\Psi }_{2310}^{(1168)}+\widetilde{\Psi }_{2310}^{(1212)}+\widetilde{\Psi }_{2310}^{(1252)}-\widetilde{\Psi}_{2310}^{(1322)}+\widetilde{\Psi }_{2310}^{(1758)}	\\
&-\widetilde{\Psi }_{2310}^{(1828)}-\widetilde{\Psi }_{2310}^{(1868)}-\widetilde{\Psi }_{2310}^{(1912)}+\widetilde{\Psi}_{2310}^{(1938)}+\widetilde{\Psi }_{2310}^{(1982)}+\widetilde{\Psi }_{2310}^{(2022)}-\widetilde{\Psi }_{2310}^{(2092)}	\,	,
\end{split}	\\
\begin{split}
Z_1 =  Z_2	
=& -\widetilde{\Psi }_{2310}^{(218)}-2 \widetilde{\Psi }_{2310}^{(288)}+\widetilde{\Psi }_{2310}^{(328)}-2 \widetilde{\Psi }_{2310}^{(372)}-\widetilde{\Psi}_{2310}^{(398)}-\widetilde{\Psi }_{2310}^{(442)}-\widetilde{\Psi }_{2310}^{(482)}-2 \widetilde{\Psi }_{2310}^{(552)}	\\
&-\widetilde{\Psi }_{2310}^{(988)}+\widetilde{\Psi}_{2310}^{(1058)}-2 \widetilde{\Psi }_{2310}^{(1098)}+\widetilde{\Psi }_{2310}^{(1142)}-\widetilde{\Psi }_{2310}^{(1168)}+2 \widetilde{\Psi }_{2310}^{(1212)}-\widetilde{\Psi}_{2310}^{(1252)}+\widetilde{\Psi }_{2310}^{(1322)}	\\
&+2 \widetilde{\Psi }_{2310}^{(1758)}+\widetilde{\Psi }_{2310}^{(1828)}+\widetilde{\Psi }_{2310}^{(1868)}+\widetilde{\Psi}_{2310}^{(1912)}+2 \widetilde{\Psi }_{2310}^{(1938)}-\widetilde{\Psi }_{2310}^{(1982)}+2 \widetilde{\Psi }_{2310}^{(2022)}+\widetilde{\Psi }_{2310}^{(2092)}	\,	.
\end{split}
\end{align}
$W_0$ and $W_3$ (resp. $W_1$ and $W_2$) are in the same orbit $C_0$ (resp. $C_1$) under the action of the center and as discussed in section \ref{ssec:su2-prop} they give the same contribution up to $e^{\pi i K}$.
Therefore, \eqref{3f-h6-result0} can also be written as
\begin{align}
Z_{SU(2)}(M_3)=& \frac{B}{2i} q^{-\phi_3/4} \bigg(\frac{385i}{3K} \bigg)^{1/2} 
(1+e^{ \pi i K}) \Big(Z_0 + e^{2 \pi i K \frac{1}{6}} Z_1 \Big)	\,	\bigg|_{q \searrow e^{\frac{2\pi i}{K}}}		\,	.
\end{align}
Homological blocks are given by
\begin{align}
\widehat{Z}_0 &= -\widetilde{\Psi }_{2310}^{(288)}-\widetilde{\Psi }_{2310}^{(372)}-\widetilde{\Psi }_{2310}^{(552)}+\widetilde{\Psi }_{2310}^{(1212)}	\,	,	\\
\widehat{Z}_1 &= -\widetilde{\Psi }_{2310}^{(328)}+\widetilde{\Psi }_{2310}^{(988)}+\widetilde{\Psi }_{2310}^{(1168)}+\widetilde{\Psi }_{2310}^{(1252)}-\widetilde{\Psi}_{2310}^{(1828)}-\widetilde{\Psi }_{2310}^{(1868)}-\widetilde{\Psi }_{2310}^{(1912)}-\widetilde{\Psi }_{2310}^{(2092)}	\,	,	\\
\widehat{Z}_2 &= \widetilde{\Psi }_{2310}^{(218)}+\widetilde{\Psi }_{2310}^{(398)}+\widetilde{\Psi }_{2310}^{(442)}+\widetilde{\Psi }_{2310}^{(482)}-\widetilde{\Psi}_{2310}^{(1058)}-\widetilde{\Psi }_{2310}^{(1142)}-\widetilde{\Psi }_{2310}^{(1322)}+\widetilde{\Psi }_{2310}^{(1982)}	\,	\\
\widehat{Z}_3 &= -\widetilde{\Psi }_{2310}^{(1098)}+\widetilde{\Psi }_{2310}^{(1758)}+\widetilde{\Psi }_{2310}^{(1938)}+\widetilde{\Psi }_{2310}^{(2022)}
\end{align}
where
$\widehat{Z}_0 = q^{\frac{9024}{9240}} \mathbb{Z}[[q]]$, $\widehat{Z}_1 = q^{\frac{5944}{9240}} \mathbb{Z}[[q]]$, $\widehat{Z}_2 = q^{\frac{1324}{9240}} \mathbb{Z}[[q]]$, and $\widehat{Z}_3 = q^{\frac{4044}{9240}} \mathbb{Z}[[q]]$.
So $Z_0 = Z_3 = \widehat{Z}_0 + \widehat{Z}_1 + \widehat{Z}_2 + \widehat{Z}_3$ and $Z_1 = Z_2 = 2 \widehat{Z}_0 - \widehat{Z}_1 +2 \widehat{Z}_3 - \widehat{Z}_2$. 
We put them in a form of $\frac{1}{\sqrt{3}} Z_{\dot{a}} = \frac{1}{\sqrt{3}} \begin{pmatrix} Z_0 \\ Z_1 \\ Z_3 \\ Z_2 \end{pmatrix} =\frac{1}{\sqrt{3}} \begin{pmatrix} Z_0 \\ Z_1 \\ Z_0 \\ Z_1 \end{pmatrix} = \frac{1}{\sqrt{3}} \begin{pmatrix} 1 & 1 \\ 1 & 1 \end{pmatrix} \otimes \begin{pmatrix} 1 & 1 \\ 2 & -1 \end{pmatrix} \begin{pmatrix} \widehat{Z}_0 \\ \widehat{Z}_1 \\ \widehat{Z}_3 \\ \widehat{Z}_2 \end{pmatrix}$.
The $S$-matrix $\frac{1}{\sqrt{3}}\begin{pmatrix} 1 & 1 \\ 2 & -1 \end{pmatrix}$ also appeared in the case of $H=3$ where we note that $\mathbb{Z}_6=\mathbb{Z}_2 \times \mathbb{Z}_3$.
It can also be calculated from \eqref{su2-smat0} and \eqref{su2-lk} with $C_0 = \{ W_0, W_3\}$ and $C_1 = \{ W_1, W_2 \}$.
Or $(Y \times S_{ab})_{\dot{a}\dot{b}}$ can also be calculated from \eqref{su2-smat2} and \eqref{su2-lk} where indices $a$ and $b$ in \eqref{su2-smat2} run over $0,1,3$, and $2$.
Thus, the WRT invariant can be written as
\begin{align}
Z_{SU(2)}(M_3) 
= \frac{B}{2i} q^{-\phi_3/4} \bigg(\frac{385i}{K} \bigg)^{1/2} 
\sum_{\dot{a}, \dot{b}=\dot{0}}^{\dot{3}} e^{2 \pi i K CS_{\dot{a}}} \Big( \begin{pmatrix} 1 & 1 \\ 1 & 1 \end{pmatrix} \otimes S_{ab} \Big)_{\dot{a} \dot{b}} \widehat{Z}_{\dot{b}}(q)	\,	\bigg|_{q \searrow e^{\frac{2\pi i}{K}}}	
\end{align} 
where $\widehat{Z}_{\dot{b}}=(\widehat{Z}_{0},\widehat{Z}_{1},\widehat{Z}_{3},\widehat{Z}_{2})^{T}$ and $CS_{\dot{a}} = (CS_0, CS_1, CS_3, CS_2) = (0, \frac{1}{6}, \frac{1}{2}, \frac{2}{3})$.


\subsection{Resurgent analysis}
\label{ssec:resurgent}

We discuss the resurgent analysis on the expression we obtained.
We consider the case $H=1$.
The overall factor of the analytically continued Chern-Simons partition function or the WRT invariant is proportional to $1/\sqrt{K}$.
Since the partition functions with $H=1$ are given by linear combinations of $\widetilde{\Psi}^{(l)}_{P}(q)$, we consider first the resurgent analysis for $\frac{1}{\sqrt{K}}\widetilde{\Psi}^{(l)}_{P}(q)$ \cite{Gukov-Marino-Putrov}.	\\

The perturbative expansion of $\frac{1}{\sqrt{K}}\widetilde{\Psi}^{(l)}_{P}$ is $\widetilde{\Psi}^{(l)'}_{P}:=\frac{1}{\sqrt{K}}\widetilde{\Psi}^{(l),\text{pert}}_{P} = \sum_{n=0}^{\infty} \psi^{(l)}_{2P}(n) \sum_{m=0}^{\infty} \frac{1}{m!} \Big( \frac{i\pi n^2}{2P}  \Big)^m \frac{1}{K^{m+\frac{1}{2}}}$, so its Borel transform is given by
\begin{align}
\mathcal{B} \widetilde{\Psi}^{(l)'}_{P}(\xi) &= 
\sum_{n=0}^{\infty} \psi^{(l)}_{2P}(n) \sum_{m=0}^{\infty} \frac{1}{m!} \bigg( \frac{i \pi n^2}{2 P } \bigg)^m \frac{1}{\Gamma \Big(m+\frac{1}{2}\Big)} \xi^{m-\frac{1}{2}}	\\
&= \frac{1}{\sqrt{\pi \xi}} \frac{\sinh \Big( (P-l) \big(\frac{2 i \pi \xi }{P } \big)^{\frac{1}{2}} \Big)}{\sinh \Big( P \big(\frac{2 i \pi \xi }{P } \big)^{\frac{1}{2}} \Big)}	\,	.
\end{align} 
Here, $\text{Re} \, y > 0$ and $P>0$ are assumed and
\begin{align}
\Gamma(m+\frac{1}{2}) &= \frac{\sqrt{\pi}}{2^{2m}} \frac{(2m)!}{m!}	\,	,	\\
\frac{\sinh ( P-l) y }{\sinh  P y} &= \sum_{n=0}^{\infty} \psi^{(l)}_{2  P} (n) e^{-ny}		\label{exp-for-resurg}
\end{align}
are used.
We note that we have seen \eqref{exp-for-resurg} in \eqref{3f-exp} with $\eqref{3f-chi-decomp}$.

After taking $y \rightarrow y/P$ in \eqref{exp-for-resurg} and then multiplying $e^{-\frac{K}{2\pi i} \frac{1}{P}y^2}$ at each side of the equality of \eqref{exp-for-resurg}, we integrate them over $\epsilon + i \mathbb{R}$ with $\epsilon > 0$,
\begin{align}
\int_{\epsilon + i \mathbb{R}} dy \frac{\sinh \, ( P-l) \frac{y}{P} }{\sinh  y} e^{-\frac{K}{2\pi i} \frac{1}{P} y^2}	
=
\int_{\epsilon + i \mathbb{R}} dy \sum_{n=0}^{\infty} \psi^{(l)}_{2 P} (n) e^{-n\frac{y}{P}}	e^{-\frac{K}{2\pi i} \frac{1}{P} y^2}	\,	,
\label{3f-resur0}
\end{align}
where we note that the RHS is the integral we evaluated in section \ref{ssec:calculation} and it gives $\frac{1}{\sqrt{K}} \big( 2 \pi^2 i P \big)^{\frac{1}{2}} \widetilde{\Psi}_{P}^{(l)}(q)$.
Integrating the LHS after change of variable $\xi = \frac{1}{2\pi i}\frac{1}{P}y^2$, we finally have \cite{Gukov-Marino-Putrov}
\begin{align}
\frac{1}{\sqrt{K}} \widetilde{\Psi}^{(l)}_{P}(q) 
&=\frac{1}{2} \bigg( \int_{i e^{+i\delta}\mathbb{R}_{+}} + \int_{i e^{-i\delta}\mathbb{R}_{+}}  \bigg)
\frac{d\xi}{\sqrt{\pi \xi}} \frac{\sinh \Big( (P-l) \big(\frac{2 i \pi \xi }{P } \big)^{\frac{1}{2}} \Big)}{\sinh \Big( P \big(\frac{2 i \pi \xi }{P } \big)^{\frac{1}{2}} \Big)} e^{-K \xi}	\\
&= \frac{1}{2} \bigg( \int_{i e^{+i\delta}\mathbb{R}_{+}} + \int_{i e^{-i\delta}\mathbb{R}_{+}}  \bigg) d\xi \,
\mathcal{B}\widetilde{\Psi}^{(l)'}_{P}(\xi) \, e^{-K\xi}		
\end{align}
where $\delta$ is a small real number and the direction of the contour is from the origin to infinity.
Therefore, the original $\widetilde{\Psi}^{(l)}_{P}(q)$ can be recovered as the average of the Borel sum of $\mathcal{B} \widetilde{\Psi}_{P}^{(l)'}(\xi)$.	\\

As the analytically continued CS partition function on Seifert manifolds with three singular fibers and with $H=1$ for the trivial flat connection is given by a linear combination of $\widetilde{\Psi}^{(a)}_{P}(q)$, we have
\begin{align}
Z^{\text{ab}}_{SU(2)}(M_3) = \frac{B}{2i} q^{-\phi_3/4} ( 2 i P)^{1/2} \sum_{s=0}^{3}
\frac{1}{2} \bigg( \int_{i e^{+i\delta}\mathbb{R}_{+}} + \int_{i e^{-i\delta}\mathbb{R}_{+}}  \bigg) d\xi \,
\mathcal{B}\widetilde{\Psi}^{(R_s)'}_{P}(\xi) \, e^{-K\xi}	\,	.	\,	\label{3f-resur1}
\end{align}
The residues at poles of this integral are the contribution from the non-abelian flat connections in the resurgent analysis \cite{Gukov-Marino-Putrov}.
That is, by deforming the integral contour for $K \in \mathbb{Z}_+$, the integral picks the poles and calculation of the residue agrees with the known results on the contributions from non-abelian flat connections.
Thus, the $\widetilde{\Psi}^{(l)}_{P}(q)$ or their linear combinations, which are obtained from the Borel sum of the perturbative expansion around the trivial flat connection and are convergent $q$-series, contain the contribution from non-abelian flat connections as transseries.
Hence $Z^{\text{ab}}_{SU(2)}(M_3)$ in \eqref{3f-resur1}, which is expressed in the context of the resurgent analysis, provide the full exact partition function $Z_{SU(2)}(M_3)$ for $H=1$.
Meanwhile, as we noted earlier, \eqref{3f-resur1} or \eqref{3f-resur0} in the resurgent analysis are basically the integral formula \eqref{int-formula} with $H=1$ from which we obtained \eqref{3f-a-wrt}.
Therefore, in sum, we obtained a convergent $q$-series \eqref{3f-a-wrt} from the exact expression \eqref{int-formula} for the trivial flat connection with $H=1$, which is basically \eqref{3f-resur1} or \eqref{3f-resur0} in the resurgent analysis, and the resurgent analysis tells that \eqref{3f-a-wrt} is indeed the full exact partition function.
In the context of the exact formula of the WRT invariants in \cite{Lawrence-Rozansky}, contributions from non-abelian flat connections also appear as the residue part and we can see, for example, that the residues in \cite{Lawrence-Rozansky} agree with the result of \cite{Gukov-Marino-Putrov,Chun:2017dbf} on $F=3$ and $H=1$.\footnote{In \cite{Lawrence-Rozansky}, though the residue part looks different as it contains the additional factor in the denominator, they agree with the result of \cite{Gukov-Marino-Putrov,Chun:2017dbf} on $F=3$ and $H=1$.
More specifically, when the number of singular fibers is three, we can show that $\sum_{m=1}^{2P-1}\Big( \sum_{j=1}^{3}\frac{1}{P_j} \cot (\frac{m \pi}{P_j} ) \Big) \prod_{j=1}^{3} \sin (\frac{m \pi}{P_j})$ in \cite{Lawrence-Rozansky} vanishes, while the other part $\Big(\frac{m}{P}-\frac{1}{H} \Big) \prod_{j=1}^{3} \sin (\frac{m \pi}{P_j})$ in the residue is nonzero. 
When $H=1$, this agrees with the residue calculation in \cite{Gukov-Marino-Putrov,Chun:2017dbf} on the Brieskorn spheres.
We found that recently resurgent analysis on arbitrary $F$ with $H=1$ was also considered in \cite{2018arXiv181105376E}.
 }

Also when $H \geq 2$, residues are interpreted as the contribution from non-abelian flat connections in \cite{Lawrence-Rozansky}.
We expect that the residues corresponding to non-abelian flat connections attached to abelian flat connections in the resurgent analysis agree with the residues calculated in \cite{Lawrence-Rozansky}.
In other words, we expect that \eqref{3f-result} with $H \geq 2$ also provide the full exact partition function.
Similarly, we also expect that the partition functions that we calculate for arbitrary number of singular fibers $F$, $H$, and genus $g$ in section \ref{sec:su2f4} and section \ref{sec:su2fg} provide the full exact partition function.


\subsection{Other ranges of $P$ and $H$ and reversed orientation of $M_3$}
\label{ssec:other-ranges}
So far, we have considered the case that $P$ and $H$ are both positive.
We also assumed that $\text{Im} \, K <0 \ \Leftrightarrow \ |q|<1$.
We consider other ranges of their values.

In the expansion \eqref{3f-exp}, it was assumed that $\text{Re}(y/P)>0$.
This is obtained when $\text{Re}\, y>0$ and $P>0$ and also $\text{Re}\, y<0$ and $P<0$.
The former case with $H>0$ has been considered in the previous sections with $|q|<1$.

When $H<0$ with $\text{Re}\, y>0$ and $P>0$, the coefficient of $y^2$, which is $-\frac{K}{2\pi i} \frac{H}{P}$, in the exponential factor in \eqref{3f-integral} needs to be positive so that we can choose a contour that passes through a fixed $\text{Re}\, y$ and extends along the imaginary axis of the $y$-plane for the consistency of the calculation.
This requires that $\text{Im}\, K>0 \, \Leftrightarrow \, |q|>1$.
Then the integral gives $q^{\frac{n^2}{4HP}}$ in the expression and $|q^{\frac{1}{4HP}}|<1$.
Therefore, $\sum_{n=0}^{\infty}\psi^{(a)}_{2P} q^{\frac{n^2}{4HP}}$ is convergent.
In sum, if $\text{Re}\, y>0$ and $P>0$, the calculation is consistent when $H>0$ and $\text{Im}\, K<0 \Leftrightarrow |q|<1$ or $H<0$ and $\text{Im}\, K>0 \Leftrightarrow |q|>1$.

We can do a similar analysis with $\text{Re}\, y < 0$ and $P<0$.
The calculation is consistent if $H>0$ and $\text{Im}\, K>0 \Leftrightarrow |q|>1$ or $H<0$ and $\text{Im}\, K<0 \Leftrightarrow |q|<1$.

When $\text{Re}(y/P)<0$, the RHS of the expansion \eqref{3f-exp} should be $\sum_{n=0}^{\infty} \chi_{2P}(n) e^{\frac{n}{P}y}$.
This change only leads to the change of the exponent, $e^{2\pi i n \frac{t}{H}} \rightarrow e^{-2\pi i n \frac{t}{H}}$ in \eqref{3f-a-wrt}.
But as we saw in \eqref{3f-odd} and \eqref{3f-even}, this doesn't affect the final result.
This case $\text{Re}(y/P)<0$ is obtained when $\text{Re}\, y>0$ and $P<0$ and also when $\text{Re}\, y<0$ and $P>0$.
Summarizing the analysis, if $\text{Re}\, y>0$ and $P<0$, $H\gtrless0$ and $\text{Im}\, K \gtrless 0 \Leftrightarrow |q| \gtrless 1$, respectively, provide consistent calculations.
Similarly, if $\text{Re}\, y<0$ and $P>0$, the calculation is consistent when $H\gtrless0$ and $\text{Im}\, K \lessgtr 0 \Leftrightarrow |q| \lessgtr 1$, respectively.

In summary, whatever $\text{Re} \, y$ is, the calculation is consistent when
\begin{alignat}{6}
P>0	\,	,	&\	H>0	\,	,	&&\	\text{Im} \, K < 0	\,	\Leftrightarrow	\,	|q|<1	\,	,	&&\quad		P>0	\,	,	&&\	H<0	\,	,	&&\	\text{Im} \, K > 0	\,	\Leftrightarrow	\,	|q|>1	\,	,	\label{range-cases1} 	\\
P<0	\,	,	&\	H>0	\,	,	&&\	\text{Im} \, K > 0	\,	\Leftrightarrow	\,	|q|>1	\,	, 	&&\quad		P<0	\,	,	&&\	H<0	\,	,	&&\	\text{Im} \, K < 0	\,	\Leftrightarrow	\,	|q|<1 \,	.	\label{range-cases2}
\end{alignat}
\vspace{0mm}

The WRT invariant stays the same in those ranges just with a few differences.
For example, when $P>0$ and $H<0$ so $\text{Im}\,K>0$, we have
\begin{align}
Z_{SU(2)}(M_3)
&= \frac{B}{2 i} q^{-\phi_3/4} \bigg( \frac{2i}{K} \frac{P}{H} \bigg)^{1/2} \sum_{t=0}^{|H|-1} e^{2 \pi i K \frac{P}{H} t^2} \
\sum_{n=0}^{\infty} \chi_{2P}(n) e^{2 \pi i \frac{t}{H} n } q^{\frac{n^2}{4HP}}		\bigg|_{q \nearrow e^{\frac{2 \pi i}{K}}}	\label{PposHneg}
\end{align}
and we see that this is the same with the case of $P>0$, $H>0$, and $\text{Im}\,K<0$, if the sign of $\text{Re} \, K$ is also reversed, up to a numerical factor and $q^{-\phi_3/4}$ that depends on the values of $Q_j$'s.

This has an implication on the reverse of the orientation of $M_3$ and the sign of $K$.
In the calculation above and also below where we have set $b$ of the Seifert invariants to be zero, the reverse of the orientation of the Seifert manifold $M_3 = X(P_1/Q_1, \cdots, P_F/Q_F)$ is realized by the change of signs of all $Q_j$'s, 
\begin{align}
M_3 = X(P_1/Q_1, \cdots, P_F/Q_F)	\quad	\rightarrow	\quad	-M_3 = X(-P_1/Q_1, \cdots, -P_F/Q_F)	\,	.
\end{align}
This also leads to $H \rightarrow -H$ since $H:=P\sum_{j=1}^{F}\frac{Q_j}{P_j}$.
We see from \eqref{range-cases1} and \eqref{range-cases2} that given a $P$ when $H$ changes its sign, then also the sign of $\text{Im}\, K$ should be changed.
Since the Dedekind sum in $\phi_3$ satisfies $s(-Q,P)=-s(Q,P)$, when the orientation of $M_3$ is reversed to $-M_3$ and $q$ is inverted to $q^{-1}$, we see that the partition function stays the same, \textit{i.e.} $Z_G(M_3,K) = Z_{G}(-M_3,-K)$ up to a numerical factor, which is consistent with expectation \cite{Gukov-Pei-Putrov-Vafa, Cheng:2018vpl}.


\section{Higher rank gauge group}
\label{sec:higher-rank}

The expression of the WRT invariants of Seifert manifolds for the ADE gauge group with arbitrary number of singular fibers $F$ was obtained in \cite{Marino2004}.
The contributions from the reducible flat connections for the gauge group $G$ are given by
\begin{align}
\begin{split}
\hspace{-10mm}\frac{(-1)^{|\Delta_+|}}{|\mathcal{W}| (2 \pi i)^r} \bigg( \frac{\text{Vol } \Lambda_{\text{w}} }{\text{Vol } \Lambda_{\text{r}}} \bigg) \frac{\text{sign}(P)^{|\Delta_+|} }{|P|^{r/2}}
e^{\frac{\pi i d}{4} \text{sign}(H/P) - \frac{\pi i d \check{c}_{\mathfrak{g}}}{12 K} \phi }	
\sum_{{\bf t} \in \Lambda_{\text{r}} / H \Lambda_{\text{r}}} \int_{\Gamma^r} d {\pmb \beta} \, e^{-\frac{K}{4 \pi i} \frac{H}{P} {\pmb \beta}^2 - K {\bf t} \cdot {\pmb \beta}} \frac{\prod_{j=1}^{F} \prod_{{\pmb \alpha} > 0} 2 \sinh \frac{{\pmb \beta} \cdot {\pmb \alpha}}{2P_j} }{\prod_{{\pmb \alpha} > 0} \Big( 2 \sinh \frac{{\pmb \beta} \cdot {\pmb \alpha}}{2} \Big)^{F-2}}
\end{split}	\label{wrt-gen}
\end{align}
where elements ${\bf t} \in \Lambda_r / H \Lambda_r$ correspond to the reducible flat connections where those related by Weyl reflections are the equivalent reducible flat connections.\footnote{As will be discussed in section \ref{sec:su-n-wrt}, we expect that the Gaussian integral parts of \eqref{wrt-gen} without residue parts are for the contributions from the completely reducible flat connections, \textit{i.e.} the abelian flat connections, and the Weyl orbit of $\mathbf{t} \in \Lambda_r / H \Lambda_r$ corresponds to the abelian flat connection of the gauge group $G$.} 
Also, the case ${\bf t}=0$ corresponds to the trivial flat connection.
Here, $r$, $d$, and $\check{c}_{\mathfrak{g}}$ denote the rank, the dimension of group $G$, and the dual Coxeter number of the Lie algebra $\mathfrak{g}$ of $G$, respectively.
Also $|\Delta_+|$, $\mathcal{W}$, $\Lambda_{\text{r}}$, and $\Lambda_{\text{w}}$ denote, respectively, the number of positive roots, the Weyl group, the root and the weight lattice.
$K$ is the quantum-corrected CS level $K=k+\check{c}_{\mathfrak{g}}$.
The integration contour $\Gamma^r = \Gamma \times \cdots \times \Gamma$ is a multiple contour of $\Gamma$ in $\mathbb{C}^r$ where $\Gamma$ is the contour $\Gamma_0$ discussed in the previous section.
As done in the $SU(2)$ case, we would like to express \eqref{wrt-gen} in terms of the $q$-series with integer coefficients and integer powers.
In this paper, we consider the case $G=SU(N)$.
We leave the cases of other gauge groups as a future work.


\subsection{Calculation of the partition function}
\label{sec:su-n-wrt}

The integral part of \eqref{wrt-gen} for $G=SU(N)$ is given by
\begin{align}
\begin{split}
\sum_{{\bf t} \in \Lambda_{\text{r}} / H \Lambda_{\text{r}}}	\int_{\Gamma_{\mathbf{t}}^{N-1}} d\beta_1 \ldots d\beta_{N-1} 
& \, e^{-\frac{K}{2 \pi i} \frac{H}{P} (\sum_{i=1}^{N-1} \beta_i^2 + \sum_{i<j}^{N-1} \beta_i \beta_j) - K (2 \sum_{i=1}^{N-1} t_i \beta_i + \sum_{i \neq j}^{N-1} t_i \beta_j )}	\\
& \qquad	\qquad	\qquad \times \frac{\prod_{f=1}^{F} \prod_{i<j}^{N} 2 \sinh \frac{1}{2 P_f} (\beta_i - \beta_j) }{\prod_{i<j}^{N} \big(2 \sinh \frac{1}{2} (\beta_i - \beta_j) \big)^{F-2}}
\end{split}	\label{su-n-int1}
\end{align}
up to the overall factor\footnote{\label{ov-factor} When $G=SU(N)$, the overall factor to the integral is
\begin{align}
\frac{(-1)^{(N-1)N/2}}{(2 \pi i)^{N-1} N!} \frac{1}{N} \frac{\text{sign}(P)^{\frac{N(N-1)}{2}} }{|P|^{\frac{N-1}{2}}} e^{\frac{\pi i (N^2-1)}{4} \text{sign}(H/P) - \frac{\pi i}{12 K} (N^2-1) N \phi_F }	\,	.	
\end{align}
} 
where $\beta_N = - \sum_{i=1}^{N-1} \beta_i$.
Here, $\Gamma_{\mathbf{t}}^{N-1}$ denotes a integration contour that passes a stationary phase point given a $\mathbf{t}$ as in \eqref{int-formula} and the residue parts are dropped.
We consider the case $F=3$ with $\sum_{j} \frac{1}{P_j} <1$ for simplicity.
For larger number of singular fibers, higher genus, or $\sum_{j} \frac{1}{P_j} >1$, we can use the formula in section \ref{sec:su2fg} in the following calculations.	\\

Given a pair $i$ and $j$ with $i<j$ and $j \neq N$, we have
\begin{align}
\frac{\prod_{f=1}^{3} \Big( e^{ \frac{1}{P_f}\frac{1}{2} (\beta_i - \beta_j)} - e^{-\frac{1}{P_f}\frac{1}{2} (\beta_i - \beta_j)} \Big)}{\Big( e^{\frac{1}{2} (\beta_i - \beta_j)} - e^{-\frac{1}{2} (\beta_i - \beta_j)} \Big)}
= \sum_{n_{i,j}=0}^{\infty} \chi_{2P}(n_{i,j}) \, e^{\frac{1}{2P} n_{i,j} (\beta_{i} - \beta_{j})}	\,	,	\label{su-n-exp1}
\end{align}
and for $j=N$,
\begin{align}
\frac{\prod_{f=1}^{3} \Big( e^{ \frac{1}{P_f}\frac{1}{2} (\beta_i - \beta_N)} - e^{-\frac{1}{P_f}\frac{1}{2} (\beta_i - \beta_N)} \Big)}{\Big( e^{\frac{1}{2} (\beta_i - \beta_N)} - e^{-\frac{1}{2} (\beta_i - \beta_N)} \Big)}
= \sum_{n_{i,N}=0}^{\infty} \chi_{2P}(n_{i,N}) \, e^{-\frac{1}{2P} n_{i,N} (\beta_{i} + \sum_{l=1}^{N-1} \beta_{l})}	\,		\label{su-n-exp2}
\end{align}
where we chose $0 < \text{Re} \, \beta_1 < \text{Re} \, \beta_2 < \cdots < \text{Re} \, \beta_{N-1}$ and $P>0$ for convergence.\footnote{We can choose other orderings of $\text{Re} \, \beta_i$ in the calculations, which makes some of expansions in \eqref{su-n-exp1}
to have $-n_{i,j}$. But by renaming $n_{i,j}$'s and considering $\sum_{\mathbf{t} \in \Lambda_r/H\Lambda_r}$ we can see that the final expressions are all the same.} 
Then, \eqref{su-n-int1} is expressed as
\begin{align}
\begin{split}
\hspace{-5mm} \sum_{{\bf t} \in \Lambda_{\text{r}} / H \Lambda_{\text{r}}}
\int_{\Gamma^{N-1}} \, d\beta _1 d\beta_2 \cdots d\beta_{N-1}
&\exp \bigg(-\frac{K}{ 2 \pi i} \frac{H}{P} \Big(\sum _{i=1}^{N-1} \beta _i^2+\sum _{i<j}^{N-1} \beta _i \beta _j \Big) \bigg)	
\exp \bigg(- K  \Big(2 \sum _{i=1}^{N-1} \beta _i t_i+\sum _{i\neq j}^{N-1} t_i \beta _j \Big) \bigg)	\\
&\hspace{-20mm}\times \sum_{\substack{n_{i,j}=0 \\ 1 \leq i < j \leq N}}^{\infty} \Big( \prod_{i<j} \chi_{2P}(n_{i,j}) \Big)
\exp \bigg(\frac{1}{2} \sum _{i=1}^{N-1} \Big( -\sum _{j=1}^{i-1} n_{j,i} + \sum _{j=i+1}^{N-1} n_{i,j} -n_{i,N} - \sum_{j=1}^{N-1} n_{j,N} \Big) \frac{\beta _i}{P} \bigg) 	\,	.
\end{split}	\label{su-n-int2}
\end{align}
We do a similar calculation as in section \ref{ssec:calculation}. 
We analytically continue $K$ to a complex number with $\text{Im} \, K <0$, take the integral contour as $\gamma_j$, $j=1,\ldots, N-1$ that passes through $\text{Re} \, \beta_j > 0$ and extends parallel to the imaginary axis of the $\beta_j$-plane, and perform the Gaussian integral.
Then we obtain the partition function of the analytically continued $SU(N)$ CS theory,
\begin{align}
\begin{split}
\hspace{-3mm}\frac{2^{\frac{N-1}{2}}\pi^{N-1}}{\sqrt{N}} \left(\frac{2i}{K} \frac{P}{H}\right)^{\frac{N-1}{2}} 
\sum_{{\bf t} \in \Lambda_{N-1} / H \Lambda_{N-1}}
\Bigg[
&\exp \bigg(\frac{2 \pi i K P}{H} \Big(\sum _{m=1}^{N-1} t_m^2+\sum _{m<n}^{N-1} t_m t_n \Big) \bigg)		\\
&\hspace{-53mm} \times \sum_{\substack{n_{i,j}=0 \\ 1 \leq i < j \leq N}}^{\infty} \Big( \prod_{i<j} \chi_{2P}(n_{i,j}) \Big) \exp \bigg(-\frac{ \pi i }{H} \sum _{m=1}^{N-1} t_m \Big(  -\sum _{j=1}^{m-1} n_{j,m} + \sum _{j=m+1}^{N-1} n_{m,j} -n_{m,N} - \sum_{j=1}^{N-1} n_{j,N} \Big)	\bigg)	\\
&\hspace{-35mm} \times \exp \bigg(
\frac{ \pi i }{2 K H P} \Big(\sum _{1\leq i<j \leq N} n_{i,j}^2 + \sum_{1 \leq i<j<l \leq N-1} n_{i,j} n_{i,l} -\sum_{1 \leq i<j< N} n_{i,j} n_{i,N} 	\\
&\hspace{-5mm}- \sum _{1 \leq i<j<l \leq N} n_{i,j} n_{j,l} + \sum _{1 \leq i<j< N} n_{i,j} n_{j,N} + \sum _{1 \leq i<j<l \leq N} n_{i,l} n_{j,l} \Big)
\bigg)
\Bigg]	\,	.	\label{sun-a-wrt}
\end{split}
\end{align}
Or this can be written as
\begin{align}
\begin{split}
\frac{2^{\frac{N-1}{2}}}{\sqrt{N}} \left(-2 i K \frac{P}{H}\right)^{\frac{N-1}{2}} \left( \frac{\pi i}{K} \right)^{N-1}
\sum_{{\bf t} \in \Lambda_{N-1} / H \Lambda_{N-1}}
\Bigg[
&e^{ \pi i K \frac{P}{H} \sum_{m=1}^{N} t_m^2 }		\\
&\hspace{-35mm} \times \sum_{\substack{n_{i,j}=0 \\ 1 \leq i < j \leq N}}^{\infty} \Big( \prod_{1 \leq i < j \leq N} \chi_{2P}(n_{i,j}) \Big) \,
e^{-\frac{ \pi i }{H} \sum _{m=1}^{N} t_m b_m(\vec{n})}	\,
q^{ \frac{1}{8 H P} \sum_{m=1}^{N} b_m(\vec{n})^2}
\Bigg]	\,	.	\label{sun-a-wrt2}
\end{split}
\end{align}
Here, $b_i(\vec{n}) := -\sum _{j=1}^{i-1} n_{j,i}+\sum _{j=i+1}^{N-1} n_{i,j} - n_{i,N}$ for  $i=1, \ldots, N-1$ and $b_N(\vec{n}) = -\sum_{i=1}^{N-1}b_i(\vec{n})=\sum_{j=1}^{N-1} n_{j,N}$ with $\vec{n}:= (n_{1,2}, n_{1,3}, \ldots, n_{2,3}, \ldots, n_{N-1,N})$ and $t_N = -\sum_{j=1}^{N-1} t_j$.\footnote{The result is similar for $G=U(N)$. Instead of above $b_i(\vec{n})$'s and $t_i$'s, if we choose $b_i(\vec{n}) = -\sum _{j=1}^{i-1} n_{j,i} + \sum _{j=i+1}^N n_{i,j}$ and take $t_N$ to be an independent variable, we obtain the partition function of the analytically continued $G=U(N)$ CS theory up to an overall factor.
}
When $N=2$, we can recover \eqref{3f-odd} and \eqref{3f-even} from \eqref{sun-a-wrt2}.	\\

As will be discussed below, we expect that the Gaussian integral parts of \eqref{wrt-gen} or \eqref{sun-a-wrt2} correspond to the contributions from the abelian flat connections and the residues correspond to the contributions from the non-abelian flat connections as in the case of $G=SU(2)$.
Since all non-abelian flat connections are attached to the abelian flat connections as transseries in the case of $G=SU(2)$ (when $H=1$), we may also expect that similar phenomena happen in the $G=SU(N)$ case.
So we expect that the partition function \eqref{sun-a-wrt2} provides the full exact partition function.	\\

Regarding other possible ranges of $P$, $H$, $\text{Im} \, K$, we can do a similar analysis as we discussed in section \ref{ssec:other-ranges}, and obtain the same result also for the case of $G=SU(N)$.
So we will only consider the case $P>0$, $H>0$, and $\text{Im} \, K < 0$.

\subsubsection*{$\mathbf{t} \in \Lambda_r / H \Lambda_r$ and flat connections}
\label{sssec:flat-connections}

We discuss the $G=SU(N)$ flat connections on Seifert manifolds corresponding to $\mathbf{t} \in \Lambda_{N-1} / H \Lambda_{N-1}$ in \eqref{su-n-int1} or \eqref{sun-a-wrt2}.

If the stabilizer subgroup of $G$ acting on $\text{Hom}(\pi_1(M_3), G)/\text{conj.}$ or on the holonomy $\text{Hol}_A$ of the flat connection $A$ is a center of $G$, it is called the irreducible flat connection.
If not, it is called the reducible flat connection.
In the case of $G=SU(2)$, reducible and irreducible flat connections are the abelian and the non-abelian flat connection, respectively.
So abelian or non-abelian flat connections were used interchangeably with reducible or irreducible flat connections, respectively, in section \ref{sec:su2f3}.		\\

In the higher rank case, reducible flat connections are not necessarily abelian flat connections in general but there are reducible flat connections that are non-abelian.
For example, when $G=SU(3)$, the holonomy of the reducible flat connection can be $U(1)\times U(1)$ or $S(U(2) \times U(1) )$.

In order to sort out the type of flat connections that contribute to the partition function \eqref{sun-a-wrt2} in the $G=SU(N)$ case, we briefly review the case of $G=SU(2)$ \cite{Rozansky-residue, Rozansky-trivial1, Rozansky-largeK, Lawrence-Rozansky}.
The Seifert manifold $M_3$ is obtained by the $(P_j,Q_j)$ surgery, $j=1, \ldots, F$, on the link $p_j \times S^1$ in $\Sigma_g \times S^1$ where $p_j$'s are points on $\Sigma_g$.
Let $h$ be the loop that wraps $S^1$, $x_j$ be the loops around each punctures $p_j$, and $c_l$, $d_l$, $l=1,\ldots, g$, be the standard generators of $\pi_1(\Sigma_g)$.
Then the fundamental group $\pi_1(M_3)$ of the Seifert manifold is generated by $h$, $x_j$, $c_l$, and $d_l$ that satisfy $x_j^{P_j} h^{Q_j}=1$ and $x_1 \ldots x_F [c_1, d_1] \ldots [c_g, d_g] = 1$ where $h$ commutes with all $x_j$'s, $c_l$'s, and $d_l$'s,
\begin{align}
\pi_1(M_3) = \{ h, x_j\text{'s}, c_l\text{'s}, d_l\text{'s} \, \big| \,  x_j^{P_j} h^{Q_j}=1, \, \prod_{j=1}^{F} x_j \prod_{l=1}^{g} [c_l,d_l]=1, \, [h,x_j]=[h,c_l]=[h,d_l]=1 \}	\,	.
\end{align}
Introducing a map $\phi : \pi_1 \rightarrow [0,\frac{1}{2}]$ such that $e^{2\pi i \sigma_3 \phi(h)}$ is in the same conjugacy class of $\text{Hol}_A(h)$ in $SU(2)$, one can see that holonomies $\text{Hol}_A(h), \text{Hol}_A(x_1), \ldots, \text{Hol}_A(x_F)$ depend on the choice of $\phi(h)$ and some other integers that satisfy certain conditions.
When $\phi(h)\neq 0, \frac{1}{2}$, $\text{Hol}_A(h)$ is $U(1)$, and since $h$ commutes with $x_j$'s, $\text{Hol}_A(x_j)$'s are also $U(1)$. Similarly, $\text{Hol}_A(c_l)$ and $\text{Hol}_A(d_l)$ are also $U(1)$. 
The stabilizer subgroup for them is $U(1)$, so this choice $\phi(h)\neq 0, \frac{1}{2}$ gives the reducible flat connection, which is abelian.
In the notation of \eqref{int-formula}, $e^{2\pi i \sigma_3 \phi(h)} \cong \text{Hol}_{A}(h)$ is identified with $e^{2\pi i \sigma_3 \frac{P}{H}t}$, $1 \leq t \leq H-1$ \cite{Lawrence-Rozansky} up to the Weyl action and the conjugation.
In addition, we see from several examples that the Weyl orbit of such a $t$ corresponds to one abelian flat connection whose holonomy $\text{Hol}_A(h)$ is a conjugacy class of $e^{2\pi i \sigma_3 \frac{P}{H}t}$.
When $\phi(h) = 0, \frac{1}{2}$, $\text{Hol}_A(h)$ is a center of $SU(2)$, so flat connections are irreducible in general.
However, in the special case of this choice, we can have the abelian flat connection where one can also see that $\phi(h)=0$ corresponds to the trivial flat connection and $\phi(h)=\frac{1}{2}$ corresponds to the central flat connection when $H$ is even.
And the contributions to the WRT invariant from the trivial and the central flat connection corresponding to $\phi(h)=0$ and $\frac{1}{2}$ are indeed the integral \eqref{int-formula} on $t=0$ and $\frac{H}{2}$, respectively.
Therefore, a Weyl orbit of $\mathbf{t} \in \Lambda_1/H \Lambda_1$ in the integral formula \eqref{int-formula} corresponds to an abelian flat connection of $SU(2)$ whose holonomy $\text{Hol}_A(h)$ is a conjugacy class of $e^{2\pi i \sigma_3 \frac{P}{H}t}$. 	\\

We perform a similar analysis in the higher rank case.
As in the case of $G=SU(2)$, $\mathbf{t} \in \Lambda_r/H\Lambda_r $ determines $\text{Hol}_A(h) \cong e^{2\pi i \frac{P}{H} \text{diag} \, (t_1, \ldots, t_N)}$ up to the conjugation and we consider the type of flat connections from a given $\mathbf{t}$. 
There are the cases that entries of $e^{2\pi i \frac{P}{H} \text{diag} \, (t_1, \ldots, t_N)}$ are all distinct or the cases that some of entries are coincident.
For the former, we have the completely reducible flat connection, \textit{i.e.} the abelian flat connection $U(1)^{N-1}$ of $G=SU(N)$ and we expect that a Weyl orbit of such a $\mathbf{t}$ corresponds to an abelian flat connection whose holonomy $\text{Hol}_A(h)$ is a conjugacy class of $e^{2\pi i \frac{P}{H} \text{diag} \, (t_1, \ldots, t_N)}$ as in the case of $G=SU(2)$.
For the latter, such $\mathbf{t}$'s seem to lead to the non-abelian reducible flat connections in general, but we expect that a Weyl orbit of such a $\mathbf{t} \in \Lambda_r / H \Lambda_r$ in \eqref{sun-a-wrt2} corresponds to an abelian flat connection.
For example, we consider the case of $G=SU(3)$.
When $\text{Hol}_A(h)$ takes a form of $\text{diag} \, (v, v, v^{-2}) \in SU(3)$, $v \in \mathbb{T}$, all other holonomies can take a form of $S(U(2) \times U(1))$ in general where the non-abelian $U(2)$ part of $S(U(2)\times U(1))$ arises due to $(v,v)$ part of $(v, v, v^{-2})$.
Meanwhile, physical consideration such as the M2-M5 system \eqref{M-conf} where the massless BPS particles are naturally charged under $U(1)^{N-1}/S_{N-1}$ magnetically indicates that the abelian flat connections are more appropriate.
Also, for example, in the integral expression \eqref{su-n-int1} for $G=SU(3)$ on $\mathbf{t} = (2,-2,0) \in \Lambda_2 / 4 \Lambda_2$, we consider its $SU(2)$ part on $(2,-2) \in \Lambda_1 / 4 \Lambda_1$, which is also a Gaussian integral type. 
If $(2,-2)$ part of $(2,-2,0)$ led to the non-abelian reducible flat connection, the $SU(2)$ part of the integral on $(2,-2)$ would be given by the residue of the integral, which is not the case.
Rather, since $(2,-2) \in \Lambda_1/4\Lambda_1$ corresponds to an abelian flat connection in the $SU(2)$ case, $\mathbf{t} = (2,-2,0) \in \Lambda_2/4\Lambda_2$ would correspond to an abelian flat connection of $SU(3)$.
More generally, we expect that a Weyl orbit of $\mathbf{t} \in \Lambda_2 / H \Lambda_2 $ in \eqref{sun-a-wrt2} which gives $\text{Hol}_A(h) \cong \text{diag} \, (v, v, v^{-2})$ corresponds to an abelian flat connection.
The case $\mathbf{t}=0$, which is the case that $\text{Hol}_A(h)$ is the identity, corresponds to the trivial flat connection \cite{Marino2004}.
A $\mathbf{t}$ such that $\text{Hol}_A(h)$ is a center would correspond to a central flat connection as in the case of $SU(2)$.
Thus, in sum, we expect that a Weyl orbit of $\mathbf{t} \in \Lambda_r / H \Lambda_r$ in \eqref{sun-a-wrt2} corresponds to an abelian flat connection of $G=SU(N)$ whose holonomy $\text{Hol}_A(h)$ is a conjugacy class of $e^{2\pi i \frac{P}{H} \text{diag} \, (t_1, \ldots, t_N)}$.

\subsection{Properties of the formula}
\label{ssec:sun-prop}

We discuss some properties of \eqref{sun-a-wrt2}.

\subsubsection*{\underline{\textnormal{Shift by $H$}}}
Given a $\mathbf{t} = (t_1, t_2, \ldots, t_N)$ with $t_N = -\sum_{j=1}^{N-1}t_j$, any shifts of $t_j$'s, $j=1, \ldots, N-1$, by (a multiple of) $H$ don't affect $e^{\pi i K \frac{P}{H} \sum_{m=1}^{N} t_m^2}$ when $K \in \mathbb{Z}$ and $e^{-\frac{\pi i}{H} \sum_{m=1}^{N} t_m b_m(\vec{n})}$.
It is easy to see that such a shift only leads to a change of $\sum_{m=1}^{N} t_m^2$ by a multiple of $2H$. 
For $\sum_{m=1}^{N} t_m b_m(\vec{n})$, we first note that values that $n_{i,j}$'s take are all odd or all even, which depends on $P_j$'s, $j=1, \ldots, F$. 
We also note that $b_m(\vec{n})$ contains $N-1$ $n_{i,j}$'s.
If we take $t_1 \rightarrow t_1+H$, the difference in $\sum_{m=1}^{N} t_m b_m(\vec{n})$ is $H(2b_1(\vec{n})+b_2(\vec{n}) + \cdots + b_{N-1}(\vec{n}))$, which contains even number of $n_{i,j}$'s.
So whatever $N$ is even or odd, the difference is a multiple of $2H$.
Hence, such a shift by $H$ doesn't affect $e^{\pi i K \frac{P}{H} \sum_{m=1}^{N} t_m^2}$ when $K \in \mathbb{Z}$ and $e^{-\frac{\pi i}{H} \sum_{m=1}^{N} t_m b_m(\vec{n})}$.

\subsubsection*{\underline{\textnormal{The orbit under the action of the center}}}
We will see in examples below that there are cases that some elements $\mathbf{t} \in \Lambda_r/H\Lambda_r$ are related by the action of the center $e^{2\pi i \frac{c}{N}} I_N$, $c=0,1, \ldots, N-1 \ \text{mod} \, N$, of $G=SU(N)$ as in the $SU(2)$ case.
As elements $\mathbf{t} \in \Lambda_r/ H\Lambda_r$ give the holonomy $\text{diag} \, (e^{2\pi i \frac{P}{H}t_1}, \ldots, e^{2\pi i \frac{P}{H}t_N})$, the center that can relate elements in $\Lambda_r/ H\Lambda_r$ take a form of $e^{2\pi i \frac{P}{H} \mathbf{c}}$, for example, with $\mathbf{c}:=(\underbrace{c, \ldots , c}_{N-m}, \underbrace{c-H, \ldots c-H}_{m})$, $c \in \mathbb{Z}_H$ where $m$ is an integer such that $Nc=mH$ which comes from the condition $\sum_{i=1}^{N} c_i=0$.
For instance, when $N=2$ discussed in the previous section, elements can be related by the nontrivial center when $H$ is an even number, and given such an $H$ we can see from $Nc=mH$ that $c$ can be $0$ or $\frac{H}{2}$.
Also, when $N=2$ and if $H$ is an odd number, only possible $c$ that satisfies $Nc=mH$ is zero, which gives the identity $I_2$.

Given a $\mathbf{t} = (t_1, t_2, \ldots, t_N)$, the action of the center, for example, $\mathbf{c}$ on $\mathbf{t}$ doesn't affect $e^{-\frac{\pi i}{H} \sum_{j=1}^{N} t_j b_j(\vec{n})}$.
The difference of $e^{-\frac{\pi i}{H} \sum_{j=1}^{N} t_j b_j(\vec{n})}$ between with $\mathbf{t}+\mathbf{c}=(t_1+c, \ldots, t_{N-m}+c, t_{N-m+1}+c-H, \ldots, t_N+c-H)$ and with $\mathbf{t}=(t_1, \ldots, t_N)$ is $e^{m \pi i \sum_{j=N-m+1}^{N} b_j(\vec{n})}$.
We check whether $m \sum_{j=N-m+1}^{N} b_j(\vec{n})$ is even.
As noted earlier, given an $F$ and $P_j$'s, $n_{i,j}$'s take value in all odd numbers or all even numbers, and the total number of $n_{i,j}$'s in $b_j(\vec{n})$ for $G=SU(N)$ is $N-1$.
Therefore, when $N$ is odd, $N-1$ is even, so $b_j(\vec{n})$'s are all even.
Thus, $m \sum_{j=N-m+1}^{N} b_j(\vec{n})$ is even when $N$ is odd.
Meanwhile, when $N$ is even, $N-1$ is odd.
When $H$ is even, from the conditions on $P_j$'s and $Q_j$'s, $P_j$'s should be all odd, which makes all $n_{i,j}$'s to take values in even numbers when considering $P\alpha = P \sum_{j=1}^{F} \frac{\epsilon_j}{P_j}$, \eqref{sinh-gen}, and \eqref{cosh-gen} in section \ref{sec:su2fg}. 
Therefore $b_j(\vec{n})$'s are all even, so $m \sum_{j=N-m+1}^{N} b_j(\vec{n})$ is even.
For the case that $H$ is odd, from $Nc=mH$, as the LHS is even but $H$ is odd $m$ should be even.
Accordingly, $m \sum_{j=N-m+1}^{N} b_j(\vec{n})$ is even.
Therefore, whatever $N$, $H$, $F$, $P_j$'s, and $Q_j$'s are, elements that are related by the action of the center give the same $e^{-\frac{\pi i}{H} \sum_{j=1}^{N} t_j b_j(\vec{n})}$.
This implies that the elements or the abelian flat connections related by the action of the center give the same contribution to the analytically continued CS partition function or the WRT invariant up to the $e^{\pi i K \frac{P}{H} \sum_{j=1}^{N} t_j^2}$ factor.

Meanwhile, such an action can change $e^{\pi i K \frac{P}{H} \sum_{j=1}^{N} t_j^2}$ for some cases.
Here, we consider the case with $K \in \mathbb{Z}$.
After some calculations, possible difference is given by $e^{\pi i K P(N+m)c}$.
If $P(N+m)c$ is even, then the elements related by such an action, which give the same contribution to the WRT invariant, have the same factor $e^{\pi i K \frac{P}{H} \sum_{j=1}^{N} t_j^2}$.
However, if $P(N+m)c$ is odd, they have different coefficients by $e^{\pi i K}$.
When $N$ is odd, in order for $P(N+m)c$ to be odd, both $c$ and $P$ should be odd and $m$ should be even.
However, this leads that, from $Nc=mH$, the LHS is odd but the RHS is even, which is a contradiction.
Thus, $N$ cannot be odd for $P(N+m)c$ to be odd and the extra factor $e^{\pi i K}$ doesn't appear in the case of odd $N$.
Meanwhile, if $N$ is even, both $c$ and $m$ should be odd to make $P(N+m)c$ to be odd.
From $Nc=mH$, we see that $H$ should be even.
Therefore, if there are even $N$, even $H$, odd $m$, and odd $c$ that satisfy $Nc=mH$, there is an extra factor $e^{\pi i K}$.
When $N=2$, we saw in the previous section that the extra factor $e^{\pi i K}$ appeared when $H$ is a multiple of 2 but not a multiple of 4, which is consistent with the above discussion applied to the $SU(2)$ case.
When $N$ is even and $H$ is odd, there is no extra factor $e^{\pi i K}$.

As a remark, since it is expected that contributions from all non-abelian flat connections can be obtained from the homological blocks in the resurgent analysis, the symmetry above would imply that the contributions from non-abelian flat connections that are related by the action of the center are also the same.
As this is a symmetry at the level of full quantum contributions to the partition function with respect to flat connections, this would also imply that there is a symmetry of the moduli space of all connections under the action of the center.\footnote{This was also discussed in \cite{Cheng:2018vpl} for $G=SU(2)$.}

\subsubsection*{\underline{\textnormal{Weyl orbit of $\mathbf{t}$ and of $-\mathbf{t}$}}}

There are cases that element $\mathbf{t}$ and $-\mathbf{t}$, which are complex conjugate to each other at the level of holonomy, are not in the same Weyl orbit.
For example, when $N=3$ and $H=4$, $(2,-1,-1)$ and $(-2,1,1)$ are in different Weyl orbits.
We can see that the contributions from the Weyl orbit of $\mathbf{t}$ and from the Weyl orbit of $-\mathbf{t}$ to the analytically continued CS partition function or the WRT invariant are the same.
This means that the contributions from the abelian flat connection corresponding to the Weyl orbit of $\mathbf{t}$ and from the conjugate abelian flat connection corresponding to the Weyl orbit of $-\mathbf{t}$ are the same.

In \eqref{sun-a-wrt2}, we consider renaming or permuting $n_{i,j}$'s.
If we permute indices of $n_{i,j}$'s with a convention $i<j$ in such a way that $i \leftrightarrow (N-1)-i$ for $i=1, \ldots, \frac{N-1}{2}$ and $N-1 \leftrightarrow N$ with $\frac{N-1}{2}$ staying the same when $N$ is odd, or $i \leftrightarrow (N-1)-i$ for $i=1, \ldots, \frac{N-2}{2}$ and $N-1 \leftrightarrow N$ when $N$ is even, we find $b_i \leftrightarrow -b_{(N-1)-i}$ and $b_{N-1} \leftrightarrow -b_N$.
Under this permutation, we see that $q^{\frac{1}{8HP} \sum_{m=1}^{N} b_m(\vec{n})^2}$ stays the same.
Since the Weyl orbit of $\mathbf{t}$ contains all possible permutations of $t_i$'s in $\Lambda_r/H\Lambda_r$, $e^{-\frac{\pi i}{H} \sum_{j=1}^N t_j b_j(\vec{n})}$ on a given Weyl orbit of $\mathbf{t}$ in $\Lambda_r/H\Lambda_r$ is the same as $e^{-\frac{\pi i}{H} \sum_{j=1}^N (-t_j) b_j(\vec{n})}$ on Weyl orbit of $-\mathbf{t}$ in $\Lambda_r/H\Lambda_r$ with above permutations on $n_{i,j}$'s and with $t_i \leftrightarrow t_{(N-1)-i}$ and $t_{N-1} \leftrightarrow t_N$.
Also, obviously $e^{\pi i K \frac{P}{H} \sum_{j=1}^{N} t_j^2}$ is the same for $\mathbf{t}$ and $-\mathbf{t}$.
Thus, the contributions from the abelian flat connection corresponding to the Weyl orbtit of $\mathbf{t}$ and from the conjugate abelian flat connection corresponding to the Weyl orbit of $-\mathbf{t}$ to the partition function of the analytically continued CS theory or the WRT invariant are the same.
As in the case of the action of the center, this symmetry would also imply that there is a symmetry of the moduli space of all connections under the complex conjugation.

\subsubsection*{\underline{\textnormal{General structure}}}

In order to write an expected general expression for the $S$-matrix, we introduce some notations.
From now on, we take $K \in \mathbb{Z}$.
We group elements in $\Lambda_r/H\Lambda_r$ by Weyl orbits.
Among Weyl orbits, there are cases that an orbit contains both $\mathbf{t}$ and $-\mathbf{t}$ up to $H \Lambda_r$. 
We denote such a Weyl orbit of $\mathbf{t}$ in $\Lambda_r/H\Lambda_r$ by $W_t$ where $t$ is a label for abelian flat connections whose holonomy $\text{Hol}_A(h)$ is conjugated to $e^{2\pi i \frac{P}{H} \mathbf{t}}$.
There are also cases that a Weyl orbit containing $\mathbf{t}$ is distinct from the one containing $-\mathbf{t}$.
We denote such orbits as $W_t$ and $W_{-t}$, respectively.
For example, when $N=4$ and $H=4$, which is one of examples discussed below, the Weyl orbit of $(1, -1, 0, 0)$ contains $(-1, 1, 0, 0)$ up to $4 \Lambda_3$ so we denote the orbit by $W_1$ with a label $t=1$.
Meanwhile, the Weyl orbits of $(2, -1, -1, 0)$ and $(-2, 1, 1, 0)$ are not the same up to $4 \Lambda_3$, so we denote them by $W_3$ and $W_{-3}$ with a label $t=3$, respectively.

In some cases, it happens that some elements $\mathbf{t}$'s are related by the action of the center.
As they give the same contributions to the WRT invariant up to an overall coefficient $e^{\pi i K}$, we group the Weyl orbits by the orbits of them under the action of the centers.
We denote such orbits by $C_a$ where $a$ is a label for the abelian flat connections corresponding to $W_t$'s in $C_a$. 
There are cases that $W_t$ and $W_{-t}$ are in the same $C_a$. 
For example, when $N=4$ and $H=4$, $C_0 = \{ W_0, W_2, W_7, W_{-7} \}$ where $W_0, W_2, W_7$, and $W_{-7}$ denote Weyl orbits of $(0,0,0,0)$, $(2,-2,2,-2)$, $(3,-1,-1,-1)$, and $(-3,1,1,1)$ in $\Lambda_3/4\Lambda_3$, respectively.
There are also cases that orbits $W_t$ and $W_{-t}$ are not related by the action of the center.
As we discussed above, the contributions from $W_t$ and from $W_{-t}$ to the WRT invariant are the same, so we put them in the same class with their orbits under the action of the centers, which we denote by $C_{\pm a}$.
For instance, when $N=4$ and $H=6$, which is one of examples discussed below, $W_{12}$ and $W_{14}$ denote Weyl orbits of $(3,-1,-1,-1)$ and $(0,2,2,-4)$ in $\Lambda_3/6\Lambda_3$, respectively, and they are related by the action of the center, \textit{i.e.} $(-3,3,3,-3)$ in $\Lambda_3/6\Lambda_3$.
We see that $\{ W_{12}, W_{14} \} \neq \{ W_{-12}, W_{-14}\}$ but they are complex conjugate to each other at the level of holonomy.	
We put them together and denote $``C_8" = C_{\pm 8} = \{W_{12}, W_{14}, W_{-12}, W_{-14}\}$.
\\

We denote elements in the Weyl orbit $W_t$ in $\Lambda_r/H\Lambda_r$ by $\tilde{t}$.
Also, we denote any representative of any of $W_t$ in $C_b$ or $C_{\pm b}$ by $\tilde{b}$.
Then from examples we worked out, we expect that when $G=SU(N)$ the $S$-matrix for Seifert manifolds is given by
\begin{align}
S_{ab} = \frac{1}{\sqrt{\text{gcd}(N,H)}} \sum_{W_t \in C_a} \frac{\sum_{\tilde{t} \in W_t} e^{2\pi i \, lk (\tilde{t}, \tilde{b})} }{|\text{Tor} \, H_1(M_3,\mathbb{Z})|^{\frac{N-1}{2}}}	\,	.	\label{sun-smat}
\end{align}
and 
\begin{align}
lk(a,b) = \frac{P}{H} \sum_{j=1}^{N} a_j b_j	\label{sun-lk}
\end{align}
with $a_N = -\sum_{j=1}^{N-1} a_j$ and similarly for $b_N$.
Here, $C_a$ in the summation can be $C_a$'s or $C_{\pm a}$'s depending on the $a$ under consideration.
For example, as mentioned above, ``$C_8$" in the case of $N=4$ and $H=6$ is $C_{\pm 8}$ and $W_t \in C_{\pm 8}$ are $W_{12}$, $W_{14}$, $W_{-12}$, and $W_{-14}$ in $\Lambda_3/6\Lambda_3$.
The overall normalization was chosen in such a way that $S_{ab}$ satisfies $S^2=I$, which is from the interpretation of the $S$-matrix as the $S$-duality of Type IIB string theory \cite{Gukov-Pei-Putrov-Vafa}.
We expect that \eqref{sun-smat} holds for large classes of closed 3-manifolds.	\\

Given the $S$-matrix above, from examples we considered, a general form of the WRT invariant for the Seifert manifolds is expected to take
\begin{align}
Z_{SU(N)}(M_3) = \mathcal{C} \, \text{gcd}(N,H)^\frac{1}{2} H^{\frac{N-1}{2}} \sum_{a, b} e^{\pi i K lk(a,a)} S_{ab} \widehat{Z}_b(q)	\,	\Big|_{q \searrow e^{\frac{2\pi i}{K}}}	\label{sun-wrt0}
\end{align}
when $N$ is odd or $N$ is even such that there is no odd $m$, odd $c$, and even $H$ that satisfy $Nc=mH$.
Here the overall factor $\mathcal{C}$ is
\begin{align}
\hspace{-5mm}\mathcal{C} = (N K^{N-1} i^{N(N-1)})^{\frac{1}{2} b_1(M_3)} \frac{1}{N!} \frac{1}{N^{3/2}} (-1)^{\frac{(N-1)(N-2)}{2}} \Big( \frac{i}{K} \frac{P}{H}\Big)^{\frac{N-1}{2}}  \frac{\text{sign}(P)^{\frac{N(N-1)}{2}} }{|P|^{\frac{N-1}{2}}} e^{\frac{\pi i (N^2-1)}{4} \text{sign}(H/P)} q^{-\frac{1}{24} (N^2-1) N \phi_F }	\label{overall-factor}
\end{align}
where $b_1(M_3)$ is the first Betti number of $M_3$.
Also, $\frac{1}{2}lk(a,a)=CS_a$ is interpreted as the Chern-Simons invariant for the abelian flat connection $a$ and we note that the factor $e^{\pi i K \frac{P}{H} \sum_{j=1}^{N} t_j^2}$ in \eqref{sun-a-wrt2} can be expressed as $e^{\pi i K lk(t,t)}$ where $t$ denotes $(t_1, \ldots, t_N)$.

Meanwhile, when $N$ is even and there are odd $m$, odd $c$, and even $H$ that satisfy $Nc=mH$, the WRT invariant is expected to take a form of
\begin{align}
Z_{SU(N)}(M_3) = \frac{1}{2} \, \mathcal{C} \, \text{gcd}(N,H)^\frac{1}{2} H^{\frac{N-1}{2}} \sum_{\dot{a}, \dot{b}} e^{\pi i K lk(\dot{a},\dot{a})} (Y \otimes S_{ab})_{\dot{a} \dot{b}} \widehat{Z}_{\dot{b}}(q)	\,	\Big|_{q \searrow e^{\frac{2\pi i}{K}}}		\label{sun-wrt1}
\end{align}
where $(Y \otimes S_{ab})_{\dot{a}\dot{b}} = \begin{pmatrix} S_{ab} & S_{ab} \\ S_{ab} & S_{ab} \end{pmatrix}_{\dot{a} \dot{b}}$.
If we denote the total number of distinct orbits of $W_t$'s under the action of the center up to the complex conjugate as $L$, we have $\dot{a}, \dot{b} = 0,1, \ldots, 2L-1$, $\widehat{Z}_{\dot{b}}(q) = (\widehat{Z}_0, \ldots, \widehat{Z}_{2L-1})^{T}$ and $lk(\dot{a},\dot{a}) = (lk(0,0), \ldots, lk(L-1,L-1),lk(0,0)+1, \ldots, lk(L-1,L-1)+1)$.
We note that $Z_{\dot{a}}(q) = Z_{\dot{a}+L}(q)$, $\dot{a}=0,\ldots, L-1$ where $Z_{\dot{a}}(q) = (Y \otimes S_{ab})_{\dot{a} \dot{b}} \widehat{Z}_{\dot{b}}(q)$.
We also expect that the WRT invariant for large classes of closed 3-manifolds would take a form of \eqref{sun-wrt0} or \eqref{sun-wrt1}.	\\

In next sections, we provide a number of examples.
We omit the overall factor \eqref{overall-factor} in the following examples.


\subsection{The case $H=1$}
For integer homology Seifert manifolds, $H=1$, there is a contribution only from the trivial flat connection, $t_i=0$, $i=1,\cdots, N-1$ to the partition function of the analytically continued theory or the WRT invariant, which is
\begin{align}
Z_{SU(N)}(M_3) 
&=
\sum_{\substack{n_{i,j}=0 \\ 1 \leq i < j \leq N}}^{\infty} \Big( \prod_{1 \leq i < j \leq N} \chi_{2P}(n_{i,j}) \Big)
q^{ \frac{1}{8 P} \sum_{m=1}^{N} b_m(\vec{n})^2}
\label{sun-a-wrt-triv}
\end{align}
Or by using \eqref{3f-chi-decomp}, $Z_{SU(N)}(M_3)$ can be written\footnote{For example, when $F=4$, $Z_{SU(N)}(M_3) $ is given by
\begin{align}
Z_{SU(N)}(M_3) \simeq \sum_{\substack{n_{i,j}=0 \\ 1 \leq i < j \leq N}}^{\infty} \bigg( \prod_{1 \leq i < j \leq N} \frac{1}{P} \sum_{s_{i,j}=0}^{7} (-1)^{s_{i,j}} \big( n_{i,j} \varphi_{2P}^{(R_{s_{i,j}})}(n_{i,j}) -R_{s_{i,j}} \psi_{2P}^{(R_{s_{i,j}})}(n_{i,j})\big)   \bigg) q^{ \frac{1}{8 P} \sum_{m=1}^{N} b_m(\vec{n})^2}
\end{align}
where the definition of $\varphi^{(l)}_{2P}(n)$ is available at \eqref{varphi}.
}
 as
\begin{align}
Z_{SU(N)}(M_3) 
&=
\sum_{\substack{n_{i,j}=0 \\ 1 \leq i < j \leq N}}^{\infty} \Big( \prod_{1 \leq i < j \leq N} \sum_{s_{i,j}=0}^{3} \psi_{2P}^{(R_{s_{i,j}})}(n_{i,j}) \Big) q^{ \frac{1}{8 P} \sum_{m=1}^{N} b_m(\vec{n})^2}	\,	.
 \label{sun-a-wrt-triv2}
\end{align}
We provide examples as a series expansion to express them explicitly.
From examples, we see that the analytically continued CS partition function or the WRT invariant is indeed expressed as a single homological block $\widehat{Z}_0$
\begin{align}
Z_{SU(N)}(M_{3}) =  \widehat{Z}_0 (q)	\,	.
\end{align}
\vspace{-5mm}

\subsubsection*{$\bullet$ \textnormal{$G=SU(3)$ and $(P_1, P_2, P_3) = (2,3,7)$}}
\begin{align}
\widehat{Z}_0(q) = q^{\frac{1}{42}}( 1-2 q+2 q^3+q^4-2 q^5-2 q^8+4 q^9+2 q^{10}-4 q^{11}+2 q^{13}-6 q^{14}+ \cdots)	\,	.
\end{align}

\subsubsection*{$\bullet$ \textnormal{$G=SU(3)$ and $(P_1, P_2, P_3) = (2,5,7)$}}
\begin{align}
\widehat{Z}_0(q) = q^{1+\frac{51}{70}} ( 1-2 q^3-2 q^5+2 q^6+2 q^9+q^{12}-2 q^{14}+2 q^{15}-2 q^{18}-3 q^{20}+6 q^{21}+ \cdots)	\,	.
\end{align}

\subsubsection*{$\bullet$ \textnormal{$G=SU(3)$ and $(P_1, P_2, P_3) = (5,11,13)$}}
\begin{align}
\widehat{Z}_0(q) = q^{285+\frac{529}{715}} ( 1-2 q^{39}-2 q^{47}+2 q^{81}+2 q^{96}+2 q^{117}-2 q^{119}+2 q^{141}+2 q^{147}+2 q^{153}+ \cdots)	\,	.
\end{align}

\subsubsection*{$\bullet$ \textnormal{$G=SU(4)$ and $(P_1, P_2, P_3) = (2,5,7)$}}
\begin{align}
\widehat{Z}_0(q) = q^{4+\frac{9}{28}} ( 1-3 q^3-3 q^5+5 q^6-q^7+2 q^8+3 q^9-q^{10}-q^{12}+2 q^{13}-6 q^{14}+2 q^{15}-q^{16}+ \cdots)	\,	.
\end{align}

\subsubsection*{$\bullet$ \textnormal{$G=SU(4)$ and $(P_1, P_2, P_3) = (3,5,7)$}}
\begin{align}
\widehat{Z}_0(q) = q^{27+\frac{11}{21}} (1-3 q^7-3 q^{11}+q^{14}+4 q^{15}+2 q^{18}-q^{19}+4 q^{21}+q^{22}-4 q^{23}+3 q^{24}+ \cdots)	\,	.
\end{align}


\subsection{The case $H \geq 2$}
When $H \geq 2$, we have a sum over $\mathbf{t} \in \Lambda_r / H\Lambda_r$ in \eqref{sun-a-wrt2}.
We denote the simple roots by $\alpha_{i,i+1} = e_i-e_{i+1}$, $i=1, \ldots, N-1$, where $\{e_i\}_{i=1, \ldots, N}$ are the standard orthonormal basis.
We provide some examples for $H \geq 2$.

\subsubsection{$G=SU(3)$}
When $G=SU(3)$, we obtain
\begin{align}
\begin{split}
\sum_{{\bf t} \in \Lambda_{2} / H \Lambda_{2}}
\Bigg[
e^{2 \pi i K  \frac{P}{H} \sum_{m=1}^{N} (t_1^2 + t_2^2 + t_1 t_2) }	\sum_{\substack{n_{i,j}=0 \\ 1 \leq i < j \leq 3}}^{\infty}\Big( \prod_{1 \leq i < j \leq 3} \chi_{2P}(n_{i,j})\Big) \,
e^{-\frac{ \pi i}{H} \sum _{i=1}^{3} t_i b_i(\vec{n})}	\,
q^{ \frac{1}{8 H P} \sum_{i=1}^{3} b_i(\vec{n})^2}
\Bigg]	\label{su-3-wrt}
\end{split}
\end{align}
up to the overall factor where 
\begin{align}
b_1(\vec{n}) = n_{1,2}-n_{1,3}	\,	,	\quad	
b_2(\vec{n}) = -n_{1,2}-n_{2,3}	\,	,	\quad	
b_3(\vec{n}) = n_{1,3}+n_{2,3}	\,	,
\end{align}
and $t_3 = -t_1-t_2$.
In the following examples, we write the WRT invariant in terms of homological blocks.

\subsubsection*{\textbullet \ $H=2$}
We begin with the case $H=2$ from $(P_1,P_2,P_3) = (3,5,7)$.
There are two Weyl orbits in $\Lambda_2/2\Lambda_2$,
\begin{alignat}{3}
&W_{t}			&&\hspace{10mm}		\Lambda_2/2\Lambda_2	\nonumber	\\
\vspace{1mm}
&W_0 			&&\hspace{10mm}		\{0\}		\nonumber	\\
&W_1 			&&\hspace{10mm}		\{\alpha _{12},\alpha _{23},\alpha _{12}+\alpha _{23}\}	\nonumber
\end{alignat}
where $\alpha_{12}=(1,-1,0)$, $\alpha_{13}=(1,0,-1)$, and $\alpha_{23}=(0,1,-1)$.
$W_0$ corresponds to the trivial flat connection and $W_1$ to the abelian flat connection from the type $S(U(2)\times U(1))$.
The contributions from $W_0$ and $W_1$ are denoted by $Z_0$ and $Z_1$, which are written in terms of two homological blocks $\widehat{Z}_0$ and $\widehat{Z}_1$,
\begin{align}
Z_0&=\widehat{Z}_0+\widehat{Z}_1	\,	,	\\	
Z_1&=3 \widehat{Z}_0-\widehat{Z}_1		\	
\end{align}
where
\begin{align}
\widehat{Z}_0 &= q^{5+\frac{53}{105}} (1+2 q^{12}-q^{14}-2 q^{16}-3 q^{22}+2 q^{24}-2 q^{28}+2 q^{36}-2 q^{38}+\cdots )	\,	,	\\
\widehat{Z}_1 &= -2q^{9+\frac{1}{210}} (1+q^2-q^4-q^7+q^8-q^9-q^{10}-q^{13}+q^{14}+2 q^{17}-q^{19}+\cdots )		\,	.
\end{align}
Thus, the WRT invariant is given by
\begin{align}
Z_{SU(3)}(M_{3}) = \sum_{a,b =0}^{1} e^{2 \pi i K CS_{a}} S_{ab} \widehat{Z}_{b}(q)	\,	\Big|_{q \searrow e^{\frac{2\pi i}{K}}}
\end{align}
where $(CS_0, CS_1)=(0, \frac{1}{2})$ and 
\begin{align}
S_{ab} = \frac{1}{2} \begin{pmatrix} 1 & 1 \\ 3 & -1\end{pmatrix}	\,	.
\end{align}
This $S$-matrix can also be obtained from \eqref{sun-smat} and \eqref{sun-lk} with $C_0 = \{W_0 \}$ and $C_1=\{ W_1 \}$.

\subsubsection*{\textbullet \ $H=3$}
We consider $H=3$, for example, by choosing $(P_1,P_2,P_3)=(2,5,7)$.
In the case of $H=3$, there are four Weyl orbits
\begin{alignat}{3}
\label{su3h3a0}
&W_{t}	&&\hspace{10mm}	\Lambda_2 / 3 \Lambda_2		\nonumber	\\
\vspace{1mm}
&W_0	&&\hspace{10mm}	\{0\}						\nonumber	\\
&W_1	&&\hspace{10mm}	\{\alpha_{12},\alpha_{13},\alpha_{23}	,2\alpha_{12},2\alpha_{13},2\alpha_{23}\}	\nonumber	\\
&W_2	&&\hspace{10mm}	\{\alpha_{12}+\alpha_{13}\}	\nonumber	\\
&W_3	&&\hspace{10mm}	\{\alpha_{13}+\alpha_{23}\}	\nonumber
\end{alignat}
in $\Lambda_2 / 3 \Lambda_2$.
Here, $W_0$ corresponds to the trivial flat connection, $W_2$ and $W_3$ to the central flat connections, and $W_1$ to the abelian flat connection $U(1)^2$.
So we have $C_0 = \{ W_0, W_2, W_3\}$ and $C_1 = \{ W_1 \}$.
The contributions from $W_0$, $W_2$, and $W_3$ are all the same.
We denote the contributions from $W_0$ and $W_1$ by $Z_0$ and $Z_1$, respectively.
Then, the WRT invariant is written as
\begin{align}
Z_{SU(3)}(M_3) = 3Z_0 + e^{\frac{2\pi i K}{3}}Z_1	\,	.
\end{align}
In terms of homological blocks, these $Z_a$'s are given by
\begin{align}
Z_0 &= \widehat{Z}_0+\widehat{Z}_1	\,	,	\\
\frac{1}{3}Z_1 &= 2 \widehat{Z}_0 -\widehat{Z}_1 
\end{align}
with
\begin{align}
\widehat{Z}_0 &=  -q^{2+\frac{17}{70}}\big(2+2 q^3+3 q^5+4 q^6+4 q^9+4 q^{11}+2 q^{12}+4 q^{14}+8 q^{18}+ \cdots \big)	\,	,	\\
\widehat{Z}_1 &= q^{\frac{121}{210}} \big(1-2 q+2 q^2+2 q^3+q^4+2 q^5-2 q^6+6 q^7-2 q^8+2 q^9+2 q^{10}	 + \cdots\big)	\,	.
\end{align}
Thus, we have
\begin{align}
Z_{SU(3)}(M_3) = 3 \sqrt{3} \sum_{a,b=0}^{1} e^{2 \pi i K CS_{a}} S_{ab} \widehat{Z}_b(q)	\,	\Big|_{q \searrow e^{\frac{2\pi i}{K}}}
\end{align}
with $(CS_0,CS_1) = (0, \frac{1}{3})$ and 
\begin{align}
S_{ab}=\frac{1}{\sqrt{3}}\begin{pmatrix} 1 & 1 \\ 2 & -1 \end{pmatrix}	\,	,
\end{align}
which satisfies $S_{ab}S_{bc} =\delta_{ac}$.
The $S$-matrix can also be calculated from \eqref{sun-smat} with \eqref{sun-lk} and it agrees with the above $S$-matrix.

\subsubsection*{\textbullet \ $H=4$}
We can choose, for example, $(P_1,P_2,P_3)=(3,5,7)$ to have $H=4$.
There are five Weyl orbits for the elements in $\Lambda_2/4\Lambda_2$,
\begin{alignat}{3} 
&W_{t}	&&\hspace{10mm}	\Lambda_2/4\Lambda_2	\nonumber	\\
\vspace{1mm}
&W_0	&&\hspace{10mm}	\{0\}					\nonumber	\\
&W_1	&&\hspace{10mm}	\{\alpha _{12},\alpha _{23},\alpha _{12}+\alpha _{23},3 \alpha _{12},3 \alpha _{23},3 \alpha _{12}+3 \alpha _{23}\}	\nonumber	\\
&W_2	&&\hspace{10mm}	\{2 \alpha _{12},2 \alpha _{23},2 \alpha _{12}+2 \alpha _{23}\}						\nonumber	\\
&W_3	&&\hspace{10mm}	\{2 \alpha _{12}+\alpha _{23},3 \alpha _{12}+\alpha _{23},3 \alpha _{12}+2 \alpha _{23}\}	\nonumber	\\
&W_{-3}	&&\hspace{10mm}	\{\alpha _{12}+2 \alpha _{23},\alpha _{12}+3 \alpha _{23},2 \alpha _{12}+3 \alpha _{23}\}	\nonumber	\,	.
\end{alignat}
$W_0$ corresponds to the trivial flat connection, $W_1$ to the abelian flat connection $U(1)^2$, and $W_2$, $W_3$, and $W_{-3}$ to the abelian flat connections from the type $S(U(2)\times U(1))$.
The contributions from $W_0$, $W_1$, and $W_2$ are denoted by $Z_0$, $Z_1$, and $Z_2$, respectively.
Abelian flat connections corresponding to $W_3$ and $W_{-3}$ are complex conjugate to each other, and their contributions are the same.
We denote the sum of their contributions by $Z_3$. 
$Z_a$'s are written in terms of homological blocks
\begin{align}
Z_0&=\widehat{Z}_0+\widehat{Z}_1+\widehat{Z}_2+\widehat{Z}_3	\,	,	\\
Z_1&=6 \widehat{Z}_0-2 \widehat{Z}_1-2 \widehat{Z}_2+2 \widehat{Z}_3	\,	,	\\
Z_2&=3 \widehat{Z}_0-\widehat{Z}_1+3 \widehat{Z}_2-\widehat{Z}_3		\,	,	\\
Z_3&=6 \widehat{Z}_0+2 \widehat{Z}_1-2 \widehat{Z}_2-2 \widehat{Z}_3
\end{align}
where
\begin{align}
\widehat{Z}_0 &= -q^{6+\frac{79}{105}} (1+3 q^4+2 q^{12}+3 q^{16}+2 q^{28}+2 q^{48}+2 q^{52}+q^{64}+4 q^{68} + \cdots)	\,	,	\\
\widehat{Z}_1 &= -2 q^{8+\frac{1}{420}}(1+q+q^3-2 q^5+q^6-q^7+2 q^9+q^{10}+2 q^{12}+q^{13}-q^{14} + \cdots)			\,	,	\\
\widehat{Z}_2 &= q^{2+\frac{79}{105}}(1+2 q^6-2 q^8+2 q^{12}-2 q^{14}+2 q^{18}-2 q^{20}+q^{24}-2 q^{26}+2 q^{28} + \cdots)	\,	,	\\
\widehat{Z}_3 &= -2 q^{2+\frac{211}{420}}(1+q-q^2+q^4-q^5+q^7+q^{10}+q^{13}-2 q^{14}-q^{15}+2 q^{16} + \cdots)			\,	.
\end{align}
Therefore, the WRT invariant can be written as
\begin{align}
Z_{SU(3)}(M_3) = \sum_{a,b} e^{2\pi i K CS_a} Z_a(q) = 4 \sum_{a,b} e^{2\pi i K CS_a} S_{ab} \widehat{Z}_b(q)	\,	\Big|_{q \searrow e^{\frac{2\pi i}{K}}}
\end{align}
where 
\begin{align}
S_{ab} = \frac{1}{4} \left(
\begin{array}{cccc}
 1 & 1 & 1 & 1 \\
 6 & -2 & -2 & 2 \\
 3 & -1 & 3 & -1 \\
 6 & 2 & -2 & -2 \\
\end{array}
\right)
\end{align}
and
$(CS_0, CS_1, CS_2,CS_3)=(0,\frac{1}{2},0,\frac{1}{2})$.
This $S$-matrix can also be obtained from \eqref{sun-smat} and \eqref{sun-lk} with $C_a=\{W_a\}$ for $a=0,1,2$ and $C_{\pm 3}=\{ W_{3}, W_{-3} \}$.

We may rearrange terms in $e^{\pi i K} Z_1$ and $e^{\pi i K} Z_3$ as they have the same exponential factor.
After rearrangement, it is possible to put them in a form
\begin{align}
\left(
\begin{array}{c}
 Z_0 \\
 Z_1' \\
 Z_2 \\
 Z_3' \\
\end{array}
\right) = 
\left(
\begin{array}{cccc}
 1 & 1 & 1 & 1 \\
 3 & -1 & 3 & -1 \\
 3 & 3 & -1 & -1 \\
 9 & -3 & -3 & 1 \\
\end{array}
\right) \left(
\begin{array}{c}
 \widehat{Z}_{0} \\
 \widehat{Z}_2 \\
 \widehat{Z}_3 \\
 \widehat{Z}_1 \\
\end{array}
\right)
\end{align}
where $Z_1+Z_3 = Z_1'+Z_3'$.
The $S$-matrix above is given by a tensor product of the $S$-matrix of $H=2$, \textit{i.e.} $\begin{pmatrix} 1 & 1 \\ 3 & -1 \end{pmatrix} \otimes \begin{pmatrix} 1 & 1 \\ 3 & -1 \end{pmatrix}$.

\subsubsection*{\textbullet \ $H=5$}
We consider $H=5$, for example, from $(P_1,P_2,P_3) = (2,3,7)$.
In $\Lambda_2/5\Lambda_2$, there are seven Weyl orbits
\begin{alignat}{3}
&W_{t}		&&\hspace{10mm}	\Lambda_2/5\Lambda_2	\nonumber	\\
\vspace{1mm}
&W_0		&&\hspace{10mm}	\{0\}					\nonumber	\\
&W_1		&&\hspace{10mm}	\{\alpha _{12},\alpha _{13},\alpha _{23},4 \alpha _{12},4 \alpha _{13},4 \alpha _{23}\}			\nonumber	\\
&W_2		&&\hspace{10mm}	\{2 \alpha _{12},2 \alpha _{13},2 \alpha _{23},3 \alpha _{12},3 \alpha _{13},3 \alpha _{23}\}		\nonumber	\\
&W_3		&&\hspace{10mm}	\{\alpha _{12}+\alpha _{13},3 \alpha _{12}+\alpha _{13},\alpha _{12}+3 \alpha _{13}\}			\nonumber	\\
&W_{-3}		&&\hspace{10mm}	\{\alpha _{13}+\alpha _{23},3 \alpha _{13}+\alpha _{23},\alpha _{13}+3 \alpha _{23}\}			\nonumber	\\
&W_4		&&\hspace{10mm}	\{2 \alpha _{12}+2 \alpha _{13}, \alpha _{12}+2 \alpha _{13},2 \alpha _{12}+ \alpha _{13}\}		\nonumber	\\
&W_{-4}		&&\hspace{10mm}	\{2 \alpha _{13}+2 \alpha _{23},\alpha _{13}+2 \alpha _{23}, +2 \alpha _{13}+\alpha _{23}\}	\nonumber	\,	.	
\end{alignat}
$W_0$ corresponds to trivial flat connection, $W_1$ and $W_2$ to the abelian flat connections $U(1)^2$, and $W_{3}$, $W_{-3}$, $W_{4}$, and $W_{-4}$ to the abelian flat connections from the type $S(U(2)\times U(1))$.
The contributions from $W_0$, $W_1$, and $W_2$ are denoted by $Z_0$, $Z_1$, and $Z_2$, respectively.
Abelian flat connections corresponding to $W_3$ and $W_{-3}$ (resp. $W_4$ and $W_{-4}$) are complex conjugate to each other and contributions from them are the same.
We denote their sum by $Z_3$ (resp. $Z_4$).
Then the WRT invariant is written as
\begin{align}
Z_{SU(3)}(M_3) = Z_0 + e^{\frac{4 i \pi  K}{5}} Z_1 + e^{\frac{6 \pi i K}{5}}Z_2 + e^{\frac{2 \pi i K}{5}}Z_3 + e^{\frac{8 \pi i K}{5}}Z_4	\,	.
\end{align}
Each $Z_a$'s are given by
\begin{align}
Z_0&=\widehat{Z}_0+\widehat{Z}_1+\widehat{Z}_2+\widehat{Z}_3+\widehat{Z}_4	\,	,	\\
Z_1&=6 \widehat{Z}_0+\frac{1}{2} \left(-\sqrt{5}-3\right) \widehat{Z}_1+\frac{1}{2} \left(\sqrt{5}-3\right) \widehat{Z}_2+\left(\sqrt{5}+1\right) \widehat{Z}_3+\left(1-\sqrt{5}\right) \widehat{Z}_4	\,	,	\\
Z_2&=6 \widehat{Z}_0+\frac{1}{2} \left(\sqrt{5}-3\right) \widehat{Z}_1+\frac{1}{2} \left(-\sqrt{5}-3\right) \widehat{Z}_2+\left(1-\sqrt{5}\right) \widehat{Z}_3+\left(\sqrt{5}+1\right) \widehat{Z}_4	\,	,	\\
Z_3&=6 \widehat{Z}_0+\left(\sqrt{5}+1\right) \widehat{Z}_1+\left(1-\sqrt{5}\right) \widehat{Z}_2+\frac{1}{2} \left(\sqrt{5}-3\right) \widehat{Z}_3+\frac{1}{2} \left(-\sqrt{5}-3\right) \widehat{Z}_4	\,	,	\\
Z_4&=6 \widehat{Z}_0+\left(1-\sqrt{5}\right) \widehat{Z}_1+\left(\sqrt{5}+1\right) \widehat{Z}_2+\frac{1}{2} \left(-\sqrt{5}-3\right) \widehat{Z}_3+\frac{1}{2} \left(\sqrt{5}-3\right) \widehat{Z}_4	
\end{align}
where
\begin{align}
&\hspace{-3mm}\widehat{Z}_0 = -q^{4+\frac{17}{42}} (2+4 q^5+q^{10}+4 q^{15}+6 q^{25}+\cdots)	\,	,	\\	
&\hspace{-3mm}\widehat{Z}_1 = q^{\frac{169}{210}} (1+4 q-6 q^2+2 q^4-2 q^6-3 q^8+2 q^{10}-4 q^{11}+2 q^{12}+6 q^{13}+2 q^{14}+\cdots)	\,	,	\\
&\hspace{-3mm}\widehat{Z}_2 = q^{\frac{1}{210}} (1-2 q+2 q^2+2 q^3-q^4+2 q^5+6 q^6-2 q^7+q^8+6 q^9-2 q^{11}+\cdots)				\,	,	\\
&\hspace{-3mm}\widehat{Z}_3 = 2 q^{\frac{127}{210}} (1-q+q^2+2 q^3-2 q^4+3 q^6-q^7+2 q^9+q^{11}-q^{12}+q^{13}+q^{14}+\cdots)		\,	,	\\
&\hspace{-3mm}\widehat{Z}_4 = -2 q^{\frac{43}{210}} (1+2 q^2+q^3-2 q^4+2 q^5+q^6+2 q^8-q^9+q^{10}+3 q^{11}+3 q^{14}+\cdots)			\,	.
\end{align}
Hence, we obtain
\begin{align}
Z_{SU(3)}(M_{3}) = 5 \sum_{a,b} e^{2\pi i K CS_a} S_{ab} \widehat{Z}_b(q)		\,	\Big|_{q \searrow e^{\frac{2\pi i}{K}}}
\end{align}
with $(CS_{0},CS_{1},CS_{2},CS_{3},CS_{4}) = (0,\frac{2}{5},\frac{3}{5},\frac{1}{5},\frac{4}{5})$ and
\begin{align}
S_{ab} = \frac{1}{5} 
\left(
\begin{array}{ccccc}
 1 & 1 & 1 & 1 & 1 \\
 6 & \frac{1}{2} \left(-\sqrt{5}-3\right) & \frac{1}{2} \left(\sqrt{5}-3\right) & \sqrt{5}+1 & 1-\sqrt{5} \\
 6 & \frac{1}{2} \left(\sqrt{5}-3\right) & \frac{1}{2} \left(-\sqrt{5}-3\right) & 1-\sqrt{5} & \sqrt{5}+1 \\
 6 & \sqrt{5}+1 & 1-\sqrt{5} & \frac{1}{2} \left(\sqrt{5}-3\right) & \frac{1}{2} \left(-\sqrt{5}-3\right) \\
 6 & 1-\sqrt{5} & \sqrt{5}+1 & \frac{1}{2} \left(-\sqrt{5}-3\right) & \frac{1}{2} \left(\sqrt{5}-3\right) 
\end{array}
\right) 	\,	,
\end{align}
which satisfies $S_{ab}S_{bc} = \delta_{ac}$.
This $S$-matrix can also be obtained from \eqref{sun-smat} and \eqref{sun-lk} with $C_a =\{ W_a \}$, $a=0,1,2$, and $C_{\pm a} = \{ W_a, W_{-a}\}$, $a=3,4$.

\subsubsection*{\textbullet \ $H=6$}
It is possible to have $H=6$, for example, from $(P_1,P_2,P_3) = (5,7,11)$.
The elements in $\Lambda_2/6\Lambda_2$ are grouped by 10 Weyl orbits,
\begin{alignat}{3}
&\hspace{-3mm}W_t		&&\hspace{2mm}	\Lambda_2/6\Lambda_2	\nonumber	\\
\vspace{1mm}
&\hspace{-3mm}W_0	&&\hspace{2mm}	\{0\}					\nonumber	\\
&\hspace{-3mm}W_1	&&\hspace{2mm}	\{\alpha _{12},\alpha _{23},\alpha _{12}+\alpha _{23},5 \alpha _{12},5 \alpha _{23},5 \alpha _{12}+5 \alpha _{23}\}	\nonumber	\\
&\hspace{-3mm}W_2	&&\hspace{2mm}	\{2 \alpha _{12},2 \alpha _{23},2 \alpha _{12}+2 \alpha _{23},4 \alpha _{12},4 \alpha _{23},4 \alpha _{12}+4 \alpha _{23}\}	\nonumber	\\
&\hspace{-3mm}W_3	&&\hspace{2mm}	\{3 \alpha _{12},3 \alpha _{23},3 \alpha _{12}+3 \alpha _{23}\}	\nonumber	\\
&\hspace{-3mm}W_4	&&\hspace{2mm}	\{2 \alpha _{12}+\alpha _{23},5 \alpha _{12}+\alpha _{23},5 \alpha _{12}+4 \alpha _{23}\}	\nonumber	\\
&\hspace{-3mm}W_{-4}	&&\hspace{2mm}	\{\alpha _{12}+2 \alpha _{23},\alpha _{12}+5 \alpha _{23},4 \alpha _{12}+5 \alpha _{23}\}	\nonumber	\\
&\hspace{-3mm}W_5	&&\hspace{2mm}	\{3 \alpha _{12}+\alpha _{23},3 \alpha _{12}+2 \alpha _{23},4 \alpha _{12}+\alpha _{23},4 \alpha _{12}+3 \alpha _{23},5 \alpha _{12}+2 \alpha _{23},5 \alpha _{12}+3 \alpha _{23}\}	\nonumber	\\
&\hspace{-3mm}W_{-5}	&&\hspace{2mm}	\{\alpha _{12}+3 \alpha _{23},2 \alpha _{12}+3 \alpha _{23},\alpha _{12}+4 \alpha _{23},3 \alpha _{12}+4 \alpha _{23},2 \alpha _{12}+5 \alpha _{23},3 \alpha _{12}+5 \alpha _{23}\}	\nonumber	\\
&\hspace{-3mm}W_6	&&\hspace{2mm}	\{4 \alpha_{12}+2 \alpha_{23}\}		\nonumber	\\
&\hspace{-3mm}W_{-6}	&&\hspace{2mm}	\{2 \alpha_{12}+4 \alpha_{23}\}		\nonumber	\,	.	
\end{alignat}
$W_0$ corresponds to the trivial flat connection, $W_6$ and  $W_{-6}$ to the central flat connections, $W_1$, $W_2$, $W_5$, and $W_{-5}$ to the abelian flat connections $U(1)^2$, $W_3$, $W_4$ and $W_{-4}$ to the abelian flat connections from the type $S(U(2)\times U(1))$.
We denote the contributions from $W_0$, $W_1$, $W_2$, and $W_3$ by $Z_0'$, $Z_1'$, $Z_2'$, and $Z_3'$, respectively.
The contributions from $W_4$ and $W_{-4}$, which are in the same orbit of $W_3$ under the action of the center, are the same and equal to $Z_3'$.
Similarly, we can see that $W_5$ and $W_{-5}$ (resp. $W_6$ and $W_{-6}$) are related to $W_1$ (resp. $W_0$) by the action of the center, and their contributions are the same and equal to $Z_1'$ (resp. $Z_0'$).
In sum, we have $C_0=\{ W_0, W_6, W_{-6}\}$, $C_1 = \{ W_1, W_{5},W_{-5}\}$, $C_2 = \{ W_2 \}$, and $C_3 = \{ W_3, W_{4}, W_{-4}\}$.
Thus, the WRT invariant can be written as
\begin{align}
Z_{SU(3)}(M_3) = 3Z_0'+3e^{\frac{i \pi  k}{3}}Z_1'+e^{\frac{4 i \pi  k}{3}}Z_2'+3e^{i \pi  k}Z_3' =: 3(Z_0+e^{\frac{i \pi  k}{3}}Z_1+e^{\frac{4 i \pi  k}{3}}Z_2+e^{i \pi  k}Z_3)
\end{align}
where $Z_2'=:3Z_2$. 
Each $Z_a$'s are written in terms of homological blocks,
\begin{align}
3\left(
\begin{array}{c}
 Z_0 \\
 Z_1 \\
 Z_2 \\
 Z_3 \\
\end{array}
\right)=
3\left(
\begin{array}{cccc}
 1 & 1 & 1 & 1 \\
 6 & 1 & -3 & -2 \\
 2 & -1 & -1 & 2 \\
 3 & -1 & 3 & -1 \\
\end{array}
\right) \left(
\begin{array}{c}
 \widehat{Z}_0 \\
 \widehat{Z}_1 \\
 \widehat{Z}_2 \\
 \widehat{Z}_3 \\
\end{array}
\right)
\end{align}
where
\begin{align}
\widehat{Z}_0 &= -q^{35+\frac{349}{385}} (1+2 q^6+2 q^8+2 q^{14}+2 q^{18}+3 q^{24}+2 q^{32}+2 q^{42} + \cdots)		\,	,	\\
\widehat{Z}_1 &= -2q^{27+\frac{169}{2310}} (1-q^3-q^5-q^6-q^8-q^{13}+q^{14}-q^{17}-q^{21}-q^{22}-2 q^{23} + \cdots)	\,	,	\\
\widehat{Z}_2 &= q^{20+\frac{662}{1155}} (1+2 q^{12}+2 q^{16}-2 q^{18}+2 q^{20}-q^{26}+3 q^{48}+2 q^{52}+2 q^{56} + \cdots)	\\	
\widehat{Z}_3 &= -2 q^{24+\frac{313}{770}} (1+q^6+q^{15}+q^{16}+q^{18}+q^{21}+q^{22}+q^{24}+q^{30}+q^{31}+ \cdots)	\,	,	\,	.
\end{align}
Therefore, the WRT invariant can be expressed as
\begin{align}
Z_{SU(3)} (M_{3}) = 6 \sqrt{3} \sum_{a,b} e^{2\pi i K CS_a} S_{ab} \widehat{Z}_b(q)	\,	\Big|_{q \searrow e^{\frac{2\pi i}{K}}}
\end{align}
where $(CS_0,CS_1,CS_2,CS_3)=(0,\frac{1}{6},\frac{2}{3}, \frac{1}{2})$ and the $S$-matrix is
\begin{align}
S_{ab} = \frac{1}{2\sqrt{3}} \left(
\begin{array}{cccc}
 1 & 1 & 1 & 1 \\
 6 & 1 & -3 & -2 \\
 2 & -1 & -1 & 2 \\
 3 & -1 & 3 & -1 \\
\end{array}
\right)	\,	,
\end{align}
which satisfies $S_{ab}S_{bc}=\delta_{ac}$.
The $S$-matrix can also be calculated from \eqref{sun-smat} and \eqref{sun-lk} and it agrees with the above $S$-matrix.

We also find that the relation between $Z_a$ and $\widehat{Z}_a$ can be put in a form of
\begin{align}
\left(
\begin{array}{c}
 Z_0 \\
 Z_2 \\
 Z_3 \\
 Z_1 \\
\end{array}
\right)=\left(
\begin{array}{cccc}
 1 & 1 & 1 & 1 \\
 2 & -1 & 2 & -1 \\
 3 & 3 & -1 & -1 \\
 6 & -3 & -2 & 1 \\
\end{array}
\right) \left(
\begin{array}{c}
 \widehat{Z}_0 \\
 \widehat{Z}_2 \\
 \widehat{Z}_3 \\
 \widehat{Z}_1 \\
\end{array}
\right)	\,	.
\end{align}
The matrix above is a tensor product of the $S$-matrix of the case $H=2$ and $H=3$, \textit{i.e.} $\begin{pmatrix} 1 & 1 \\ 3 & -1 \end{pmatrix} \otimes \begin{pmatrix} 1 & 1 \\ 2 & -1 \end{pmatrix}$ where we note that $\mathbb{Z}_6 = \mathbb{Z}_2 \times \mathbb{Z}_3$.

\subsubsection*{\textbullet $H=7$}
We take $(P_1,P_2,P_3)=(2,5,11)$ for $H=7$ as an example.
In $\Lambda_2/ 7 \Lambda_2$, there are 12 Weyl orbits
\begin{alignat}{3}
&\hspace{-1mm}W_{t}	&&\hspace{3mm}	\Lambda_2/ 7 \Lambda_2		\nonumber	\\
\vspace{1mm}
&\hspace{-1mm}W_0	&&\hspace{3mm}	\{0\}						\nonumber	\\
&\hspace{-1mm}W_1	&&\hspace{3mm}	\{\alpha _{12},\alpha _{13},\alpha _{23},6 \alpha _{12},6 \alpha _{13},6 \alpha _{23}\}			\nonumber	\\
&\hspace{-1mm}W_2	&&\hspace{3mm}	\{2 \alpha _{12},2 \alpha _{13},2 \alpha _{23},5 \alpha _{12},5 \alpha _{13},5 \alpha _{23}\}		\nonumber	\\
&\hspace{-1mm}W_3	&&\hspace{3mm}	\{3 \alpha _{12},3 \alpha _{13},3 \alpha _{23},4 \alpha _{12},4 \alpha _{13},4 \alpha _{23}\}		\nonumber	\\
&\hspace{-1mm}W_4	&&\hspace{3mm}	\{\alpha _{12}+\alpha _{13},5 \alpha _{12}+\alpha _{13},\alpha _{12}+5 \alpha _{13}\}			\nonumber	\\
&\hspace{-1mm}W_{-4}	&&\hspace{3mm}	\{\alpha _{13}+\alpha _{23},5 \alpha _{13}+\alpha _{23},\alpha _{13}+5 \alpha _{23}\}			\nonumber	\\
&\hspace{-1mm}W_5	&&\hspace{3mm}	\{2 \alpha _{12}+2 \alpha _{13},3 \alpha _{12}+2 \alpha _{13},2 \alpha _{12}+3 \alpha _{13}\}	\nonumber	\\
&\hspace{-1mm}W_{-5}	&&\hspace{3mm}	\{2 \alpha _{13}+2 \alpha _{23},3 \alpha _{13}+2 \alpha _{23},2 \alpha _{13}+3 \alpha _{23}\}	\nonumber	\\
&\hspace{-1mm}W_6	&&\hspace{3mm}	\{3 \alpha _{12}+3 \alpha _{13},\alpha _{12}+3 \alpha _{13},3 \alpha _{12}+\alpha _{13}\}		\nonumber	\\
&\hspace{-1mm}W_{-6}	&&\hspace{3mm}	\{3 \alpha _{13}+3 \alpha _{23},\alpha _{13}+3 \alpha _{23},3 \alpha _{13}+\alpha _{23}\}		\nonumber	\\
&\hspace{-1mm}W_7	&&\hspace{3mm}	\{2 \alpha _{12}+\alpha _{13},\alpha _{12}+2 \alpha _{13},4 \alpha _{12}+2 \alpha _{13},4 \alpha _{12}+\alpha _{13},\alpha _{12}+4 \alpha _{13},2 \alpha _{12}+4 \alpha _{13}\}		\nonumber	\\
&\hspace{-1mm}W_{-7}	&&\hspace{3mm}	\{2 \alpha _{13}+\alpha _{23},\alpha _{13}+2 \alpha _{23},4 \alpha _{13}+2 \alpha _{23},4 \alpha _{13}+\alpha _{23},\alpha _{13}+4 \alpha _{23},2 \alpha _{13}+4 \alpha _{23}\}		\nonumber	\,	.
\end{alignat}
$W_0$ corresponds to the trivial flat connection, $W_1$, $W_2$, $W_3$, $W_7$, and $W_{-7}$ to the abelian flat connections $U(1)^2$, $W_{4}$, $W_{-4}$, $W_{5}$, $W_{-5}$, $W_{6}$, and $W_{-6}$ to the abelian flat connections from the type $S(U(2)\times U(1))$.
The contributions from $W_0$, $W_1$, $W_2$, and $W_3$ are denoted by $Z_0$, $Z_1$, $Z_2$, and $Z_3$, respectively.
Also, the contributions from abelian flat connections corresponding to $W_4$ and $W_{-4}$, which are complex conjugate to each other, are the same and the sum of their contributions is denoted by $Z_4$.
This is similar to $W_5$ and $W_{-5}$, $W_6$ and $W_{-6}$, and $W_7$ and $W_{-7}$, and we denote their sum by $Z_5$, $Z_6$, and $Z_7$, respectively.
Then the WRT invariant is written as
\begin{align}
Z_{SU(3)}(M_{3}) = \sum_{a} e^{\pi i K S_a} Z_a = 7\sum_{a,b} e^{2\pi i K CS_a} S_{ab} \widehat{Z}_{b}
\end{align}
where
\begin{align}
\widehat{Z}_0 &= q^{22+\frac{107}{110}}  (1+2 q^7+4 q^{21}+4 q^{42}+2 q^{63}+4 q^{70}-q^{84}+\cdots)	\,	,	\\
\widehat{Z}_1 &= q^{\frac{529}{770}} (1+2 q^6-4 q^9+4 q^{10}-2 q^{18}+6 q^{19}+2 q^{30}-2 q^{33}+4 q^{39}-q^{40}+\cdots)	\,	,	\\
\widehat{Z}_2 &= -q^{5+\frac{639}{770}} (3-2 q+2 q^3-q^4+4 q^9+2 q^{12}-2 q^{15}+2 q^{19}+2 q^{21}+2 q^{28}+\cdots)		\,	,	\\
\widehat{Z}_3 &= -q^{2+\frac{309}{770}} (1-4 q^7+4 q^{12}-2 q^{15}+4 q^{21}+2 q^{25}+q^{28}-2 q^{30}-2 q^{33}+\cdots)		\,	,	\\
\widehat{Z}_4 &= -2q^{1+\frac{89}{770}} (1-q+q^3-2 q^7+2 q^9+2 q^{12}+q^{33}-q^{34}+q^{36}+q^{37}+\cdots)	\,	,	\\
\widehat{Z}_5 &= 2q^{4+\frac{419}{770}} (1-q^3+2 q^{10}+2 q^{13}+q^{19}+q^{22}+2 q^{33}+q^{39}+2 q^{45}+2 q^{48}+\cdots)	\,	,	\\
\widehat{Z}_6 &= -2q^{3+\frac{199}{770}} (1-q+2 q^3+q^{21}-q^{22}+q^{24}-q^{33}-q^{36}-q^{37}+2 q^{39}+\cdots)			\,	,	\\
\widehat{Z}_7 &= 2q^{2+\frac{107}{110}} (1-q^2+2 q^3-q^5+2 q^9+2 q^{12}-2 q^{14}-3 q^{17}+q^{18}+q^{23}+\cdots)			\,	,	
\end{align}
$CS_a = (0,\frac{5}{7},\frac{6}{7},\frac{3}{7},\frac{1}{7},\frac{4}{7},\frac{2}{7},0)$, and the $S$-matrix is
\begin{align}
\hspace{-20mm}
{\fontsize{5pt}{12pt} \selectfont
S_{ab}
=\frac{1}{7}
\hspace{-1mm}
\left(
\begin{array}{@{\mkern-1mu} c @{\mkern0mu} c @{\mkern0mu} c @{\mkern0mu} c @{\mkern0mu} c @{\mkern0mu} c @{\mkern0mu} c @{\mkern0mu} c@{\mkern-1mu}}
 1 & 1 & 1 & 1 & 1 & 1 & 1 & 1 \\
 6 & -2 \left(\cos \left(\frac{\pi }{7}\right)+2 \sin \left(\frac{\pi }{14}\right)\right) & 2 \sin \left(\frac{3 \pi }{14}\right)-4 \cos \left(\frac{\pi }{7}\right) & 4 \sin \left(\frac{3 \pi }{14}\right)-2 \sin \left(\frac{\pi }{14}\right) & 4 \sin \left(\frac{3 \pi }{14}\right)+2 & 2-4 \sin \left(\frac{\pi }{14}\right) & 2-4 \cos \left(\frac{\pi }{7}\right) & -1 \\
 6 & 2 \sin \left(\frac{3 \pi }{14}\right)-4 \cos \left(\frac{\pi }{7}\right) & 4 \sin \left(\frac{3 \pi }{14}\right)-2 \sin \left(\frac{\pi }{14}\right) & -2 \left(\cos \left(\frac{\pi }{7}\right)+2 \sin \left(\frac{\pi }{14}\right)\right) & 2-4 \sin \left(\frac{\pi }{14}\right) & 2-4 \cos \left(\frac{\pi }{7}\right) & 4 \sin \left(\frac{3 \pi }{14}\right)+2 & -1 \\
 6 & 4 \sin \left(\frac{3 \pi }{14}\right)-2 \sin \left(\frac{\pi }{14}\right) & -2 \left(\cos \left(\frac{\pi }{7}\right)+2 \sin \left(\frac{\pi }{14}\right)\right) & 2 \sin \left(\frac{3 \pi }{14}\right)-4 \cos \left(\frac{\pi }{7}\right) & 2-4 \cos \left(\frac{\pi }{7}\right) & 4 \sin \left(\frac{3 \pi }{14}\right)+2 & 2-4 \sin \left(\frac{\pi }{14}\right) & -1 \\
 6 & 4 \sin \left(\frac{3 \pi }{14}\right)+2 & 2-4 \sin \left(\frac{\pi }{14}\right) & 2-4 \cos \left(\frac{\pi }{7}\right) & 4 \sin \left(\frac{3 \pi }{14}\right)-2 \sin \left(\frac{\pi }{14}\right) & -2 \left(\cos \left(\frac{\pi }{7}\right)+2 \sin \left(\frac{\pi }{14}\right)\right) & 2 \sin \left(\frac{3 \pi }{14}\right)-4 \cos \left(\frac{\pi }{7}\right) & -1 \\
 6 & 2-4 \sin \left(\frac{\pi }{14}\right) & 2-4 \cos \left(\frac{\pi }{7}\right) & 4 \sin \left(\frac{3 \pi }{14}\right)+2 & -2 \left(\cos \left(\frac{\pi }{7}\right)+2 \sin \left(\frac{\pi }{14}\right)\right) & 2 \sin \left(\frac{3 \pi }{14}\right)-4 \cos \left(\frac{\pi }{7}\right) & 4 \sin \left(\frac{3 \pi }{14}\right)-2 \sin \left(\frac{\pi }{14}\right) & -1 \\
 6 & 2-4 \cos \left(\frac{\pi }{7}\right) & 4 \sin \left(\frac{3 \pi }{14}\right)+2 & 2-4 \sin \left(\frac{\pi }{14}\right) & 2 \sin \left(\frac{3 \pi }{14}\right)-4 \cos \left(\frac{\pi }{7}\right) & 4 \sin \left(\frac{3 \pi }{14}\right)-2 \sin \left(\frac{\pi }{14}\right) & -2 \left(\cos \left(\frac{\pi }{7}\right)+2 \sin \left(\frac{\pi }{14}\right)\right) & -1 \\
 12 & -2 & -2 & -2 & -2 & -2 & -2 & 5 \\
\end{array}
\right)	\,	,
}
\end{align}
which satisfies $S_{ab}S_{bc} = \delta_{ac}$.
This $S$-matrix can also be obtained from \eqref{sun-smat} and \eqref{sun-lk} with $C_a=\{ W_a \}$, $a=0,1,2,3$, and $C_{\pm a} = \{ W_{a}, W_{-a}\}$, $a=4,5,6,7$.

\subsubsection{$G=SU(4)$}
When $G=SU(4)$, \eqref{sun-a-wrt2} becomes
\begin{align}
\begin{split}
\sum_{{\bf t} \in \Lambda_{3} / H \Lambda_{3}}
\Bigg[
e^{2 \pi i K  \frac{P}{H} (t_1^2 + t_2^2 + t_2^3 + t_1 t_2 + t_1 t_3 + t_2 t_3) }	
\sum_{\substack{n_{i,j}=0 \\ 1 \leq i < j \leq 4}}^{\infty} 
\Big( \prod_{1 \leq i < j \leq 4} \chi_{2P}(n_{i,j}) \Big)
e^{-\frac{ i \pi }{H} \sum _{i=1}^{4} t_i b_i(\vec{n})}	\,
q^{ \frac{1}{8 H P} \sum_{i=1}^{4} b_i(\vec{n})^2}
\Bigg]	\label{su-4-wrt}
\end{split}
\end{align}
up to the overall factor where 
\begin{align}
\begin{split}
b_1(\vec{n}) &= n_{1,2}+n_{1,3}-n_{1,4}	\,	,	\quad	
b_2(\vec{n}) = -n_{1,2}+n_{2,3}-n_{2,4}	\,	,	\quad	\\
b_3(\vec{n}) &= -n_{1,3}-n_{2,3}-n_{3,4}	\,	,	\quad
b_4(\vec{n}) = n_{1,4}+n_{2,4}+n_{3,4}	\,	,
\end{split}
\end{align}
and $t_4 = -t_1-t_2-t_3$.
In the following examples, we express the WRT invariant in terms of homological blocks.

\subsubsection*{\textbullet \ $H=2$}
We consider an example for $H=2$.
This case can be obtained, for example, from $(P_1,P_2,P_3)=(3,5,7)$.
The elements in $\Lambda_3/2\Lambda_3$ can be put in three Weyl orbits,
\begin{alignat}{3}
&W_{t}	&&\hspace{10mm}	\Lambda_3/2\Lambda_3		\nonumber	\\
\vspace{1mm}
&W_0	&&\hspace{10mm}	\{0\}						\nonumber	\\
&W_1	&&\hspace{10mm}	\{\alpha _{12},\alpha _{23},\alpha _{34},\alpha _{12}+\alpha _{23},\alpha _{23}+\alpha _{34},\alpha _{12}+\alpha _{23}+\alpha _{34}\}		\nonumber	\\
&W_2	&&\hspace{10mm}	\{\alpha _{12}+\alpha _{34}\}	\nonumber
\end{alignat}
where $W_0$ corresponds to the trivial flat connection, $W_2$ to the central flat connection, and $W_1$ to the abelian flat connection from the type $S(U(2) \times U(2))$.
So we have $C_0 =\{ W_0, W_2\}$ and $C_1 =\{ W_1\}$.
The contributions from $W_0$ and $W_2$ are the same, and we denote their sum by $2Z_0$.
The contribution from $W_1$ is denoted by $Z_1$.

These $Z_a$'s are expressed as
\begin{align}
Z_0=\widehat{Z}_0+\widehat{Z}_1	\,	,	\\
\frac{1}{2}Z_1=3 \widehat{Z}_0-\widehat{Z}_1	
\end{align}
where
\begin{align}
\widehat{Z}_0 &= q^{13+\frac{16}{21}} (1+q^7+2 q^9+q^{11}+3 q^{12}-2 q^{13}-2 q^{14}+2 q^{15}-4 q^{16}-2 q^{17} +\cdots)	\,	,	\\
\widehat{Z}_1 &= q^{17+\frac{11}{42}} (3+3 q^2-4 q^4+q^6-4 q^7+4 q^8-4 q^9-3 q^{10}+4 q^{11}+5 q^{12} +\cdots)	\,	.
\end{align}
Thus, the WRT invariant can be written as
\begin{align}
Z_{SU(3)} (M_{3}) = 4 \sum_{a,b} e^{2 \pi i K CS_a} S_{ab} \widehat{Z}_{b}(q)	\,	\Big|_{q \searrow e^{\frac{2\pi i}{K}}}
\end{align}
where $CS_a = (0, \frac{1}{2})$ and the $S$-matrix is 
\begin{align}
S_{ab} = \frac{1}{2} \left(
\begin{array}{cc}
 1 & 1 \\
 3 & -1 \\
\end{array}
\right)	\,	,
\end{align}
which satisfies $S_{ab}S_{bc}=\delta_{ac}$.
This $S$-matrix can also be obtained from \eqref{sun-smat} and \eqref{sun-lk}.

\subsubsection*{\textbullet \ $H=3$}
We consider, for example, $(P_1,P_2,P_3)=(2,5,7)$ to have $H=3$.
In $\Lambda_3/3\Lambda_3$, there are five Weyl orbits
\begin{alignat}{3}
&W_{t}	&&	\hspace{5mm}	\Lambda_3/3\Lambda_3	\nonumber	\\
&W_0	&&	\hspace{5mm}	\{0\}					\nonumber	\\
&W_1	&&	\hspace{5mm}	\{\alpha_{12},\alpha_{23},\alpha_{34},\alpha_{12}+\alpha_{23},\alpha_{23}+\alpha_{34},\alpha_{12}+\alpha_{23}+\alpha_{34},	\nonumber	\\
&		&&	\hspace{42mm}2 \alpha_{12},2 \alpha_{23},2 \alpha_{34},2 \alpha_{12}+2 \alpha_{23},2 \alpha_{23}+2\alpha_{34},2 \alpha_{12}+2 \alpha_{23}+2 \alpha_{34}\}	\,	\nonumber	\\
&W_2	&&	\hspace{5mm}	\{\alpha_{12}+\alpha_{34},\alpha_{12}+2 \alpha_{23}+\alpha_{34},2 \alpha_{12}+\alpha_{34},\alpha_{12}+2 \alpha_{34},2 \alpha_{12}+2 \alpha_{34},2 \alpha_{12}+\alpha_{23}+2 \alpha_{34}\}	\nonumber	\\
&W_3	&&	\hspace{5mm}	\{2 \alpha_{12}+\alpha_{23},2 \alpha_{23}+\alpha_{34},2 \alpha_{12}+\alpha_{23}+\alpha_{34},2 \alpha_{12}+2 \alpha_{23}+\alpha_{34}\}	\nonumber	\\
&W_{-3}	&&	\hspace{5mm}	\{\alpha_{12}+2 \alpha_{23},\alpha_{23}+2 \alpha_{34},\alpha_{12}+\alpha_{23}+2 \alpha_{34},\alpha_{12}+2 \alpha_{23}+2 \alpha_{34}\}		\,	.	\nonumber
\end{alignat}
Here, $W_0$ corresponds to the trivial flat connection, $W_1$ and $W_2$ to the abelian flat connections from the type $S(U(2) \times U(1) \times U(1))$ and $S(U(2) \times U(2))$, respectively, and $W_3$ and $W_{-3}$ to the abelian flat connections from the type $S(U(3) \times U(1))$.
We denote the contributions from $W_0$, $W_1$, and $W_2$ by $Z_0$, $Z_1$, and $Z_2$, respectively.
Abelian flat connections corresponding to $W_3$ and $W_{-3}$ are complex conjugate to each other and their contributions are the same.
We denote their sum by $Z_3$.
These $Z_a$'s are written in terms of homological blocks
\begin{align}
Z_0&=\widehat{Z}_0+\widehat{Z}_1+\widehat{Z}_2+\widehat{Z}_3	\,	,	\\
Z_1&=12 \widehat{Z}_0-3 \widehat{Z}_1+0 \widehat{Z}_2+3 \widehat{Z}_3	\,	,	\\
Z_2&=6 \widehat{Z}_0+0 \widehat{Z}_1+3 \widehat{Z}_2-3 \widehat{Z}_3		\,	,	\\
Z_3&=8 \widehat{Z}_0+2 \widehat{Z}_1-4 \widehat{Z}_2-\widehat{Z}_3
\end{align}
with
\begin{align}
\widehat{Z}_0 &= -q^{3+\frac{3}{28}} (1+2 q^3+2 q^6+3 q^9-2 q^{12}-10 q^{15}+7 q^{18}-4 q^{21}-14 q^{24}+\cdots)	\,	,	\\
\widehat{Z}_1 &= q^{1+\frac{37}{84}} (1-3 q+5 q^2+3 q^3-q^4+2 q^5-4 q^6+19 q^7-9 q^8-q^{10}+\cdots)		\,	,	\\
\widehat{Z}_2 &= -q^{3+\frac{65}{84}} (1+q-2 q^2+q^3+2 q^4+2 q^5-2 q^6-7 q^7+4 q^8-q^{10}-5 q^{11}+\cdots)	\,	,	\\
\widehat{Z}_3 &= -2q^{3+\frac{3}{28}} (1-q+2 q^3+2 q^5+q^6-3 q^7-q^8+3 q^9-q^{10}+5 q^{11}+\cdots)		\,	.
\end{align}
Thus, the WRT invariant can be written as
\begin{align}
Z_{SU(4)}(M_{3}) = 3^{3/2} \sum_{a,b} e^{2\pi i K CS_a} S_{ab} \widehat{Z}_{b}(q)	\,	\Big|_{q \searrow e^{\frac{2\pi i}{K}}}
\end{align}
where $CS_a = (0, \frac{1}{3}, \frac{2}{3}, 0)$ and $S_{ab}$ is 
\begin{align}
S_{ab} = \frac{1}{3^{3/2}}
\left(
\begin{array}{cccc}
 1 & 1 & 1 & 1 \\
 12 & -3 & 0 & 3 \\
 6 & 0 & 3 & -3 \\
 8 & 2 & -4 & -1 \\
\end{array}
\right)	\,	,
\end{align}
which satisfies $S_{ab}S_{bc}=\delta_{ac}$.
This $S$-matrix can also be obtained from \eqref{sun-smat} and \eqref{sun-lk} with $C_a = \{ W_a\}$, $a=0,1,2$, and $C_{\pm3} = \{ W_3, W_{-3}\}$.

\subsubsection*{\textbullet \ $H=4$}
The case $H=4$ can be obtained, for example, from $(P_1,P_2,P_3)=(3,5,7)$.
The elements in $\Lambda_3/4\Lambda_3$ are put in 10 Weyl orbits,
\begin{align}
&W_{t}	&&\hspace{2mm}	\Lambda_3/4\Lambda_3	\nonumber\\
\vspace{1mm}
&W_0	&&\hspace{2mm}	\{0\}					\nonumber\\	
&W_1	&&\hspace{2mm}	\{\alpha _{12},\alpha _{23},\alpha _{34},\alpha _{12}+\alpha _{23},\alpha _{23}+\alpha _{34},\alpha _{12}+\alpha _{23}+\alpha _{34},3 \alpha _{12},3 \alpha _{23},3 \alpha _{34},	\nonumber\\
&		&&\hspace{68mm}3 \alpha _{12}+3 \alpha _{23},3 \alpha _{23}+3 \alpha _{34},3 \alpha _{12}+3 \alpha _{23}+3 \alpha _{34}\}		\nonumber\\
&W_2	&&\hspace{2mm}	\{2 \alpha _{12},2 \alpha _{23},2 \alpha _{34},2 \alpha _{12}+2 \alpha _{23},2 \alpha _{23}+2 \alpha _{34},2 \alpha _{12}+2 \alpha _{23}+2 \alpha _{34}\}	\nonumber\\
&W_3	&&\hspace{2mm}	\{2 \alpha _{12}+\alpha _{23},2 \alpha _{23}+\alpha _{34},2 \alpha _{12}+\alpha _{23}+\alpha _{34},2 \alpha _{12}+2 \alpha _{23}+\alpha _{34},3 \alpha _{12}+\alpha _{23},	\nonumber\\
&		&&\hspace{25mm}	3 \alpha _{23}+\alpha _{34},3 \alpha _{12}+2 \alpha _{23},3 \alpha _{23}+2 \alpha _{34},3 \alpha _{12}+2 \alpha _{23}+2 \alpha _{34},		\nonumber\\
&		&&\hspace{48mm}3 \alpha _{12}+\alpha _{23}+\alpha _{34}, 3 \alpha _{12}+3 \alpha _{23}+\alpha _{34},3 \alpha _{12}+3 \alpha _{23}+2 \alpha _{34}\}	\nonumber\\	
&W_{-3}	&&\hspace{2mm}	\{\alpha _{12}+2 \alpha _{23},\alpha _{23}+2 \alpha _{34},\alpha _{12}+\alpha _{23}+2 \alpha _{34},\alpha _{12}+2 \alpha _{23}+2 \alpha _{34},\alpha _{12}+3 \alpha _{23},	\nonumber\\
&		&&\hspace{25mm}	\alpha _{23}+3 \alpha _{34},2 \alpha _{12}+3 \alpha _{23},2 \alpha _{23}+3 \alpha _{34},2 \alpha _{12}+2 \alpha _{23}+3 \alpha _{34},	\nonumber\\
&		&&\hspace{48mm}\alpha _{12}+\alpha _{23}+3 \alpha _{34},\alpha _{12}+3 \alpha _{23}+3 \alpha _{34},2 \alpha _{12}+3 \alpha _{23}+3 \alpha _{34}\}	\nonumber\\
&W_4	&&\hspace{2mm}	\{\alpha _{12}+\alpha _{34},\alpha _{12}+2 \alpha _{23}+\alpha _{34},3 \alpha _{12}+3 \alpha _{34},3 \alpha _{12}+\alpha _{34},\alpha _{12}+3 \alpha _{34},3 \alpha _{12}+2 \alpha _{23}+3 \alpha _{34}\}	\nonumber	\\
&W_5	&&\hspace{2mm}	\{2 \alpha _{12}+\alpha _{34},\alpha _{12}+2 \alpha _{34},2 \alpha _{12}+\alpha _{23}+2 \alpha _{34},2 \alpha _{12}+3 \alpha _{23}+\alpha _{34},	\nonumber\\
&		&&\hspace{12mm}	\alpha _{12}+3 \alpha _{23}+2 \alpha _{34},2 \alpha _{12}+3 \alpha _{23}+2 \alpha _{34},\alpha _{12}+3 \alpha _{23}+\alpha _{34},3 \alpha _{12}+2 \alpha _{34},	\nonumber\\
&		&&\hspace{22mm}	2 \alpha _{12}+3 \alpha _{34},3 \alpha _{12}+\alpha _{23}+3 \alpha _{34},2 \alpha _{12}+\alpha _{23}+3 \alpha _{34},3 \alpha _{12}+\alpha _{23}+2 \alpha _{34}\}	\nonumber	\\
&W_6	&&\hspace{2mm}	\{2 \alpha _{12}+2 \alpha _{34}\}			\nonumber
\end{align}
\begin{align}	
&\hspace{-50.3mm}W_7	&&\hspace{-47.3mm}\hspace{2mm}	\{3 \alpha _{12}+2 \alpha _{23}+\alpha _{34}\}	\nonumber\\
&\hspace{-50.3mm}W_{-7}	&&\hspace{-47.3mm}\hspace{2mm}	\{\alpha _{12}+2 \alpha _{23}+3 \alpha _{34}\}	\nonumber	\,	.
\end{align}
Here, $W_0$ corresponds to the trivial flat connection, $W_6$, $W_7$, and $W_{-7}$ to the central flat connections, $W_{1}$, $W_{3}$, $W_{-3}$, and $W_{5}$ to the abelian flat connections from the type $S(U(2) \times U(1) \times U(1))$, and $W_{2}$ and $W_{4}$ to the abelian flat connections from the type $S(U(2) \times U(2))$.
The contributions from $W_0$, $W_6$, $W_7$, and $W_{-7}$, which are in the same orbit by the action of the center, are the same but have different overall exponential factors by $e^{\pi i K}$.
We denote their contributions with overall exponential factors as $Z_0$, $Z_0$, $e^{\pi i K} Z_0$, and $e^{\pi i K} Z_0$, respectively.
Similarly, the contributions from $W_1$, $W_3$, $W_{-3}$, and $W_5$ (resp. from $W_2$ and $W_4$), which are related by the action of the center, are the same up to an overall exponential factor, and they are denoted by $e^{\frac{\pi i K}{2}} Z_1$, $e^{\frac{3\pi i K}{2}} Z_1$, $e^{\frac{3\pi i K}{2}} Z_1$, and $e^{\frac{\pi i K}{2}} Z_1$ (resp. $2 Z_2$ and $2 e^{\pi i K} Z_2$), respectively.
So here we have $C_0 = \{ W_0, W_6, W_7, W_{-7}\}$, $C_1= \{ W_1, W_3, W_{-3}, W_5 \}$, $C_2 = \{ W_2, W_4\}$.
Then, the WRT invariant can be written as
\begin{align}
Z_{SU(4)}(M_3) &= 2(1+ e^{\pi i K}) Z_0 + 2(1+ e^{\pi i K}) e^{\frac{\pi i}{2} K} Z_1 + 2(1+ e^{\pi i K}) Z_2 	\\
&= 2(1+e^{\pi i K}) (Z_0 + e^{\frac{\pi i}{2}K}Z_1 + Z_2)	\	\,
\end{align}
and $Z_a$'s are expressed in terms of homological blocks,
\begin{align}
Z_0&=\left(\widehat{Z}_0+\widehat{Z}_3\right)+\left(\widehat{Z}_1+\widehat{Z}_4\right)+\left(\widehat{Z}_2+\widehat{Z}_5\right)	\,	,	\\
Z_1&=12 \left(\widehat{Z}_0+\widehat{Z}_3\right)-4 \left(\widehat{Z}_2+\widehat{Z}_5\right)	\,	,	\\
Z_2&=3 \left(\widehat{Z}_0+\widehat{Z}_3\right)-\left(\widehat{Z}_1+\widehat{Z}_4\right)+3 \left(\widehat{Z}_2+\widehat{Z}_5\right)
\end{align}
where
\begin{align}
\widehat{Z}_0 &= -2q^{13+\frac{37}{42}} (1+3 q^4+2 q^8+6 q^{12}+6 q^{16}+4 q^{20}+9 q^{24}+9 q^{28}+11 q^{32} +\cdots)	\,	,	\\
\widehat{Z}_1 &= -q^{8+\frac{53}{84}} (3+3 q-4 q^2+q^3+4 q^4-3 q^5+5 q^6+8 q^7-2 q^9+7 q^{10} +\cdots)	\,	,	\\
\widehat{Z}_2 &= q^{6+\frac{37}{42}} (1+3 q^6-4 q^8+7 q^{12}-2 q^{14}-q^{16}+6 q^{18}+3 q^{20}+7 q^{22}+4 q^{24} +\cdots)	\,	,	\\
\widehat{Z}_3 &= q^{10+\frac{8}{21}} (1+q^2+2 q^4+3 q^8+q^{14}+6 q^{16}+8 q^{18}+6 q^{20}+3 q^{22}+11 q^{24} +\cdots)	\,	,	\\
\widehat{Z}_4 &= 2q^{12+\frac{11}{84}} (2+2 q-2 q^2+2 q^3+q^4-2 q^5+6 q^6+2 q^7+8 q^8+6 q^9 +\cdots)	\,	,	\\	
\widehat{Z}_5 &= q^{11+\frac{8}{21}} (2-2 q^2-2 q^4-3 q^6-2 q^8-5 q^{10}-14 q^{12}-5 q^{14}-4 q^{16} +\cdots)	\,	.	
\end{align}
Therefore, the WRT invariant can be written as
\begin{align}
Z_{SU(4)}(M_{3}) = \sum_{\dot{a}} e^{2 \pi i K CS_{\dot{a}}}  Z_{\dot{a}} =  8\sum_{\dot{a},\dot{b}} e^{2 \pi i K CS_{\dot{a}}} \big( Y \otimes S_{ab} \big)_{\dot{a}\dot{b}} \widehat{Z}_{\dot{b}}(q)	\,	\Big|_{q \searrow e^{\frac{2\pi i}{K}}}
\end{align}
where $Z_{\dot{a}} = 2(Z_0,Z_1,Z_2,Z_0,Z_1,Z_2)^{T}$, $CS_{\dot{a}}=(0,\frac{1}{4},0,\frac{1}{2},\frac{3}{4},\frac{1}{2})$, and $S_{ab}$ is
\begin{align}
S_{ab} = \frac{1}{4}
\left(
\begin{array}{ccc}
 1 & 1 & 1 \\
 12 & 0 & -4 \\
 3 & -1 & 3 \\
\end{array}
\right)	\,	,
\end{align}
which satisfies $S_{ab}S_{bc}=\delta_{ac}$.
The $S$-matrix $S_{ab}$ can also be calculated from \eqref{sun-smat} and \eqref{sun-lk} and it agrees with the above $S$-matrix.

\subsubsection*{\textbullet \ $H=5$}
The case $H=5$ can be obtained, for example, from $(P_1,P_2,P_3)=(2,3,7)$.
Elements in $\Lambda_3/5\Lambda_3$ are grouped by 14 Weyl orbits
\begin{align}
&\hspace{0mm}W_{t}	&&\hspace{0mm}	\Lambda_3/5\Lambda_3	\nonumber\\
\vspace{1mm}
&\hspace{0mm}W_0		&&\hspace{0mm}	\{ 0 \}					\nonumber\\
&\hspace{0mm}W_1		&&\hspace{0mm}	\{\alpha _{12},\alpha _{23},\alpha _{34},\alpha _{12}+\alpha _{23},\alpha _{23}+\alpha _{34},\alpha _{12}+\alpha _{23}+\alpha _{34},	\nonumber\\
&					&&\hspace{35mm}	4 \alpha _{12},4 \alpha _{23},4 \alpha _{34},4 \alpha _{12}+4 \alpha _{23},4 \alpha _{23}+4 \alpha _{34},4 \alpha _{12}+4 \alpha _{23}+4 \alpha _{34}\}	\nonumber\\
&\hspace{0mm}W_2		&&\hspace{0mm}	\{2 \alpha _{12},2 \alpha _{23},2 \alpha _{34},2 \alpha _{12}+2 \alpha _{23}+2 \alpha _{34},2 \alpha _{12}+2 \alpha _{23},2 \alpha _{23}+2 \alpha _{34},		\nonumber\\
&					&&\hspace{35mm}	3 \alpha _{12},3 \alpha _{23},3 \alpha _{34},3 \alpha _{12}+3 \alpha _{23},3 \alpha _{23}+3 \alpha _{34},3 \alpha _{12}+3 \alpha _{23}+3 \alpha _{34}\}	\nonumber\\
&\hspace{0mm}W_3		&&\hspace{0mm}	\{2 \alpha _{12}+\alpha _{23},2 \alpha _{23}+\alpha _{34},2 \alpha _{12}+\alpha _{23}+\alpha _{34},2 \alpha _{12}+2 \alpha _{23}+\alpha _{34},4 \alpha _{12}+\alpha _{23},		\nonumber\\
&					&&\hspace{20mm}	4 \alpha _{23}+\alpha _{34},4 \alpha _{12}+3 \alpha _{23},4 \alpha _{23}+3 \alpha _{34},4 \alpha _{12}+3 \alpha _{23}+3 \alpha _{34},	\nonumber\\
&					&&\hspace{40mm}	4 \alpha _{12}+\alpha _{23}+\alpha _{34},4 \alpha _{12}+4 \alpha _{23}+3 \alpha _{34},4 \alpha _{12}+4 \alpha _{23}+\alpha _{34}\}	\nonumber\\
&\hspace{0mm}W_{-3}	&&\hspace{0mm}	\{\alpha _{12}+2 \alpha _{23},\alpha _{23}+2 \alpha _{34},\alpha _{12}+\alpha _{23}+2 \alpha _{34},\alpha _{12}+2 \alpha _{23}+2 \alpha _{34},\alpha _{12}+4 \alpha _{23},	\nonumber\\
&					&&\hspace{20mm}	\alpha _{23}+4 \alpha _{34},3 \alpha _{12}+4 \alpha _{23},3 \alpha _{23}+4 \alpha _{34},3 \alpha _{12}+3 \alpha _{23}+4 \alpha _{34},	\nonumber\\
&					&&\hspace{40mm}	\alpha _{12}+\alpha _{23}+4 \alpha _{34},3 \alpha _{12}+4 \alpha _{23}+4 \alpha _{34}, \alpha _{12}+4 \alpha _{23}+4 \alpha _{34}\}	\nonumber\\
&\hspace{0mm}W_4		&&\hspace{0mm}	\{\alpha _{12}+\alpha _{34},\alpha _{12}+2 \alpha _{23}+\alpha _{34},4 \alpha _{12}+\alpha _{34},\alpha _{12}+4 \alpha _{34},4 \alpha _{12}+3 \alpha _{23}+4 \alpha _{34},4 \alpha _{12}+4 \alpha _{34}\}	\nonumber\\
&\hspace{0mm}W_5		&&\hspace{0mm}	\{2 \alpha _{12}+2 \alpha _{34},2 \alpha _{12}+4 \alpha _{23}+2 \alpha _{34},3 \alpha _{12}+2 \alpha _{34},2 \alpha _{12}+3 \alpha _{34},3 \alpha _{12}+\alpha _{23}+3 \alpha _{34},3 \alpha _{12}+3 \alpha _{34}\}	\nonumber\\
&\hspace{0mm}W_6		&&\hspace{0mm}	\{2 \alpha _{12}+\alpha _{34},\alpha _{12}+2 \alpha _{34},2 \alpha _{12}+\alpha _{23}+2 \alpha _{34},\alpha _{12}+3 \alpha _{23}+\alpha _{34},	\nonumber\\
&					&&\hspace{3mm}	2 \alpha _{12}+3 \alpha _{23}+\alpha _{34},\alpha _{12}+3 \alpha _{23}+2 \alpha _{34},2 \alpha _{12}+3 \alpha _{23}+2 \alpha _{34},3 \alpha _{12}+\alpha _{34},	\nonumber\\
&					&&\hspace{6mm}	\alpha _{12}+3 \alpha _{34},3 \alpha _{12}+2 \alpha _{23}+3 \alpha _{34},3 \alpha _{12}+4 \alpha _{23}+3 \alpha _{34},3 \alpha _{12}+4 \alpha _{23}+\alpha _{34},	\nonumber\\
&					&&\hspace{9mm}	\alpha _{12}+4 \alpha _{23}+3 \alpha _{34},\alpha _{12}+4 \alpha _{23}+\alpha _{34},4 \alpha _{12}+2 \alpha _{34},2 \alpha _{12}+4 \alpha _{34},	\nonumber\\
&					&&\hspace{12mm}	4 \alpha _{12}+2 \alpha _{23}+4 \alpha _{34},4 \alpha _{12}+3 \alpha _{34},3 \alpha _{12}+4 \alpha _{34},4 \alpha _{12}+\alpha _{23}+4 \alpha _{34},	\nonumber\\
&					&&\hspace{15mm}	4 \alpha _{12}+2 \alpha _{23}+3 \alpha _{34},4 \alpha _{12}+\alpha _{23}+2 \alpha _{34},3 \alpha _{12}+2 \alpha _{23}+4 \alpha _{34},2 \alpha _{12}+\alpha _{23}+4 \alpha _{34}\}	\nonumber\\
&\hspace{0mm}W_7		&&\hspace{0mm}	\{3 \alpha _{12}+\alpha _{23}, 3\alpha _{23}+\alpha _{34},3 \alpha _{12}+2 \alpha _{23},3 \alpha _{23}+2 \alpha _{34},3 \alpha _{12}+\alpha _{23}+\alpha _{34},	\nonumber\\
&					&&\hspace{15mm}	3 \alpha _{12}+2 \alpha _{23}+2 \alpha _{34},3 \alpha _{12}+3 \alpha _{23}+\alpha _{34},3 \alpha _{12}+3 \alpha _{23}+2 \alpha _{34},	\nonumber\\
&					&&\hspace{30mm}	4 \alpha _{12}+2 \alpha _{23},4 \alpha _{23}+2 \alpha _{34},4 \alpha _{12}+2 \alpha _{23}+2 \alpha _{34},4 \alpha _{12}+4 \alpha _{23}+2 \alpha _{34}\}	\nonumber\\
&\hspace{0mm}W_{-7}	&&\hspace{0mm}	\{\alpha _{12}+3 \alpha _{23},\alpha _{23}+3 \alpha _{34},2 \alpha _{12}+3 \alpha _{23},2 \alpha _{23}+3 \alpha _{34},\alpha _{12}+\alpha _{23}+3 \alpha _{34},	\nonumber\\
&					&&\hspace{15mm}	2 \alpha _{12}+2 \alpha _{23}+3 \alpha _{34},\alpha _{12}+3 \alpha _{23}+3 \alpha _{34},2 \alpha _{12}+3 \alpha _{23}+3 \alpha _{34},	\nonumber\\
&					&&\hspace{30mm}	2 \alpha _{12}+4 \alpha _{23},2 \alpha _{23}+4 \alpha _{34},2 \alpha _{12}+2 \alpha _{23}+4 \alpha _{34},2 \alpha _{12}+4 \alpha _{23}+4 \alpha _{34}\}	\nonumber	\\
&W_8		&&\hspace{0mm}	\{3 \alpha _{12}+2 \alpha _{23}+\alpha _{34},4 \alpha _{12}+3 \alpha _{23}+2 \alpha _{34},4 \alpha _{12}+3 \alpha _{23}+\alpha _{34},4 \alpha _{12}+2 \alpha _{23}+\alpha _{34}\}			\nonumber\\
&W_{-8}	&&\hspace{0mm}	\{\alpha _{12}+2 \alpha _{23}+3 \alpha _{34},2 \alpha _{12}+3 \alpha _{23}+4 \alpha _{34},\alpha _{12}+3 \alpha _{23}+4 \alpha _{34},\alpha _{12}+2 \alpha _{23}+4 \alpha _{34}\}	\nonumber\\
&W_9		&&\hspace{0mm}	\{3 \alpha _{12}+\alpha _{23}+2 \alpha _{34},3 \alpha _{12}+4 \alpha _{23}+2 \alpha _{34},\alpha _{12}+4 \alpha _{23}+2 \alpha _{34},3 \alpha _{12}+\alpha _{23}+4 \alpha _{34}\}	\nonumber\\
&W_{-9}	&&\hspace{0mm}	\{2 \alpha _{12}+\alpha _{23}+3 \alpha _{34},2 \alpha _{12}+4 \alpha _{23}+3 \alpha _{34},2 \alpha _{12}+4 \alpha _{23}+\alpha _{34},4 \alpha _{12}+\alpha _{23}+3 \alpha _{34}\}	\,	.	\nonumber	
\end{align}
Here, $W_0$ corresponds to the trivial flat connection, $W_{6}$ to the abelian flat connection $U(1)^3$, $W_{1}$, $W_{2}$, $W_{3}$, $W_{-3}$, $W_{7}$, and $W_{-7}$ to the abelian flat connections from the type $S(U(2) \times U(1) \times U(1))$,  $W_{4}$ and $W_{5}$ to the abelian flat connections from the type $S(U(2) \times U(2))$, and $W_{8}$, $W_{-8}$, $W_{9}$, and $W_{-9}$ to the abelian flat connections from the type $S(U(3) \times U(1))$.
We denote the contributions from $W_0$, $W_1$, $W_2$, $W_4$, $W_5$, and $W_6$ by $Z_0$, $Z_1$, $Z_2$, $Z_4$, $Z_5$, and $Z_6$, respectively.
The contributions from abelian flat connections corresponding to $W_3$ and $W_{-3}$, which are complex conjugate to each other, are the same and we denote their sum by $Z_3$.
The case $W_7$ and $W_{-7}$, $W_8$ and $W_{-8}$, and $W_9$ and $W_{-9}$ are also similar and we denote the sum of their contributions as $Z_7$, $Z_8$, and $Z_9$, respectively.
These $Z_a$'s can be written in terms of homological blocks $\widehat{Z}_a$ and the WRT invariant can be written as
\begin{align}
Z_{SU(4)}(M_{3}) = \sum_{a,b} e^{2 \pi i K CS_a} Z_a(q) = 5^{3/2}\sum_{a,b} e^{2 \pi i K CS_a} S_{ab} \widehat{Z}_b(q)		\,	\Big|_{q \searrow e^{\frac{2\pi i}{K}}}
\end{align}
where
\begin{align}
\widehat{Z}_0 &= -q^{\frac{1}{84}} (q+7 q^6+16 q^{11}+13 q^{16}+11 q^{21}+10 q^{26}+17 q^{31}+3 q^{36}+6 q^{41}	+\cdots )	\,	,	\\
\widehat{Z}_1 &= q^{\frac{173}{420}} (1-3 q-q^2+15 q^3+4 q^4-15 q^5+9 q^6+6 q^7+3 q^8+4 q^9	+\cdots )		\,	,	\\
\widehat{Z}_2 &= q^{\frac{257}{420}} (1-3 q+9 q^2+2 q^3-4 q^4-3 q^5+25 q^6+10 q^7-19 q^8+16 q^9	+\cdots )	\,	,	\\
\widehat{Z}_3 &= -2q^{\frac{89}{420}} (1+q+3 q^2-q^3-2 q^5+11 q^6+9 q^7-15 q^8-8 q^9+25 q^{10}	+\cdots )	\,	,	\\
\widehat{Z}_4 &= -q^{1+\frac{341}{420}} (1-q+q^2+4 q^3-3 q^4-q^5+8 q^6-3 q^7+2 q^9+5 q^{10}	+\cdots )	\,	,	\\
\widehat{Z}_5 &= -q^{\frac{89}{420}} (1-q+5 q^2+6 q^3-8 q^4+7 q^6+3 q^8+7 q^{10}+6 q^{11}-6 q^{12}	+\cdots )	\,	,	\\
\widehat{Z}_6 &= q^{\frac{1}{84}} (1-2 q+5 q^2-8 q^3+6 q^4+12 q^5+8 q^6-16 q^7+6 q^8+24 q^9	+\cdots	)	\,	,	\\
\widehat{Z}_7 &= 2q^{1+\frac{341}{420}} (8-8 q-6 q^2-3 q^3+10 q^4-9 q^5-15 q^6-3 q^7+q^8-2 q^9	+\cdots )		\,	,	\\
\widehat{Z}_8 &= q^{\frac{173}{420}} (1+2 q^2-3 q^3-2 q^4+5 q^5+2 q^6-q^7-5 q^8+2 q^9+8 q^{10}+5 q^{11}	+\cdots )	\,	,	\\
\widehat{Z}_9 &= 2q^{\frac{257}{420}} (1-q+q^2+2 q^3-5 q^4+6 q^5+6 q^6-q^7-3 q^8-2 q^9+10 q^{10}	+\cdots )	\,	,
\end{align}
and 
\begin{align}
CS_a = (0, \frac{2}{5}, \frac{3}{5}, \frac{1}{5}, \frac{4}{5}, \frac{1}{5}, 0, \frac{4}{5}, \frac{2}{5}, \frac{3}{5})	\,	,
\end{align}
\begin{align}
\hspace{-10mm}S_{ab} = \frac{1}{5^{\frac{3}{2}}}
{\tiny
\left(
\begin{array}{cccccccccc}
 1 & 1 & 1 & 1 & 1 & 1 & 1 & 1 & 1 & 1 \\
 12 & \frac{1}{2} \left(-3 \sqrt{5}-1\right) & \frac{1}{2} \left(3 \sqrt{5}-1\right) & \frac{1}{2} \left(\sqrt{5}-1\right) & 2 \left(\sqrt{5}+1\right) & 2-2 \sqrt{5} & -3 & \frac{1}{2} \left(-\sqrt{5}-1\right) & \frac{1}{2} (-3) \left(\sqrt{5}-3\right) & \frac{3}{2} \left(\sqrt{5}+3\right) \\
 12 & \frac{1}{2} \left(3 \sqrt{5}-1\right) & \frac{1}{2} \left(-3 \sqrt{5}-1\right) & \frac{1}{2} \left(-\sqrt{5}-1\right) & 2-2 \sqrt{5} & 2 \left(\sqrt{5}+1\right) & -3 & \frac{1}{2} \left(\sqrt{5}-1\right) & \frac{3}{2} \left(\sqrt{5}+3\right) & \frac{1}{2} (-3) \left(\sqrt{5}-3\right) \\
 24 & \sqrt{5}-1 & -\sqrt{5}-1 & -2 \sqrt{5}-1 & 2 \left(\sqrt{5}-3\right) & -2 \left(\sqrt{5}+3\right) & 4 & 2 \sqrt{5}-1 & \frac{1}{2} (-3) \left(\sqrt{5}-1\right) & \frac{3}{2} \left(\sqrt{5}+1\right) \\
 6 & \sqrt{5}+1 & 1-\sqrt{5} & \frac{1}{2} \left(\sqrt{5}-3\right) & \frac{1}{2} \left(7-\sqrt{5}\right) & \frac{1}{2} \left(\sqrt{5}+7\right) & 1 & \frac{1}{2} \left(-\sqrt{5}-3\right) & \frac{1}{2} (-3) \left(\sqrt{5}+1\right) & \frac{3}{2} \left(\sqrt{5}-1\right) \\
 6 & 1-\sqrt{5} & \sqrt{5}+1 & \frac{1}{2} \left(-\sqrt{5}-3\right) & \frac{1}{2} \left(\sqrt{5}+7\right) & \frac{1}{2} \left(7-\sqrt{5}\right) & 1 & \frac{1}{2} \left(\sqrt{5}-3\right) & \frac{3}{2} \left(\sqrt{5}-1\right) & \frac{1}{2} (-3) \left(\sqrt{5}+1\right) \\
 24 & -6 & -6 & 4 & 4 & 4 & -1 & 4 & -6 & -6 \\
 24 & -\sqrt{5}-1 & \sqrt{5}-1 & 2 \sqrt{5}-1 & -2 \left(\sqrt{5}+3\right) & 2 \left(\sqrt{5}-3\right) & 4 & -2 \sqrt{5}-1 & \frac{3}{2} \left(\sqrt{5}+1\right) & \frac{1}{2} (-3) \left(\sqrt{5}-1\right) \\
 8 & 3-\sqrt{5} & \sqrt{5}+3 & \frac{1}{2} \left(1-\sqrt{5}\right) & -2 \left(\sqrt{5}+1\right) & 2 \left(\sqrt{5}-1\right) & -2 & \frac{1}{2} \left(\sqrt{5}+1\right) & -\sqrt{5}-2 & \sqrt{5}-2 \\
 8 & \sqrt{5}+3 & 3-\sqrt{5} & \frac{1}{2} \left(\sqrt{5}+1\right) & 2 \left(\sqrt{5}-1\right) & -2 \left(\sqrt{5}+1\right) & -2 & \frac{1}{2} \left(1-\sqrt{5}\right) & \sqrt{5}-2 & -\sqrt{5}-2 \\
\end{array}
\right)}	\,	,
\end{align}
which satisfies $S_{ab}S_{bc} = \delta_{ac}$.
This $S$-matrix can also be obtained from \eqref{sun-smat} and \eqref{sun-lk} with $C_a=\{W_a\}$ for $a=0,1,2,4,5,6$ and $C_{\pm a}=\{ W_{a}, W_{-a} \}$ for $a=3,7,8,9$.

\subsubsection*{\textbullet \ $H=6$}
The case $H=6$ can be obtained, for example, from $(P_1,P_2,P_3)=(5,7,11)$.
In $\Lambda_3/6\Lambda_3$, there are 22 Weyl orbits,
\begin{align}
&W_{t}	&&\hspace{0mm}	\Lambda_3/6\Lambda_3	\nonumber\\
&W_0	&&\hspace{0mm}	\{0\}					\nonumber\\
&W_1	&&\hspace{0mm}	\{\alpha _{12},\alpha _{23},\alpha _{34},\alpha _{12}+\alpha _{23},\alpha _{23}+\alpha _{34},\alpha _{12}+\alpha _{23}+\alpha _{34},	\nonumber\\
&		&&\hspace{0mm}	\hspace{20mm}5 \alpha _{12},5 \alpha _{23},5 \alpha _{34},5 \alpha _{12}+5 \alpha _{23},5 \alpha _{23}+5 \alpha _{34},5 \alpha _{12}+5 \alpha _{23}+5 \alpha _{34}\}	\nonumber\\
&W_2	&&\hspace{0mm}	\{2 \alpha _{12},2 \alpha _{23},2 \alpha _{34},2 \alpha _{12}+2 \alpha _{23},2 \alpha _{23}+2 \alpha _{34},2 \alpha _{12}+2 \alpha _{23}+2 \alpha _{34},	\nonumber\\
&		&&\hspace{0mm}	\hspace{20mm}4 \alpha _{12},4 \alpha _{23},4 \alpha _{34},4 \alpha _{12}+4 \alpha _{23},4 \alpha _{23}+4 \alpha _{34},4 \alpha _{12}+4 \alpha _{23}+4 \alpha _{34}\}	\nonumber\\
&W_3	&&\hspace{0mm}	\{3 \alpha _{12},3 \alpha _{23},3 \alpha _{34},3 \alpha _{12}+3 \alpha _{23},3 \alpha _{23}+3 \alpha _{34},3 \alpha _{12}+3 \alpha _{23}+3 \alpha _{34}\}	\nonumber\\
&W_4	&&\hspace{0mm}	\{2 \alpha _{12}+\alpha _{23},2 \alpha _{23}+\alpha _{34},2 \alpha _{12}+\alpha _{23}+\alpha _{34},2 \alpha _{12}+2 \alpha _{23}+\alpha _{34},
5 \alpha _{12}+\alpha _{23},	\nonumber\\
&		&&\hspace{0mm}	\hspace{20mm}5 \alpha _{12}+4 \alpha _{23},5 \alpha _{23}+4 \alpha _{34},5 \alpha _{23}+\alpha _{34},5 \alpha _{12}+5 \alpha _{23}+\alpha _{34},	\nonumber\\
&		&&\hspace{0mm}	\hspace{30mm}5 \alpha _{12}+5 \alpha _{23}+4 \alpha _{34},5 \alpha _{12}+\alpha _{23}+\alpha _{34},5 \alpha _{12}+4 \alpha _{23}+4 \alpha _{34}\}	\nonumber\\
&W_{-4}	&&\hspace{0mm}	\{\alpha _{12}+2 \alpha _{23},\alpha _{23}+2 \alpha _{34},\alpha _{12}+\alpha _{23}+2 \alpha _{34},\alpha _{12}+2 \alpha _{23}+2 \alpha _{34},
\alpha _{12}+5 \alpha _{23},	\nonumber\\
&		&&\hspace{0mm}	\hspace{20mm}4 \alpha _{12}+5 \alpha _{23},4 \alpha _{23}+5 \alpha _{34},\alpha _{23}+5 \alpha _{34},4 \alpha _{12}+5 \alpha _{23}+5 \alpha _{34},	\nonumber\\
&		&&\hspace{0mm}	\hspace{30mm}\alpha _{12}+5 \alpha _{23}+5 \alpha _{34},\alpha _{12}+\alpha _{23}+5 \alpha _{34},4 \alpha _{12}+4 \alpha _{23}+5 \alpha _{34}\}	\nonumber\\
&W_5	&&\hspace{0mm}	\{2 \alpha _{12}+\alpha _{34},\alpha _{12}+2 \alpha _{34},2 \alpha _{12}+\alpha _{23}+2 \alpha _{34},2 \alpha _{12}+3 \alpha _{23}+\alpha _{34},	\nonumber\\
&		&&\hspace{0mm}	\hspace{2mm}\alpha _{12}+3 \alpha _{23}+2 \alpha _{34},2 \alpha _{12}+3 \alpha _{23}+2 \alpha _{34},\alpha _{12}+3 \alpha _{23}+\alpha _{34},4 \alpha _{12}+\alpha _{34},	\nonumber\\
&		&&\hspace{0mm}	\hspace{3mm}\alpha _{12}+4 \alpha _{34},4 \alpha _{12}+3 \alpha _{23}+4 \alpha _{34},5 \alpha _{12}+2 \alpha _{34},2 \alpha _{12}+5 \alpha _{34},5 \alpha _{12}+3 \alpha _{23}+5 \alpha _{34},	\nonumber\\
&		&&\hspace{0mm}	\hspace{4mm}5 \alpha _{12}+4 \alpha _{34},4 \alpha _{12}+5 \alpha _{34},5 \alpha _{12}+\alpha _{23}+5 \alpha _{34},5 \alpha _{12}+\alpha _{23}+2 \alpha _{34},	\nonumber\\
&		&&\hspace{0mm}	\hspace{5mm}5 \alpha _{12}+3 \alpha _{23}+4 \alpha _{34},2 \alpha _{12}+\alpha _{23}+5 \alpha _{34},	4 \alpha _{12}+3 \alpha _{23}+5 \alpha _{34},\alpha _{12}+5 \alpha _{23}+\alpha _{34},	\nonumber\\
&		&&\hspace{0mm}	\hspace{6mm}\alpha _{12}+5 \alpha _{23}+4 \alpha _{34},4 \alpha _{12}+5 \alpha _{23}+\alpha _{34},4 \alpha _{12}+5 \alpha _{23}+4 \alpha _{34}\}	\nonumber\\
&W_6	&&\hspace{0mm}	\{\alpha _{12}+\alpha _{34},\alpha _{12}+2 \alpha _{23}+\alpha _{34},5 \alpha _{12}+\alpha _{34},\alpha _{12}+5 \alpha _{34},	\nonumber\\
&		&&\hspace{0mm}	\hspace{65mm}5 \alpha _{12}+4 \alpha _{23}+5 \alpha _{34},5 \alpha _{12}+5 \alpha _{34}\}	\nonumber\\
&W_7	&&\hspace{0mm}	\{3 \alpha _{12}+\alpha _{34},\alpha _{12}+3 \alpha _{34},3 \alpha _{12}+2 \alpha _{23}+3 \alpha _{34},\alpha _{12}+4 \alpha _{23}+\alpha _{34},	\nonumber\\
&		&&\hspace{0mm}	\hspace{5mm}\alpha _{12}+4 \alpha _{23}+3 \alpha _{34},3 \alpha _{12}+4 \alpha _{23}+\alpha _{34},3 \alpha _{12}+4 \alpha _{23}+3 \alpha _{34},5 \alpha _{12}+3 \alpha _{34},	\nonumber\\
&		&&\hspace{0mm}	\hspace{10mm}3 \alpha _{12}+5 \alpha _{34},5 \alpha _{12}+2 \alpha _{23}+5 \alpha _{34},5 \alpha _{12}+2 \alpha _{23}+3 \alpha _{34},3 \alpha _{12}+2 \alpha _{23}+5 \alpha _{34}\}	\nonumber\\
&W_8	&&\hspace{0mm}	\{3 \alpha _{12}+2 \alpha _{34},2 \alpha _{12}+3 \alpha _{34},3 \alpha _{12}+\alpha _{23}+3 \alpha _{34},4 \alpha _{12}+3 \alpha _{34},3 \alpha _{12}+4 \alpha _{34},	\nonumber\\
&		&&\hspace{0mm}	\hspace{5mm}4 \alpha _{12}+\alpha _{23}+4 \alpha _{34},4 \alpha _{12}+\alpha _{23}+3 \alpha _{34},3 \alpha _{12}+\alpha _{23}+4 \alpha _{34},2 \alpha _{12}+5 \alpha _{23}+2 \alpha _{34},	\nonumber\\
&		&&\hspace{0mm}	\hspace{10mm}2 \alpha _{12}+5 \alpha _{23}+3 \alpha _{34},3 \alpha _{12}+5 \alpha _{23}+2 \alpha _{34},3 \alpha _{12}+5 \alpha _{23}+3 \alpha _{34}\}	\nonumber\\
&W_9	&&\hspace{0mm}	\{2 \alpha _{12}+4 \alpha _{23}+2 \alpha _{34},2 \alpha _{12}+2 \alpha _{34},4 \alpha _{12}+2 \alpha _{34},2 \alpha _{12}+4 \alpha _{34},	\nonumber\\
&		&&\hspace{0mm}	\hspace{65mm}4 \alpha _{12}+2 \alpha _{23}+4 \alpha _{34},4 \alpha _{12}+4 \alpha _{34}\}	\nonumber\\
&W_{10}	&&\hspace{0mm}	\{3 \alpha _{12}+3 \alpha _{34}\}		\nonumber
\end{align}
\begin{align}
&W_{11}	&&\hspace{0mm}	\{3 \alpha _{12}+\alpha _{23},3 \alpha _{12}+2 \alpha _{23},3 \alpha _{23}+\alpha _{34},3 \alpha _{23}+2 \alpha _{34},3 \alpha _{12}+2 \alpha _{23}+2 \alpha _{34},	\nonumber\\
&		&&\hspace{0mm}	\hspace{2mm}3 \alpha _{12}+\alpha _{23}+\alpha _{34},3 \alpha _{12}+3 \alpha _{23}+\alpha _{34},3 \alpha _{12}+3 \alpha _{23}+2 \alpha _{34},4 \alpha _{12}+\alpha _{23},	\nonumber\\
&		&&\hspace{0mm}	\hspace{4mm}4 \alpha _{12}+3 \alpha _{23},4 \alpha _{23}+\alpha _{34},4 \alpha _{23}+3 \alpha _{34},4 \alpha _{12}+4 \alpha _{23}+\alpha _{34},	\nonumber\\
&		&&\hspace{0mm}	\hspace{6mm}4 \alpha _{12}+4 \alpha _{23}+3 \alpha _{34},4 \alpha _{12}+\alpha _{23}+\alpha _{34},4 \alpha _{12}+3 \alpha _{23}+3 \alpha _{34},5 \alpha _{12}+2 \alpha _{23},	\nonumber\\
&		&&\hspace{0mm}	\hspace{8mm}5 \alpha _{12}+3 \alpha _{23},5 \alpha _{23}+2 \alpha _{34},
5 \alpha _{23}+3 \alpha _{34},5 \alpha _{12}+5 \alpha _{23}+2 \alpha _{34},	\nonumber\\
&		&&\hspace{0mm}	\hspace{10mm}5 \alpha _{12}+5 \alpha _{23}+3 \alpha _{34},5 \alpha _{12}+2 \alpha _{23}+2 \alpha _{34},5 \alpha _{12}+3 \alpha _{23}+3 \alpha _{34}\}	\nonumber\\
&W_{-11}	&&\hspace{0mm}	\{\alpha _{12}+3 \alpha _{23},2 \alpha _{12}+3 \alpha _{23},\alpha _{23}+3 \alpha _{34},2 \alpha _{23}+3 \alpha _{34},2 \alpha _{12}+2 \alpha _{23}+3 \alpha _{34},	\nonumber\\
&		&&\hspace{2mm}	\alpha _{12}+\alpha _{23}+3 \alpha _{34},\alpha _{12}+3 \alpha _{23}+3 \alpha _{34},2 \alpha _{12}+3 \alpha _{23}+3 \alpha _{34},\alpha _{12}+4 \alpha _{23},	\nonumber\\
&		&&\hspace{4mm}	3 \alpha _{12}+4 \alpha _{23},\alpha _{23}+4 \alpha _{34},3 \alpha _{23}+4 \alpha _{34},\alpha _{12}+4 \alpha _{23}+4 \alpha _{34},3 \alpha _{12}+4 \alpha _{23}+4 \alpha _{34},	\nonumber\\
&		&&\hspace{6mm}	\alpha _{12}+\alpha _{23}+4 \alpha _{34},3 \alpha _{12}+3 \alpha _{23}+4 \alpha _{34},2 \alpha _{12}+5 \alpha _{23},3 \alpha _{12}+5 \alpha _{23},	\nonumber\\
&		&&\hspace{8mm}	2 \alpha _{23}+5 \alpha _{34},3 \alpha _{23}+5 \alpha _{34},2 \alpha _{12}+5 \alpha _{23}+5 \alpha _{34},3 \alpha _{12}+5 \alpha _{23}+5 \alpha _{34},	\nonumber\\
&		&&\hspace{10mm}	2 \alpha _{12}+2 \alpha _{23}+5 \alpha _{34},3 \alpha _{12}+3 \alpha _{23}+5 \alpha _{34}\}	\,	\nonumber\\
&W_{12}	&&\hspace{0mm}	\{3 \alpha _{12}+2 \alpha _{23}+\alpha _{34},5 \alpha _{12}+2 \alpha _{23}+\alpha _{34},5 \alpha _{12}+4 \alpha _{23}+\alpha _{34},5 \alpha _{12}+4 \alpha _{23}+3 \alpha _{34}\}		\nonumber\\
&W_{-12}	&&\hspace{0mm}	\{\alpha _{12}+2 \alpha _{23}+3 \alpha _{34},\alpha _{12}+2 \alpha _{23}+5 \alpha _{34},\alpha _{12}+4 \alpha _{23}+5 \alpha _{34},3 \alpha _{12}+4 \alpha _{23}+5 \alpha _{34},\}		\nonumber\\
&W_{13}	&&\hspace{0mm}	\{3 \alpha _{12}+\alpha _{23}+2 \alpha _{34},\alpha _{12}+4 \alpha _{23}+2 \alpha _{34},3 \alpha _{12}+4 \alpha _{23}+2 \alpha _{34},	\nonumber\\
&		&&\hspace{5mm}	4 \alpha _{12}+\alpha _{23}+2 \alpha _{34},4 \alpha _{12}+2 \alpha _{23}+3 \alpha _{34},\alpha _{12}+5 \alpha _{23}+2 \alpha _{34},	\nonumber\\
&		&&\hspace{10mm}	\alpha _{12}+5 \alpha _{23}+3 \alpha _{34},4 \alpha _{12}+5 \alpha _{23}+2 \alpha _{34},4 \alpha _{12}+5 \alpha _{23}+3 \alpha _{34},	\nonumber\\
&		&&\hspace{15mm}	3 \alpha _{12}+\alpha _{23}+5 \alpha _{34},4 \alpha _{12}+2 \alpha _{23}+5 \alpha _{34},
4 \alpha _{12}+\alpha _{23}+5 \alpha _{34}\}	\nonumber\\
&W_{-13}	&&\hspace{0mm}	\{2 \alpha _{12}+\alpha _{23}+3 \alpha _{34},2 \alpha _{12}+4 \alpha _{23}+\alpha _{34},2 \alpha _{12}+4 \alpha _{23}+3 \alpha _{34},	\nonumber\\
&		&&\hspace{5mm}	2 \alpha _{12}+\alpha _{23}+4 \alpha _{34},3 \alpha _{12}+2 \alpha _{23}+4 \alpha _{34},2 \alpha _{12}+5 \alpha _{23}+\alpha _{34},		\nonumber\\
&		&&\hspace{10mm}	2 \alpha _{12}+5 \alpha _{23}+4 \alpha _{34},3 \alpha _{12}+5 \alpha _{23}+\alpha _{34},3 \alpha _{12}+5 \alpha _{23}+4 \alpha _{34},	\nonumber\\
&		&&\hspace{15mm}	5 \alpha _{12}+\alpha _{23}+3 \alpha _{34},5 \alpha _{12}+2 \alpha _{23}+4 \alpha _{34},5 \alpha _{12}+\alpha _{23}+4 \alpha _{34}\}	\nonumber\\
&W_{14}	&&\hspace{0mm}	\{4 \alpha _{12}+2 \alpha _{23},4 \alpha _{23}+2 \alpha _{34},4 \alpha _{12}+4 \alpha _{23}+2 \alpha _{34},4 \alpha _{12}+2 \alpha _{23}+2 \alpha _{34}\}		\nonumber\\
&W_{-14}	&&\hspace{0mm}	\{2 \alpha _{12}+4 \alpha _{23},2 \alpha _{23}+4 \alpha _{34},2 \alpha _{12}+4 \alpha _{23}+4 \alpha _{34},2 \alpha _{12}+2 \alpha _{23}+4 \alpha _{34}\}		\nonumber\\
&W_{15}	&&\hspace{0mm}	\{4 \alpha _{12}+2 \alpha _{23}+\alpha _{34},4 \alpha _{12}+3 \alpha _{23}+\alpha _{34},4 \alpha _{12}+3 \alpha _{23}+2 \alpha _{34},	\nonumber\\
&		&&\hspace{10mm}	5 \alpha _{12}+3 \alpha _{23}+\alpha _{34},5 \alpha _{12}+3 \alpha _{23}+2 \alpha _{34},5 \alpha _{12}+4 \alpha _{23}+2 \alpha _{34}\}	\nonumber\\
&W_{-15}	&&\hspace{0mm}	\{\alpha _{12}+2 \alpha _{23}+4 \alpha _{34},\alpha _{12}+3 \alpha _{23}+4 \alpha _{34},2 \alpha _{12}+3 \alpha _{23}+4 \alpha _{34},	\nonumber\\
&		&&\hspace{10mm}	\alpha _{12}+3 \alpha _{23}+5 \alpha _{34},2 \alpha _{12}+3 \alpha _{23}+5 \alpha _{34},2 \alpha _{12}+4 \alpha _{23}+5 \alpha _{34}\}	\,	.	\nonumber
\end{align}
Here, $W_0$ corresponds to the trivial flat connection, $W_{10}$ to the central flat connection, $W_{5}$, $W_{11}$, and $W_{-11}$ to the abelian flat connection $U(1)^3$, $W_{1}$, $W_{2}$, $W_{4}$, $W_{-4}$, $W_{7}$, $W_{8}$, $W_{13}$, and $W_{-13}$ to the abelian flat connections from the type $S(U(2) \times U(1) \times U(1))$, $W_{3}$, $W_{6}$, $W_{9}$, $W_{15}$, and $W_{-15}$ to the abelian flat connections from the type $S(U(2) \times U(2))$, and $W_{12}$, $W_{-12}$, $W_{14}$ and $W_{-14}$ to the abelian flat connections from the type $S(U(3) \times U(1))$.
We can see that the contributions from abelian flat connections corresponding to $W_t$ and $W_{-t}$, $t=4,11,12,13,14,15$, which are complex conjugate to each other, are separately the same.
Sums of their contributions are denoted by $Z'_t$, $t=4,11,12,13,14,15$, respectively.
The contributions from $W_t$, $t=0,1,2, 3, 5,6,7,8,9,10$, are denoted by $Z'_t$'s, respectively. 
In addition, $W_0$ and $W_{10}$ are related by the action of the center, and their contributions are the same, $Z'_{0}=Z'_{10}$.
This is similar to $W_1$ and $W_8$, $W_2$ and $W_7$, $W_4$ and $W_{13}$, $W_6$ and $W_9$, and $W_{12}$ and $W_{14}$, and their contributions are the same, $Z'_{1}=Z'_{8}$, $Z'_{2}=Z'_{7}$, $Z'_{4}=Z'_{13}$, $Z'_{6}=Z'_{9}$, and $Z'_{12}=Z'_{14}$, respectively.
Thus, we have $C_0 = \{ W_0, W_{10} \}$, $C_1 = \{ W_1, W_8 \}$, $C_2=\{ W_2, W_7 \}$, $C_3 = \{ W_3 \}$, $C_{\pm 4} = \{ W_{4}, W_{-4}, W_{13}, W_{-13} \}$, $C_5 = \{ W_5 \}$, $C_6 = \{ W_6, W_9 \}$, $C_{\pm 7} = \{ W_{11}, W_{-11} \}$, $C_{\pm 8} = \{ W_{12}, W_{-12}, W_{14}, W_{-14} \}$, $C_{\pm 9} = \{ W_{15}, W_{-15} \}$.
After renaming $Z_a'$'s as $Z_t=Z_t'$, $t=0, 1, \ldots, 6$, $Z_{7}=Z_{11}'$, $Z_{8}=Z_{12}'$, and $Z_{9}=Z_{15}'$,
the WRT invariant can be written as
\begin{align}
Z_{SU(4)} (M_{3}) =\sum_{a} e^{2\pi i K CS_a} Z_a
=
\sqrt{2} \, 6^{\frac{3}{2}} \sum_{a,b} e^{2\pi i K S_a} S_{ab} \widehat{Z}_b		\,	\Big|_{q \searrow e^{\frac{2\pi i}{K}}}
\end{align}
where
\begin{align}
\widehat{Z}_0 &= -q^{74+\frac{59}{77}} (2+2 q^6-q^{15}-q^{18}-3 q^{21}+2 q^{24}-q^{27}-4 q^{30}-4 q^{33}	+\cdots)		\,	,	\\
\widehat{Z}_1 &= -q^{57+\frac{431}{462}} (1-2 q^3+q^6-2 q^8+2 q^{11}-2 q^{13}+q^{16}-2 q^{19}-4 q^{23}	+\cdots)			\,	,	\\
\widehat{Z}_2 &= q^{51+\frac{100}{231}} (1+3 q^{12}+q^{13}+3 q^{16}-4 q^{18}-2 q^{19}+3 q^{20}-2 q^{23}+2 q^{24}	+\cdots)	\,	,	\\
\widehat{Z}_3 &= -q^{55+\frac{41}{154}} (1+q^6+2 q^{15}+2 q^{18}+2 q^{21}+2 q^{24}+2 q^{30}+2 q^{33}+3 q^{36}	+\cdots)	\,	,	\\
\widehat{Z}_4 &= -2q^{55+\frac{41}{154}} (1+q^6+q^{12}+q^{15}+3 q^{16}+2 q^{18}-3 q^{19}+q^{20}-q^{21}	+\cdots)		\,	,	\\
\widehat{Z}_5 &= -q^{66+\frac{277}{462}} (2+3 q^4+2 q^5+q^6-2 q^{11}-4 q^{13}-4 q^{15}+q^{18}+2 q^{19}+q^{20}	+\cdots)	\,	,	\\
\widehat{Z}_6 &= q^{59+\frac{23}{231}} (1+2 q^6+q^{12}-2 q^{14}+2 q^{16}-q^{18}+q^{19}+q^{23}+2 q^{24}-q^{26}	+\cdots)	\,	,	\\
\widehat{Z}_7 &= -2q^{57+\frac{431}{462}} (1-q^3-2 q^5-2 q^6-q^8+q^{12}-q^{13}+2 q^{14}-q^{15}+2 q^{16}	+\cdots)		\,	,	\\
\widehat{Z}_8 &= 2q^{61+\frac{59}{77}} (1-q^5-2 q^{11}-q^{13}-q^{18}-q^{21}-3 q^{22}-2 q^{23}+2 q^{24}-3 q^{26}	+\cdots)	\,	,	\\
\widehat{Z}_9 &= -2q^{68+\frac{277}{462}} (2+q-q^3-q^8-2 q^9-2 q^{13}-q^{14}+q^{16}+q^{17}-q^{19}	+\cdots)			\,	,
\end{align}
and
\begin{align}
CS_a= (0,\frac{1}{6},\frac{2}{3}, \frac{1}{2}, \frac{1}{2}, \frac{5}{6},\frac{1}{3},\frac{1}{6},0,\frac{5}{6})	\,	,
\end{align}
\begin{align}
S_{ab} = \frac{1}{2 \cdot 3^{\frac{3}{2}}}\left(
\begin{array}{cccccccccc}
 1 & 1 & 1 & 1 & 1 & 1 & 1 & 1 & 1 & 1 \\
 12 & 5 & -3 & -4 & -1 & -4 & 0 & -1 & 3 & 8 \\
 12 & -3 & -3 & 12 & 3 & 0 & 0 & -3 & 3 & 0 \\
 3 & -1 & 3 & -1 & -1 & -1 & 3 & -1 & 3 & -1 \\
 24 & -2 & 6 & -8 & 1 & 4 & -12 & -2 & -3 & 4 \\
 12 & -4 & 0 & -4 & 2 & -4 & 6 & 2 & -6 & 2 \\
 6 & 0 & 0 & 6 & -3 & 3 & 3 & 0 & -3 & 3 \\
 24 & -2 & -6 & -8 & -2 & 4 & 0 & 4 & 6 & -8 \\
 8 & 2 & 2 & 8 & -1 & -4 & -4 & 2 & -1 & -4 \\
 6 & 4 & 0 & -2 & 1 & 1 & 3 & -2 & -3 & -5 \\
\end{array}
\right)	\,	,
\end{align}
which satisfies $S_{ab}S_{bc} = \delta_{ac}$.
This $S$-matrix can also be obtained from \eqref{sun-smat} and \eqref{sun-lk}.


\section{Discussion}

We discussed how to calculate homological blocks for general $SU(N)$ gauge group and for general Seifert manifolds with $P_j$'s being pairwise coprime from the exact expression of Lawrence, Rozansky, and Mari\~no.
We firstly calculated the partition function of the analytically continued $SU(N)$ Chern-Simons theory or the WRT invariant and obtained the expression for $Z_a$'s that are labelled by abelian flat connections.
From $Z_a$'s, we extracted homological blocks $\widehat{Z}_b$, which are related to $Z_a$ by the $S$-matrix.
We proposed a formula to calculate the $S$-matrix in general cases and checked that the $S$-matrix calculated from the formula by using the linking form agrees with the result that is obtained from the calculation of $Z_a$'s and $\widehat{Z}_b$'s.
Also, we discussed general forms of the WRT invariant in terms of homological blocks and checked that examples that we discussed fit into the expected forms.
In addition, we discussed some properties of the $SU(N)$ WRT invariant.
We found a symmetry that the abelian flat connections in the same orbit under the action of the center give the same contribution $Z_a$ to the WRT invariant up to the overall exponential factor $e^{\pi i K}$.
We also discussed a symmetry that the contributions from the abelian flat connection and from the conjugate abelian flat connection are the same.	
In addition, we also saw that the exact expression of Lawrence and Rozansky can be understood in the context of the resurgent analysis when $H=1$.
\\

There are several interesting directions to consider.
In \cite{Gukov-Pei-Putrov-Vafa}, the superconformal index and the topologically twisted index can be calculated from homological blocks via 
\begin{align}
\mathcal{I}(q,t) &= \sum_{a} |\text{Stab}_{S_N}(a) | \widehat{Z}_a(q, t) \widehat{Z}_a(q^{-1},t^{-1}) 	\,	,	\label{sc-index}	\\
\mathcal{I}_{\text{top}}(q,t) &= \sum_{a} |\text{Stab}_{S_N}(a) | \widehat{Z}_a(q, t) \widehat{Z}_a(q^{-1},t)	\,	, 	\label{t-index}
\end{align}
and several examples supported them.
In the calculation discussed above, it was not obvious how to calculate $\widehat{Z}(q^{-1})$ on $M_3$ with $|q|<1$, so we didn't calculate \eqref{sc-index} or \eqref{t-index} with $t=1$.
It would be interesting to calculate the indices for general Seifert manifolds.	\\

We have obtained various homological blocks.
When $G=SU(2)$ and the number of singular fibers $F$ is 3 or 4 with the genus of the base surface being zero, homological blocks are expressed in terms of the false theta functions which are known in literature.
By varying the number of singular fibers or the genus, we found that homological blocks are given by false theta functions $\widetilde{\Phi}^{(l)}_{2HP}(q)$ and $\widetilde{\Psi}^{(l)}_{2HP}(q)$, and their derivatives.
Moreover, considering the higher rank case $SU(N)$, $N \geq 3$, it is expected that the homological blocks in those cases would provide new false theta functions.\footnote{For reference, certain higher rank false theta functions similar to the expressions for the higher rank case with $F=3$ in section \ref{sec:higher-rank} have been discussed in \cite{CREUTZIG2017203, Bringmann2017}.}
It would be interesting to study modular properties of homological blocks obtained in this paper including their properties upon $q \rightarrow q^{-1}$ \cite{Cheng:2018vpl}.	\\

Resurgent analysis was discussed in the case of the $SU(2)$ Chern-Simons theory with $H=1$ in literature.
We considered it in the context of the exact formula in \cite{Lawrence-Rozansky} when $H=1$.
It would be interesting to study the resurgent analysis for more general cases.		\\

A mathematical construction or a definition for a Khovanov-type homology for closed 3-manifolds is not available yet.
But if it is constructed, from the analogy, it would be expected that there is also a homology theory for closed 3-manifolds analogous to the HOMFLY homology for knots and links.
It will be also interesting to study properties of such an expected homology theory from homological blocks for $G=SU(N)$ obtained in this paper.

\acknowledgments{I would like to thank Sergei Gukov for valuable discussions and also helpful remarks on the manuscript.
I also would like to thank Sungbong Chun, Sunghyuk Park, and Nikita Sopenko for useful discussions.
I am also grateful to the Korea Institute for Advanced Study (KIAS) for hospitality at several stages of this work.
}


\begin{appendices}


\section{$G=SU(2)$ and four singular fibers}
\label{sec:su2f4}

We can proceed similarly also for the case of four singular fibers as in the previous sections.
The contribution of abelian flat connections to the WRT invariant of Seifert manifolds with four singular fibers is
\begin{align}
Z^{\text{ab}}_{SU(2)}(M_{3}) = \frac{B}{2 \pi i} q^{-\phi_4/4} \sum_{t=0}^{H-1} \int_{\Gamma_t} dy \ e^{-\frac{K}{2 \pi i} \frac{H}{P} y^2 - 2 K t y} \
\frac{\prod_{j=1}^{4} e^{\frac{y}{P_j}} - e^{-\frac{y}{P_j}} }{(e^{y} - e^{-y})^{2}}
\label{int-formula4}
\end{align}
where
\begin{align}
P = \prod_{j=1}^{4} P_j	\,	,	\hspace{5mm}	H= \sum_{j=1}^{4} \frac{Q_j}{P_j}	\,	,	\hspace{5mm}	
\phi_4	&=	3 \, \text{sign}\left( \frac{H}{P} \right) + \sum_{j=1}^{4} \left( 12 s(Q_j, P_j) - \frac{Q_j}{P_j} \right)	
\end{align}
with the integration cycle $\Gamma_t$ as in section \ref{sec:su2f3}.
The last factor can be expressed as
\begin{align}
\frac{\prod_{j=1}^{4} e^{\frac{y}{P_j}} - e^{-\frac{y}{P_j}} }{(e^{y} - e^{-y})^{2}}
= \frac{\sum_{s=0}^{7}{(-1)^s \cosh \frac{R_s}{P} y}}{2 (\sinh y)^{2}}	
\label{cosh4}
\end{align}
where $R_s$'s with even (resp. odd) $s$ are $P\Big(\sum_{j=1}^{4} \frac{\epsilon_j}{P_j}\Big)$ with $\prod_{j=1}^{4} \epsilon_j = 1$ (resp. $-1$) up to an overall sign where $\epsilon_j = \pm 1$.\footnote{For example, 

\hspace{-23mm}\begin{tabular}{c c c c}
$\begin{aligned}[t]
R_0&=P\Big(\frac{1}{P_1}+\frac{1}{P_2}+\frac{1}{P_3}+\frac{1}{P_4}\Big)	\,	,	\\
R_4&=P\Big(\frac{1}{P_1}-\frac{1}{P_2}+\frac{1}{P_3}-\frac{1}{P_4}\Big)	\,	,
\end{aligned}$ 	& 
$\begin{aligned}[t]
R_1&=P\Big(\frac{1}{P_1}+\frac{1}{P_2}+\frac{1}{P_3}-\frac{1}{P_4}\Big)	\,	,	\\
R_5&=P\Big(\frac{1}{P_1}-\frac{1}{P_2}+\frac{1}{P_3}+\frac{1}{P_4}\Big)	\,	,
\end{aligned}$ 	& 
$\begin{aligned}[t]
R_2&=P\Big(\frac{1}{P_1}+\frac{1}{P_2}-\frac{1}{P_3}-\frac{1}{P_4}\Big)	\,	,	\\
R_6&=P\Big(\frac{1}{P_1}-\frac{1}{P_2}-\frac{1}{P_3}+\frac{1}{P_4}\Big)	\,	,
\end{aligned}$	&
$\begin{aligned}[t]
R_3&=P\Big(\frac{1}{P_1}+\frac{1}{P_2}-\frac{1}{P_3}+\frac{1}{P_4}\Big)	\,	,	\\
R_7&=P\Big(\frac{1}{P_1}-\frac{1}{P_2}-\frac{1}{P_3}-\frac{1}{P_4}\Big)	\,	.
\end{aligned}$
\end{tabular}
}

Let $\varphi_{2P}^{(l)}(n)$ be the periodic function
\begin{align}
\varphi^{(l)}_{2P}(n) = 
\begin{cases}
1	\quad	\text{if } n \equiv \pm l	\quad	\text{mod } 2P	\\
0	\quad	\text{otherwise}	\,	,
\end{cases}	\label{varphi}
\end{align}
so it satisfies $\varphi^{(l)}_{2P}(n) = \varphi^{(2P-l)}_{2P}(n)$.
Then we have
\begin{align}
\frac{\cosh (P+l)y}{\sinh Py} = \sum_{m=0}^{M_l} (e^{(l-2mP)y} - e^{-(l-2mP)y}) + \sum_{n=0}^{\infty} \varphi^{(l-2M_l P)}_{2P}(n) e^{-ny}	\,	
\label{conhin4f}
\end{align}
where $M_l=-1, 0, 1, \cdots$ is such that $2M_l P < l < 2(M_l+1)P$.
By taking derivative of \eqref{conhin4f} with respect to $y$ and using 
\begin{align}
\frac{\sinh (P+l)y}{\sinh Py} = \sum_{m=0}^{M_l} (e^{(l-2mP)y} - e^{-(l-2mP)y}) - \sum_{n=0}^{\infty} \psi^{(l-2M_l P)}_{2P}(n) e^{-ny}	\,	
\end{align}
where $M_l=-1, 0, 1, \cdots$ is such that $2M_l P < l < 2(M_l+1)P$, we obtain
\begin{align}
\frac{\cosh ly}{(\sinh Py)^2} 
= \frac{1}{P} \bigg( \sum_{m=0}^{M_l} 2mP (e^{(l-2mP)y} + e^{-(l-2mP)y}) + \sum_{n=0}^{\infty} n \varphi^{(l-2M_l P)}_{2P}(n) e^{-ny} - l \sum_{n=0}^{\infty} \psi^{(l-2M_l P)}_{2P}(n) e^{-ny} \bigg)	\,	.
\end{align}
We plug \eqref{cosh4} into \eqref{int-formula4} and calculate the integral as in section \ref{ssec:calculation}.
Evaluating the integral with the contour $\gamma$ extending along the imaginary axis of the $y$-plane that passes through $\text{Re} \, y > 0$ with an analytically continued $K$ such that $\text{Im} \, K<0$, we have the partition function of the analytically continued $SU(2)$ theory
\begin{align}
\begin{split}
Z_{SU(2)}(M_3) &= \frac{B}{4i} q^{-\phi_4/4} \Big( \frac{2i}{K}\frac{P}{H} \Big)^{\frac{1}{2}}
\sum_{t=0}^{H-1} 
e^{2\pi i K \frac{P}{H} t^2} 	\\
&\hspace{30mm} \times \sum_{s=0}^{7} (-1)^s 
\Bigg(
\sum_{m=0}^{M_{R_s}} 2m \big( e^{-\frac{2 \pi i t}{H}(R_s-2mP)}+e^{\frac{2 \pi i t}{H}(R_s-2mP)} \big) q^{\frac{1}{4 H P} (R_s-2mP)^2}
\\
&\hspace{45mm}+\frac{1}{P}\sum_{n=0}^{\infty} \big( n \varphi^{(R_s-2M_{R_s}P)}_{2P}(n) - R_s \psi^{(R_s-2M_{R_s}P)}_{2P}(n) \big) e^{2\pi i \frac{t}{H}n}  q^{\frac{n^2}{4HP} \frac{1}{K}} \Bigg)
\end{split}	\label{su2-f4-partftn}
\end{align}
where $M_{R_s}=0,1,\cdots$ satisfy $2M_{R_s}P < R_s < 2(M_{R_s}+1)P$ given an $R_s$ for each $s$.
As $P_j$'s are coprime in our setup, the maximum value that $R_s$ can take is less than $2P$.
Thus, when $F=4$, $M_{R_s}$ is $0$, so the terms in the second line of \eqref{su2-f4-partftn} vanish.
Therefore, we obtain
\begin{align}
\begin{split}
Z_{SU(2)}(M_3) &= \frac{B}{4i} q^{-\phi_4/4} \Big( \frac{2i}{K} \frac{P}{H} \Big)^{\frac{1}{2}} \frac{1}{P}
\sum_{t=0}^{H-1} 
e^{2\pi i K \frac{P}{H} t^2} 	\\
&\hspace{30mm} \times \sum_{s=0}^{7} (-1)^s 
\sum_{n=0}^{\infty} \big( n \varphi^{(R_s)}_{2P}(n) - R_s \psi^{(R_s)}_{2P}(n) \big) e^{2\pi i \frac{t}{H}n}  q^{\frac{n^2}{4PH}}	\,	.
\end{split}
\end{align}


\subsection{The case $H=1$}

For the integer Seifert homology sphere, the partition function is given by
\begin{align}
\begin{split}
Z_{SU(2)}(M_3) = \frac{B}{4i} q^{-\phi_4/4} \Big( \frac{2i}{K}P \Big)^{\frac{1}{2}} \frac{1}{P}
\sum_{s=0}^{7} (-1)^s 
\sum_{n=0}^{\infty} \big( n \varphi^{(R_s)}_{2P}(n) - R_s \psi^{(R_s)}_{2P}(n) \big) q^{\frac{n^2}{4P}}	
\label{4f-h1}
\end{split}	\,	.
\end{align}
Denoting 
\begin{align}
\widetilde{\Phi}^{(l)}_{HP}(q)
:= \sum_{n=0}^{\infty} n\varphi^{(l)}_{2HP}(n) q^{\frac{n^2}{4PH}}	\,	,
\end{align}
\eqref{4f-h1} can be written as
\begin{align}
Z_{SU(2)}(M_3) = \frac{B}{4i} q^{-\phi/4} \Big( \frac{2i}{K}P \Big)^{\frac{1}{2}} \frac{1}{P} 
\sum_{s=0}^{7} (-1)^s \big( \widetilde{\Phi}^{(R_s)}_{P}(q) - R_s \widetilde{\Psi}^{(R_s)}_{P}(q) \big)	\,	. 	\label{4f-h1-2}
\end{align}
It is known that $\widetilde{\Phi}_{HP}^{(l)}(q)$ is a false theta function, which is the Eichler integral of the modular form $\Phi^{(l)}_{HP}(q) := \sum_{n=0}^{\infty} \varphi^{(l)}_{2HP}(n) q^{\frac{n^2}{4PH}}$ of weight $1/2$ \cite{Zagier-identity}.
\eqref{4f-h1-2} agrees with the result in \cite{Hikami-lattice1} where only the case $H=1$ has been discussed.\footnote{Recently, the case of $F=4$ with $H \geq 2$ was also discussed in \cite{Cheng:2018vpl}.}


\subsection{The case $H\geq 2$}
When $H\geq 2$, $\varphi^{(l)}_{2P}(n)$ can also be decomposed in terms of $\varphi^{(l)}_{2HP}(n)$
\begin{align}
\varphi^{(l)}_{2P}(n) = \sum_{h=0}^{\big\lceil\frac{H}{2}-1\big\rceil} \varphi^{(2hP+l)}_{2HP}(n) + \sum_{h=0}^{\big\lfloor\frac{H}{2}-1\big\rfloor} \varphi^{(2(h+1)P-l)}_{2HP}(n)	\,	.	\label{phi-decomp}
\end{align}
Therefore, by using \eqref{psi-decomp} and \eqref{phi-decomp} the partition function can be written as
\begin{align}
\begin{split}
Z_{SU(2)}(M_3) &= \frac{B}{4i} q^{-\phi/4} \Big( \frac{2i}{K} \frac{P}{H} \Big)^{\frac{1}{2}} \frac{1}{P} 
\Bigg[
\sum_{s=0}^{7} (-1)^s \sum_{n=0}^{\infty} \bigg( n \big( \sum_{h=0}^{\big\lceil\frac{H}{2}-1\big\rceil} \varphi^{(2hP+R_s)}_{2HP}(n) + \sum_{h=0}^{\big\lfloor\frac{H}{2}-1\big\rfloor} \varphi^{(2(h+1)P-R_s)}_{2HP}(n) \big) \\
&\hspace{55mm}- R_s  \big( \sum_{h=0}^{\big\lceil\frac{H}{2}-1\big\rceil} \psi^{(2hP+R_s)}_{2HP}(n) - \sum_{h=0}^{\big\lfloor\frac{H}{2}-1\big\rfloor} \psi^{(2(h+1)P-R_s)}_{2HP}(n) \big) \bigg) q^{\frac{n^2}{4HP}}	\\
&\hspace{10mm} + \sum_{t=1}^{H-1} 
e^{2\pi i K \frac{P}{H} t^2} 
\sum_{s=0}^{7} (-1)^s
\Bigg(
\sum_{n=0}^{\infty} \bigg( n \big( \sum_{h=0}^{\big\lceil\frac{H}{2}-1\big\rceil} \varphi^{(2hP+R_s)}_{2HP}(n) + \sum_{h=0}^{\big\lfloor\frac{H}{2}-1\big\rfloor} \varphi^{(2(h+1)P-R_s)}_{2HP}(n) \big)	\\
&\hspace{40mm} - R_s  \big( \sum_{h=0}^{\big\lceil\frac{H}{2}-1\big\rceil} \psi^{(2hP+R_s)}_{2HP}(n) - \sum_{h=0}^{\big\lfloor\frac{H}{2}-1\big\rfloor} \psi^{(2(h+1)P-R_s)}_{2HP}(n) \big) \bigg) e^{2\pi i \frac{t}{H}n} q^{\frac{n^2}{4HP}}
\Bigg)
\Bigg]	\,	.
\end{split}
\end{align}
As before, when $H$ is odd, the WRT invariant is given by
\begin{align}
\begin{split}
& Z_{SU(2)}(M_3) = \frac{B}{4i} q^{-\phi_4/4} \Big( \frac{2i}{K}\frac{P}{H} \Big)^{\frac{1}{2}} 	\frac{1}{P} \\
&\hspace{5mm}\times
\Bigg[
\sum_{s=0}^{7} (-1)^s \Bigg( \sum_{h=0}^{\frac{H-1}{2}} ( \widetilde{\Phi}^{(2hP+R_s)}_{HP} - R_s  \widetilde{\Psi}^{(2hP+R_s)}_{HP} ) + \sum_{h=0}^{\frac{H-1}{2}-1} (\widetilde{\Phi}^{(2(h+1)P-R_s)}_{HP} + R_s \widetilde{\Psi}^{(2(h+1)P-R_s)}_{HP} )	\Bigg)	\\
&\hspace{10mm}+\sum_{s=0}^{7} (-1)^s \Bigg( \sum_{t=1}^{\frac{H-1}{2}} e^{2\pi i K \frac{P}{H} t^2} 
\Big( \sum_{h=0}^{\frac{H-1}{2}} (e^{-2\pi i \frac{t}{H}(2hP+R_s)} + e^{2\pi i \frac{t}{H}(2hP+R_s)}) (\widetilde{\Phi}^{(2hP+R_s)}_{HP} - R_s  \widetilde{\Psi}^{(2hP+R_s)}_{HP} ) 	\\
&\hspace{15mm}+ \sum_{h=0}^{\frac{H-1}{2}-1} (e^{-2\pi i \frac{t}{H}(2(h+1)P-R_s)} + e^{2\pi i \frac{t}{H}(2(h+1)P-R_s)})  (\widetilde{\Phi}^{(2(h+1)P-R_s)}_{HP} + R_s \widetilde{\Psi}^{(2(h+1)P-R_s)}_{HP} ) \Big)	\Bigg)
\Bigg]	\,	\Bigg|_{q \searrow e^{\frac{2\pi i}{K}}}	\,	.
\end{split}
\end{align}
When $H$ is even,
\begin{align}
\begin{split}
&Z_{SU(2)}(M_3) = \frac{B}{4i} q^{-\phi_4/4} \Big( \frac{2i}{K}\frac{P}{H} \Big)^{\frac{1}{2}} \frac{1}{P}  	\\
&\hspace{3mm}\times
\Bigg[
\sum_{s=0}^{7} (-1)^s \Bigg( \sum_{h=0}^{\frac{H}{2}-1} ( \widetilde{\Phi}^{(2hP+R_s)}_{HP} - R_s  \widetilde{\Psi}^{(2hP+R_s)}_{HP} ) + (\widetilde{\Phi}^{(2(h+1)P-R_s)}_{HP} + R_s \widetilde{\Psi}^{(2(h+1)P-R_s)}_{HP} )	\Bigg)	\\
&\hspace{5mm}+\sum_{s=0}^{7} (-1)^s \Bigg( \sum_{h=0}^{\frac{H}{2}-1}
\bigg( \sum_{t=1}^{\frac{H-2}{2}} e^{2\pi i K \frac{P}{H} t^2} 
\Big( (e^{-2\pi i \frac{t}{H}(2hP+R_s)} + e^{2\pi i \frac{t}{H}(2hP+R_s)}) (\widetilde{\Phi}^{(2hP+R_s)}_{HP} - R_s  \widetilde{\Psi}^{(2hP+R_s)}_{HP} ) 	\\
&\hspace{45mm}+ (e^{-2\pi i \frac{t}{H}(2(h+1)P-R_s)} + e^{2\pi i \frac{t}{H}(2(h+1)P-R_s)})  (\widetilde{\Phi}^{(2(h+1)P-R_s)}_{HP} + R_s \widetilde{\Psi}^{(2(h+1)P-R_s)}_{HP} ) \Big) 	\\
&\hspace{30mm} +e^{\frac{\pi i}{2} K P H} \Big( e^{\pi i (2hP+R_s)} (\widetilde{\Phi}^{(2hP+R_s)}_{HP} - R_s  \widetilde{\Psi}^{(2hP+R_s)}_{HP} ) 	\\
&\hspace{50mm}
+ e^{\pi i (2(h+1)P-R_s)}  (\widetilde{\Phi}^{(2(h+1)P-R_s)}_{HP} + R_s \widetilde{\Psi}^{(2(h+1)P-R_s)}_{HP} ) \Big) \bigg)	\Bigg)
\Bigg]	\,	\Bigg|_{q \searrow e^{\frac{2\pi i}{K}}}	\,	.
\end{split}
\end{align}
We note that the structure of the WRT invariant, \textit{i.e.} the coefficients to the $q$-series $\pm \widetilde{\Phi}^{(l)}_{HP} - R_s  \widetilde{\Psi}^{(l)}_{HP}$ are the same as in the case \eqref{3f-odd} or \eqref{3f-even} of $F=3$.
Therefore the WRT invariant for Seifert manifolds with four singular fibers are written similarly as in section \ref{sec:su2f3}.
We provide one example.

\subsubsection*{\textbullet \ \textnormal{$F=4$ and $H=3$}}
The structure is the same as in the case of $F=3$.
$H=3$ can be obtained, for example, from $(P_1,P_2,P_3,P_4)=(2,5,7,11)$ and $(Q_1,Q_2,Q_3,Q_4)=(-1,2,2,-2)$.
In this case, the homological blocks are given by
\begin{align}
\begin{split}
\widehat{Z}_0 = &	\,
-\widetilde{\Phi }_{2310}^{(51)}+51 \widetilde{\Psi }_{2310}^{(51)}
-\widetilde{\Phi }_{2310}^{(411)}+411 \widetilde{\Psi}_{2310}^{(411)}
-\widetilde{\Phi }_{2310}^{(579)}+579 \widetilde{\Psi }_{2310}^{(579)}		\\
&-\widetilde{\Phi}_{2310}^{(1041)}-499 \widetilde{\Psi }_{2310}^{(1041)}	
+\widetilde{\Phi }_{2310}^{(1269)}+271 \widetilde{\Psi}_{2310}^{(1269)}
+\widetilde{\Phi }_{2310}^{(1731)}-191 \widetilde{\Psi }_{2310}^{(1731)}	\\
&+\widetilde{\Phi}_{2310}^{(1899)}-359 \widetilde{\Psi }_{2310}^{(1899)}
+\widetilde{\Phi }_{2310}^{(2259)}-719 \widetilde{\Psi}_{2310}^{(2259)}	\,	,
\end{split}	\\
\begin{split}
\widehat{Z}_1 = & 	\,
\widetilde{\Phi }_{2310}^{(191)}-191\widetilde{\Psi }_{2310}^{(191)}
+\widetilde{\Phi }_{2310}^{(271)}-271 \widetilde{\Psi }_{2310}^{(271)}
+\widetilde{\Phi }_{2310}^{(359)}-359 \widetilde{\Psi }_{2310}^{(359)}
-\widetilde{\Phi}_{2310}^{(499)}+499 \widetilde{\Psi}_{2310}^{(499)}		\\
&+\widetilde{\Phi }_{2310}^{(719)}-719 \widetilde{\Psi }_{2310}^{(719)}
+\widetilde{\Phi }_{2310}^{(821)}+719 \widetilde{\Psi }_{2310}^{(821)}
-\widetilde{\Phi}_{2310}^{(961)}-579 \widetilde{\Psi}_{2310}^{(961)}
-\widetilde{\Phi }_{2310}^{(1129)}-411 \widetilde{\Psi }_{2310}^{(1129)}	\\
&+\widetilde{\Phi }_{2310}^{(1181)}+359 \widetilde{\Psi }_{2310}^{(1181)}
+\widetilde{\Phi}_{2310}^{(1349)}+191 \widetilde{\Psi}_{2310}^{(1349)}
-\widetilde{\Phi }_{2310}^{(1489)}-51 \widetilde{\Psi }_{2310}^{(1489)}
-\widetilde{\Phi }_{2310}^{(1591)}+51 \widetilde{\Psi }_{2310}^{(1591)}	\\
&+\widetilde{\Phi}_{2310}^{(1811)}-271 \widetilde{\Psi}_{2310}^{(1811)}
-\widetilde{\Phi }_{2310}^{(1951)}+411 \widetilde{\Psi }_{2310}^{(1951)}
-\widetilde{\Phi }_{2310}^{(2039)}+499 \widetilde{\Psi }_{2310}^{(2039)}
-\widetilde{\Phi }_{2310}^{(2119)}+579 \widetilde{\Psi}_{2310}^{(2119)}
\end{split}
\end{align}
where $\widehat{Z}_0 = q^{\frac{867}{3080}} \mathbb{Z}[[q]]$ and $\widehat{Z}_1 = q^{\frac{8761}{9240}} \mathbb{Z}[[q]]$.
In terms of homological blocks, the WRT invariant is written as
\begin{align}
Z_{SU(2)}(M_3) = \frac{B}{1540i} q^{-\phi_4/4} \Big( \frac{770i}{K} \Big)^{\frac{1}{2}} \sum_{a,b=0}^{1} e^{2 \pi i K CS_a} S_{ab} \widehat{Z}_b(q)  	\,	\Bigg|_{q \searrow e^{\frac{2\pi i}{K}}}
\end{align}
with $(CS_0, CS_1)=(0,\frac{1}{3})$ and $S_{ab} = \frac{1}{\sqrt{3}} \begin{pmatrix} 1 & 1 \\ 2 & -1 \end{pmatrix}$.
This $S$-matrix can also be obtained from \eqref{su2-smat0} and \eqref{su2-lk}.


\section{$G=SU(2)$, more singular fibers, and the base with higher genus}
\label{sec:su2fg}
We would like to consider Seifert manifolds with more singular fibers and with higher genus of the base surface.
The case of $H=1$ with arbitrary number of singular fibers and with genus zero was discussed in \cite{Hikami-lattice2}.
Here we consider general Seifert manifolds with more than 4 singular fibers, with a general $H$, and with higher genus case where $P_j$'s are pairwise coprime. 	\\

As before, we consider the integral expression for the WRT invariant of Seifert manifolds with $F$ singular fibers and with genus $g$ base surface according to \cite{Lawrence-Rozansky, Rozansky-residue},
\begin{align}
Z_{SU(2)}(M_3) = \frac{B}{2 \pi i} (-2K)^g q^{-\phi_F/4} \sum_{t=0}^{H-1} \int_{\Gamma_t} dy \ e^{-\frac{K}{2 \pi i} \frac{H}{P} y^2 - 2 K t y} \
\frac{\prod_{j=1}^{F} e^{\frac{y}{P_j}} - e^{-\frac{y}{P_j}} }{(e^{y} - e^{-y})^{F+2g-2}}
\label{int-formulaF}
\end{align}
where
\begin{align}
P = \prod_{j=1}^{F} P_j	\,	,	\hspace{5mm}		H=\sum_{j=1}^{F} \frac{Q_j}{P_j}	\,	,	\hspace{5mm}
\phi_F	&=	3 \, \text{sign}\left( \frac{H}{P} \right) + \sum_{j=1}^{F} \left( 12 s(Q_j, P_j) - \frac{Q_j}{P_j} \right)	
\end{align}
with the same integration cycle as in section \ref{sec:su2f3}.

We will see that the calculation for arbitrary genus $g$ can be done easily if the calculation for the genus zero case with arbitrary number of singular fibers is done.
So from now on, we proceed calculation with $g=0$.	\\

As in the previous sections, we expand the last factor in \eqref{int-formulaF}.
The numerator can be expressed in terms of the hyperbolic sine or cosine function when $F$ is odd or even, respectively,
\begin{align}
\frac{\prod_{j=1}^{F} e^{\frac{y}{P_j}} - e^{-\frac{y}{P_j}} }{(e^{y} - e^{-y})^{F-2}}
= 
\begin{cases}
\frac{1}{2^{F-1}} \frac{\sum_{ \epsilon_j = \pm 1}' \, \sinh \big( \sum_{j=1}^{F} \frac{\epsilon_j}{P_j} \big) y}{{(\sinh y)^{F-2}}}				&	\text{if $F$ is odd}	\\	
\frac{1}{2^{F-1}} \frac{\sum_{\epsilon_j = \pm 1} \frac{\epsilon}{2} \, \cosh \big( \sum_{j=1}^{F} \frac{\epsilon_j}{P_j} \big) y}{(\sinh y)^{F-2}}	&	\text{if $F$ is even}	
\end{cases}
\end{align}
where $\epsilon = \prod_{j=1}^{F} \epsilon_j$.
By $\sum'_{\epsilon_j=\pm1}$, we mean the summation over $\epsilon_j =\pm 1$ such that $\epsilon=+1$.
We find that \eqref{int-formulaF} can be expressed as the false theta functions $\widetilde{\Psi}_{HP}^{(a)}(q)$, $\widetilde{\Phi}_{HP}^{(a)}(q)$, and their derivatives.

For arbitrary number of singular fibers $F$, we see that types of functions appearing in the integrand are $\frac{\sinh \alpha y}{(\sinh y)^{2m+1}}$ or $\frac{\cosh \alpha y}{(\sinh y)^{2m}}$ where $\alpha = \sum_{j=1}^{F} \frac{\epsilon_j}{P_j}$.
After some calculations, we have
\begin{align}
\frac{\sinh \alpha y}{(\sinh y)^{2m+1}} &= \frac{1}{2m(2m-1)} \Big( \frac{d^2}{dy^2} - (2m-1-\alpha)^2 \Big) \frac{\sinh \alpha y}{(\sinh y)^{2m-1}} 
+ \frac{2\alpha}{2m} \frac{\cosh (1-\alpha) y}{(\sinh y)^{2m}}	\,	,	\label{sinh-der}	\\
\frac{\cosh \alpha y}{(\sinh y)^{2m}} &= \frac{1}{(2m-1)(2m-2)} \Big( \frac{d^2}{dy^2} - (2m-2-\alpha)^2 \Big) \frac{\cosh \alpha y}{(\sinh y)^{2m-2}} 	
- \frac{2\alpha}{2m-1} \frac{\sinh (1-\alpha) y}{(\sinh y)^{2m-1}}	\,	.	\label{cosh-der}	
\end{align}
For notational convenience, we denote
\begin{align}
S(\alpha, m) := \frac{\sinh \alpha y}{(\sinh y )^{2m+1}}	\,	,	\quad	C(\alpha, m) := \frac{\cosh \alpha y}{(\sinh y )^{2m}}
\end{align}
and
\begin{align}
s_1(\alpha, m) := \frac{1}{2m(2m-1)} \Big( \frac{d^2}{dy^2} - (2m-1-\alpha)^2 \Big)	\,	,	&\quad
s_2(\alpha, m) :=  \frac{2\alpha}{2m}	\,	, 	\\
c_1(\alpha, m) := \frac{1}{(2m-1)(2m-2)} \Big( \frac{d^2}{dy^2} - (2m-2-\alpha)^2 \Big)	\,	,	&\quad
c_2(\alpha, m) := - \frac{2\alpha}{2m-1}	\,	,
\end{align}
then \eqref{sinh-der} and \eqref{cosh-der} are written as
\begin{align}
S(\alpha, m) &= s_1(\alpha, m) S(\alpha, m-1) + s_2(\alpha, m) C(1-\alpha, m)	\,	,	\hspace{11.4mm}		m \geq 1	\,	,	\label{hsine-rel}	\\
C(\alpha, m) &= c_1(\alpha, m) C(\alpha, m-1) + c_2(\alpha, m) S(1-\alpha, m-1)	\,	,	\hspace{5mm}		m \geq 2	\,	.	\label{hcosine-rel}
\end{align}	
\vspace{0mm}

\subsubsection*{Odd $F$}

Firstly, we consider the case of odd $F$.
\eqref{hsine-rel} and \eqref{hcosine-rel} lead to a recursion relation for $S(\alpha, m)$,
\begin{align}
S(\alpha, m) &= \Big( s_1(\alpha,m) + s_2(\alpha, m) s_2(1-\alpha, m) + \frac{s_2(\alpha, m)}{s_2(\alpha, m-1)} c_1(1-\alpha, m) \Big) S(\alpha, m-1)		\nonumber\\
&\hspace{7mm}	- \frac{s_2(\alpha, m)}{s_2(\alpha, m-1)} c_1(1-\alpha, m) s_1(\alpha, m-1) S(\alpha, m-2)	\,	,	\hspace{5mm}	m \geq 2	\,	.	\label{sinh-rec}
\end{align}
Again, for simplicity, denoting the differential operators on $S(\alpha, m-1)$ and $S(\alpha, m-2)$ as 
\begin{align}
\begin{split}
X_{1}(\alpha, m-1) &:= s_1(\alpha,m) + s_2(\alpha, m) s_2(1-\alpha, m) + \frac{s_2(\alpha, m)}{s_2(\alpha, m-1)} c_1(1-\alpha, m)	\\
			&=	\frac{1}{2m(2m-1)} \bigg( 2 \frac{d^2}{dy^2} - 2(\alpha-1)^2 - 8(m-1)^2 \bigg) + \frac{\alpha (1-\alpha)}{m^2} 	\,	,
\end{split}		
		\\	
\begin{split}
X_{2}(\alpha, m-2) &:= - \frac{s_2(\alpha, m)}{s_2(\alpha, m-1)} c_1(1-\alpha, m) s_1(\alpha, m-1)		\\
			&=	-\frac{1}{2m (2m-1)(2m-2)(2m-3)} \bigg( \frac{d^4}{dy^4} - 2 \big( (2m-3)^2+\alpha^2 \big) \frac{d^2}{dy^2} + \big( (2m-3)^2 -\alpha^2 \big)^2 \bigg)	\,	,
\end{split}	
\end{align}
respectively, \eqref{sinh-rec} is written as
\begin{align}
S(\alpha, m) = X_1(\alpha, m-1) S(\alpha, m-1) + X_2(\alpha, m-2) S(\alpha, m-2)	\,	,	\quad	m \geq 2	\,	.
\end{align}
By expanding some of terms, we obtain
\begin{align}
\begin{split}
S(\alpha, m) &= \bigg( \sum_{\#s=0}^{\lfloor \frac{m-2}{2} \rfloor} \sum_{3 \leq j_1 < \ldots < j_s \leq m-1}
X_1(\alpha, m-1) \cdots X_{1,2}(\alpha, m-(j_1-2)) X_2(\alpha, m-j_1) X_1(\alpha, m-(j_1+1)) \\
&\hspace{30mm}	\cdots X_{1,2}(\alpha, m-(j_s-2)) X_2(\alpha, m-j_s) X_1(\alpha, m-(j_s+1)) \cdots X_{1,2}(\alpha, 1)	\\
&\hspace{-15mm}+ \sum_{\#s=1}^{\lceil \frac{m-2}{2} \rceil} \sum_{\substack{j_1=2, \\ 4 \leq j_2 < \ldots < j_s \leq m-1}} 
X_{2}(\alpha, m-2)  \cdots X_{1,2}(\alpha, m-(j_1-2)) X_2(\alpha, m-j_1) X_1(\alpha, m-(j_1+1)) \\ 	
&\hspace{30mm} 	\cdots X_{1,2}(\alpha, m-(j_s-2)) X_2(\alpha, m-j_s) X_1(\alpha, m-(j_s+1)) \cdots X_{1,2}(\alpha, 1)	\bigg)
S(\alpha, 1)	\\
&+\bigg(
\sum_{\#s=0}^{\lfloor \frac{m-3}{2} \rfloor} \sum_{3 \leq j_1 < \ldots < j_s \leq m-2}
X_1(\alpha, m-1) \cdots X_{1,2}(\alpha, m-(j_1-2)) X_2(\alpha, m-j_1) X_1(\alpha, m-(j_1+1)) \\
&\hspace{30mm}	\cdots X_{1,2}(\alpha, m-(j_s-2)) X_2(\alpha, m-j_s) X_1(\alpha, m-(j_s+1)) \cdots X_{1,2}(\alpha, 2)	\\
&\hspace{-15mm}+ \sum_{\#s=1}^{\lceil \frac{m-3}{2} \rceil} \sum_{\substack{j_1=2, \\ 4 \leq j_2 < \ldots < j_s \leq m-2}} 
X_{2}(\alpha, m-2)  \cdots X_{1,2}(\alpha, m-(j_1-2)) X_2(\alpha, m-j_1) X_1(\alpha, m-(j_1+1)) \\
&\hspace{18mm}	\cdots X_{1,2}(\alpha, m-(j_s-2)) X_2(\alpha, m-j_s) X_1(\alpha, m-(j_s+1)) \cdots X_{1,2}(\alpha, 2)	\bigg) X_2(\alpha, 0) S(\alpha,0)
\end{split}
\label{sinh-expansion}
\end{align}
for $m \geq 3$ where $\#s$ denotes the total number of terms $X_2(\alpha, j_s)$ appearing in the expression.
Some explanation is needed for \eqref{sinh-expansion}.
In \eqref{sinh-expansion}, $X_1(\alpha, m-1)$ and $X_2(\alpha, m-2)$ are always there if the summation containing them is valid and there is no term whose argument is greater than or equal to $m-1$ and $m-2$, respectively.
Also, the summation is such that $m-j$ in parentheses is strictly decreasing.
We mean terms $X_{1,2}(\alpha,j)$ by that they can be either $X_1$ or $X_2$ depending on the values of summation variables, and the last terms $X_{1,2}(\alpha,1)$ or $X_{1,2}(\alpha,2)$ are always there.
For example,
\begin{align}
S(\alpha, 2) 	&= 	X_1(\alpha, 1) S(\alpha, 1) + X_2(\alpha, 0) S(\alpha, 0)	\,	,	\\
S(\alpha, 3) 	&= 	\big(X_1(\alpha, 2) X_1(\alpha, 1) + X_2(\alpha, 1)\big) S(\alpha, 1) + X_1(\alpha, 2) X_2(\alpha, 0) S(\alpha, 0)		\,	,	\\
\begin{split}
S(\alpha, 4) 	&= 	\big(X_1(\alpha, 3) X_1(\alpha, 2) X_1(\alpha, 1) + X_1(\alpha, 3) X_2(\alpha, 1) + X_2(\alpha, 2) X_1(\alpha, 1)\big) S(\alpha, 1) 	\\
			&	\quad	+ \big(X_1(\alpha, 3) X_1(\alpha, 2) X_2(\alpha, 0) + X_2(\alpha, 2) X_2(\alpha, 0)\big) S(\alpha, 0)		\,	,
\end{split}			\\
\begin{split}
S(\alpha, 5)	&= 	\big(X_1(\alpha, 4) X_1(\alpha, 3) X_1(\alpha, 2) X_1(\alpha, 1) + X_1(\alpha, 4) X_1(\alpha, 3) X_2(\alpha, 1) + X_1(\alpha, 4) X_2(\alpha, 2) X_1(\alpha, 1) 	\\
			&	\quad	+ X_2(\alpha, 3) X_1(\alpha, 2) X_1(\alpha, 1) + X_2(\alpha, 3) X_2(\alpha, 1)\big) S(\alpha, 1) 
				+ \big(X_1(\alpha, 4) X_1(\alpha, 3) X_1(\alpha, 2) X_2(\alpha, 0)  	\\
			&	\quad	+ X_1(\alpha, 4) X_2(\alpha, 2) X_2(\alpha, 0) + X_2(\alpha, 3) X_1(\alpha, 2) X_2(\alpha, 0)\big) S(\alpha, 0)	\,	,
\end{split}			
\end{align}
\vspace{-7mm}
\begin{align}
\begin{split}
S(\alpha, 6)	&= 	\big(X_1(\alpha, 5) X_1(\alpha, 4) X_1(\alpha, 3) X_1(\alpha, 2) X_1(\alpha, 1) + X_1(\alpha, 5) X_1(\alpha, 4) X_1(\alpha, 3) X_2(\alpha, 1) \\
			&	\quad	+ X_1(\alpha, 5) X_1(\alpha, 4) X_2(\alpha, 2) X_1(\alpha, 1) + X_1(\alpha, 5) X_2(\alpha, 3) X_1(\alpha, 2) X_1(\alpha, 1)  	\\
			&	\quad	+ X_1(\alpha, 5) X_2(\alpha, 3) X_2(\alpha, 1)+ X_2(\alpha, 4) X_1(\alpha, 3) X_1(\alpha, 2) X_1(\alpha, 1) 	\\
			&	\quad	+ X_2(\alpha, 4) X_1(\alpha, 3) X_2(\alpha, 1) + X_2(\alpha, 4) X_2(\alpha, 2) X_1(\alpha, 1)\big) S(\alpha, 1)  	\\
			&	\quad	+ \big(X_1(\alpha, 5) X_1(\alpha, 4) X_1(\alpha, 3) X_1(\alpha, 2) X_2(\alpha, 0) + X_1(\alpha, 5) X_1(\alpha, 4) X_2(\alpha, 2) X_2(\alpha, 0)  	\\
			&	\quad	+ X_1(\alpha, 5) X_2(\alpha, 3) X_1(\alpha, 2) X_2(\alpha, 0) + X_2(\alpha, 4) X_1(\alpha, 3) X_1(\alpha, 2) X_2(\alpha, 0) 	\\
			&	\quad	+ X_2(\alpha, 4) X_2(\alpha, 2) X_2(\alpha, 0)\big) S(\alpha, 0)		\,	.
\end{split}								
\end{align}
Therefore, $S(\alpha, m)$ with $m\geq 3$ is determined from $S(\alpha,0)$ and $S(\alpha,1)$.
By using
\begin{align}
\frac{\sinh P\alpha y}{\sinh P y} = \sum_{m=0}^{M} (e^{(P\alpha - (2m+1)P)y} + e^{-(P\alpha - (2m+1)P)y}) - \sum_{n=0}^{\infty} \psi_{2P}^{(P\alpha - (2M+1)P)}(n) \, e^{-ny}
\label{sinh-gen}
\end{align}
where $M=-1,0,1, \cdots$ is such that $(2M+1)P < P \alpha < (2M+3) P$, and
\begin{align}
\frac{\cosh P \alpha y}{(\sinh P y)^2} = \sum_{m=0}^{M} 2m (e^{(P\alpha - 2mP)y} + e^{-(P\alpha - 2mP)y}) + \frac{1}{P} \sum_{n=0}^{\infty} n \varphi_{2P}^{(P\alpha - 2MP)}(n) e^{-ny} -\alpha \sum_{n=0}^{\infty} \psi_{2P}^{(P\alpha - 2MP)}(n) e^{-ny}
\label{cosh-gen}
\end{align}
where $M=0,1,\cdots$ is such that $2MP < P \alpha < (2M+2) P$, $S(\alpha,0)$ and $S(\alpha,1)$ are obtained
\begin{align}
\hspace{-5mm}S(\alpha, 0)	&= \frac{\sinh \alpha y}{\sinh y} = \sum_{m=0}^{M} (e^{(\alpha - (2m+1))y} + e^{-(\alpha - (2m+1))y}) - \sum_{n=0}^{\infty} \psi_{2P}^{(P\alpha - (2M+1)P)}(n) e^{-\frac{n}{P}y}	\,	,
\label{sinh-gen0}
\end{align}
\begin{align}
\begin{split}
\hspace{2mm}S(\alpha, 1) 	&= \frac{\sinh \alpha y}{(\sinh y)^3} = \frac{1}{2} \Big( \frac{d^2}{dy^2} - (1-\alpha)^2 \Big) \frac{\sinh \alpha y}{\sinh y} + \alpha \frac{\cosh (1-\alpha) y}{(\sinh y)^2}		\\
&= \sum_{m=0}^{M} 2m (m+1) (e^{(\alpha -(2m+1))y} + e^{-(\alpha -(2m+1))y})	\\
&\hspace{5mm} - \frac{1}{2} \sum_{n=0}^{\infty} \big( \frac{n^2}{P^2} + \alpha^2-1 \big) \psi_{2P}^{(P\alpha - (2M+1)P)}(n) e^{-\frac{n}{P}y}	
 + \frac{\alpha}{P} \sum_{n=0}^{\infty} n \varphi^{(P\alpha-(2M+1)P)}_{2P}(n) e^{-\frac{n}{P}y}
\end{split}
\label{sinh-gen1}
\end{align}
where $M=-1,0,1, \cdots$ is such that $(2M+1)P < P \alpha < (2M+3)P$.
Therefore when the number of singular fibers is odd, the partition function of the analytically continued $SU(2)$ theory is obtained from
\begin{align}
Z_{SU(2)}(M_3) = \frac{B}{2\pi i} q^{-\phi_F/4} \sum_{t=0}^{H-1} \int dy \ e^{-\frac{K}{2 \pi i} \frac{H}{P} y^2 - 2 K t y} \
\frac{1}{2^{F-1}} \sum_{\epsilon_j=\pm1} \hspace{-2mm}{}^{'}  S\Big(\sum_{j=1}^{F} \frac{\epsilon_j}{P_j}, \frac{F-3}{2}\Big)		\,	.	
\end{align}	
By doing similar calculations as done in section \ref{sec:su2f3}, we have similar results as previous cases and the difference comes from the coefficient of $e^{-ny/P}$ in the integrand.
For example, when $F=5$ and $H=1$, by integrating $S(\alpha, 1)$ from \eqref{sinh-gen1} we have
\begin{align}
\begin{split}
&4 \sum_{m=0}^{M} m(m+1) q^{\frac{1}{4P}(P\alpha - (2m+1)P)^2} 	\\
&\hspace{5mm}- \frac{1}{2} \bigg( \frac{4}{P} q\frac{\partial}{\partial q} + \frac{1}{P^2} (P\alpha)^2 -1 \bigg) \widetilde{\Psi}^{(P\alpha-(2M+1)P)}_{P}(q)
+ \frac{P\alpha}{P^2} \widetilde{\Phi}^{(P\alpha-(2M+1)P)}_{P}(q)	\,	.
\end{split}
\end{align}	
For general $H$, denoting $P\sum_{j=1}^{5} \frac{\epsilon_j}{P_j}$ with $\epsilon = +1$ by $R_s$,\footnote{When $R_s<0$, we use the formula, for example, by considering $\sinh R_s y = -\sinh (-R_s y)$.} $\psi^{(R_s-(2M+1)P)}_{2P}$ in \eqref{sinh-gen1} is decomposed to
\begin{align}
\psi^{(R_s-(2M+1)P)}_{2P} = \sum_{h=0}^{\big\lceil \frac{H}{2}-1 \big\rceil} \psi^{(2hP+R_s-(2M+1)P)}_{2HP}(n) - \sum_{h=0}^{\big\lfloor \frac{H}{2}-1 \big\rfloor} \psi^{(2(h+1)P-R_s+(2M+1)P)}_{2HP}
\end{align}
and also similarly for $\varphi^{(R_s-(2M+1)P)}_{2P}$.
Then the WRT invariant is given by
\begin{align}
\begin{split}
\hspace{-10mm}Z_{SU(2)}(M_3) &= \frac{B}{32i} q^{-\phi/4} \bigg( \frac{2i}{K} \frac{P}{H} \bigg)^{1/2} \sum_{t=0}^{H-1} e^{2\pi i K \frac{P}{H}t^2} 	\\
&\times \sum_{s=0}^{15} 
\Bigg[
2\sum_{m=0}^{M_{R_s}} m(m+1) (e^{-2 \pi i \frac{P}{H}(R_s-(2m+1)P) t} +  e^{2 \pi i \frac{P}{H}(R_s-(2m+1)P) t}) q^{\frac{1}{4HP}(R_s - (2m+1)P)^2}	\\
&\hspace{10mm}+
\Bigg( \sum_{h=0}^{\big\lceil \frac{H}{2}-1 \big\rceil}
e^{2\pi i \frac{t}{H} (2hP+R_s-(2M_s+1)P) } \bigg( - \frac{1}{2} \Big( \frac{4H}{P} q\frac{\partial}{\partial q} + \frac{R_s^2}{P^2}  -1 \Big) \widetilde{\Psi}_{HP}^{(2hP+R_s-(2M_s+1)P)}(q) \\
&\hspace{75mm}+ \frac{R_s}{P^2} \widetilde{\Phi}_{HP}^{(2hP+R_s-(2M_s+1)P)}(q) \bigg) 	\\
&\hspace{20mm}+\sum_{h=0}^{\big\lfloor \frac{H}{2}-1 \big\rfloor} e^{2\pi i \frac{t}{H} (2(h+1)P-R_s+(2M_s+1)P) } \bigg( \frac{1}{2} \Big( \frac{4H}{P} q \frac{\partial}{\partial q} + \frac{R_s^2}{P^2} -1 \Big) \widetilde{\Psi}_{HP}^{(2(h+1)P-R_s+(2M_s+1)P)}(q) 	\\
&\hspace{85mm}+ \frac{R_s}{P^2} \widetilde{\Phi}_{HP}^{((2(h+1)P-R_s+(2M_s+1)P))}(q) 
\bigg)
\Bigg]	\,	\Bigg|_{q \searrow e^{\frac{2\pi i}{K}}}
\end{split}	\label{su2f5}
\end{align}
where $M_{R_s}=-1,0,1, \ldots$ is such that $(2M_{R_s}+1)P < R_s < (2M_{R_s}+3)P$.
But when $F=5$, the maximum value that $R_s$ can have is greater than $P$ but smaller than $2P$.
So $M_{R_s}$ can be $-1$ or $0$, and terms in the second line of \eqref{su2f5} vanishes.
We provide an example for $H=3$ in section \ref{ssec:su2fg-ex}.
\vspace{5mm}

\subsubsection*{Even $F$}

We perform similar calculations for even $F$.
From \eqref{hsine-rel} and \eqref{hcosine-rel}, the recursion relation for $C(\alpha, m)$ is given by
\begin{align}
\begin{split}
C(\alpha, m) &= \bigg( c_1(\alpha, m)j + c_2(\alpha, m) s_2(1-\alpha, m-1) + \frac{c_2(\alpha, m)}{c_2(\alpha, m-1)} s_1(1-\alpha, m-1) \bigg) C(\alpha, m-1)	\\
&\hspace{7mm}	-\frac{c_2(\alpha, m)}{c_2(\alpha, m-1)} s_1(1-\alpha, m-1) c_1(\alpha, m-1) C(\alpha, m-2)	\,	,	\quad	m \geq 3	\,	.	\label{recursionC0}
\end{split}
\end{align}
Also, for simplicity, we denote the differential operators on $C(\alpha, m-1)$ and $C(\alpha, m-2)$ by
\begin{align}
\begin{split}
V_1(\alpha, m-1) 		&:= 	c_1(\alpha, m) + c_2(\alpha, m) s_2(1-\alpha,m-1) + \frac{c_2(\alpha,m)}{c_2(\alpha,m-1)} s_1(1-\alpha,m-1)	\\	
					&= \frac{1}{(2m-1)(2m-2)} \bigg( 2 \frac{d^2}{dy^2} - 4(m-1)(m-2) - 2(\alpha-1)^2 \bigg) + \frac{4\alpha(\alpha-1)}{(2m-1)(2m-2)}	\,	,	
\end{split}		
\end{align}
\begin{align}
\begin{split}
V_2(\alpha, m-2)		&:=	-\frac{c_2(\alpha, m)}{c_2(\alpha,m-1)} s_1(1-\alpha,m-1) c_1(\alpha, m-1)	\\
					&= -\frac{1}{(2m-1)(2m-2)(2m-3)(2m-4)} \bigg( \frac{d^4}{dy^4} - 2((2m-4)^2+\alpha^2) \frac{d^2}{dy^2} + ((2m-4)^2-\alpha^2)^2 \bigg)	\,	.
\end{split}					
\end{align}
Then \eqref{recursionC0} is written as
\begin{align}
C(\alpha, m) = V_1(\alpha, m-1) C(\alpha, m-1) + V_2(\alpha, m-2) C(\alpha, m-2)	\,	,	\quad	m \geq 3	\,	.
\end{align}
So we have
\begin{align}
\begin{split}
C(\alpha, m) &= \bigg( \sum_{\#s=0}^{\lfloor \frac{m-3}{2} \rfloor} \sum_{3 \leq j_1 < \ldots < j_s \leq m-2}
V_1(\alpha, m-1) \cdots V_{1,2}(\alpha, m-(j_1-2)) V_2(\alpha, m-j_1) V_1(\alpha, m-(j_1+1)) \\
&\hspace{30mm}	\cdots V_{1,2}(\alpha, m-(j_s-2)) V_2(\alpha, m-j_s) V_1(\alpha, m-(j_s+1)) \cdots V_{1,2}(\alpha, 2)	\\
&\hspace{-10mm}+ \sum_{\#s=1}^{\lceil \frac{m-3}{2} \rceil} \sum_{\substack{j_1=2 \\ 4\leq j_2 < \ldots < j_s \leq m-2}} 
V_{2}(\alpha, m-2)  \cdots V_{1,2}(\alpha, m-(j_1-2)) V_2(\alpha, m-j_1) V_1(\alpha, m-(j_1+1)) \\
&\hspace{30mm}	\cdots V_{1,2}(\alpha, m-(j_s-2)) V_2(\alpha, m-j_s) V_1(m-(j_s+1)) \cdots V_{1,2}(\alpha, 2)	\bigg) C(\alpha, 2)	\\
&+\bigg(
\sum_{\#s=0}^{\lfloor \frac{m-4}{2} \rfloor} \sum_{3 \leq j_1 < \ldots < j_s \leq m-3}
V_1(\alpha, m-1) \cdots V_{1,2}(m-(j_1-2)) V_2(\alpha, m-j_1) V_1(\alpha, m-(j_1+1)) \\
&\hspace{30mm}	\cdots V_{1,2}(\alpha, m-(j_s-2)) V_2(\alpha, m-j_s) V_1(\alpha, m-(j_s+1)) \cdots V_{1,2}(\alpha, 3)	\\
&\hspace{-10mm}+ \sum_{\#s=1}^{\lceil \frac{m-4}{2} \rceil} \sum_{\substack{j_1=2 \\  4\leq j_2 < \ldots < j_s \leq m-3}} 
V_{2}(m-2) V_1(\alpha, m-3) \cdots V_{1,2}(\alpha, m-(j_1-2)) V_2(\alpha, m-j_1) V_1(\alpha, m-(j_1+1)) \\
&\hspace{20mm}	\cdots V_{1,2}(\alpha, m-(j_s-2)) V_2(\alpha, m-j_s) V_1(\alpha, m-(j_s+1)) \cdots V_{1,2}(\alpha, 3)	\bigg) V_2(\alpha, 1) C(\alpha,1)
\end{split}
\label{cosh-expansion}
\end{align}
for $m \geq 4$. 
This is similar to the case of odd $F$ where explanations on the summation and notations are available.
We provide some examples,
\begin{align}
C(\alpha, 3) 	&= 	V_1(\alpha, 2) C(\alpha, 2) + V_2(\alpha, 1) C(\alpha, 1)	\,	,	\\
C(\alpha, 4) 	&= 	\big(V_1(\alpha, 3) V_1(\alpha, 2) + V_2(\alpha, 2)\big) C(\alpha, 2) + V_1(\alpha, 3) V_2(\alpha, 1) C(\alpha, 1)	\,	,	\\
\begin{split}
C(\alpha, 5) 	&= 	\big(V_1(\alpha, 4) V_1(\alpha, 3) V_1(\alpha, 2) + V_1(\alpha, 4) V_2(\alpha, 2) + V_2(\alpha, 3) V_1(\alpha, 2)\big) C(\alpha, 2) 	\\
			&	\quad	+ \big(V_1(\alpha, 4) V_1(\alpha, 3) V_2(\alpha, 1) + V_2(\alpha, 3) V_2(\alpha, 1)\big) C(\alpha, 1)		\,	,
\end{split}			\\
\begin{split}
C(\alpha, 6)	&= 	\big(V_1(\alpha, 5) V_1(\alpha, 4) V_1(\alpha, 3) V_1(\alpha, 2) + V_1(\alpha, 5) V_1(\alpha, 4) V_2(\alpha, 2) + V_1(\alpha, 5) V_2(\alpha, 3) V_1(\alpha, 2) 	\\
			&	\quad	+ V_2(\alpha, 4) V_1(\alpha, 3) V_1(\alpha, 2) + V_2(\alpha, 4) V_2(\alpha, 2)\big) C(\alpha, 2) 
				+ \big(V_1(\alpha, 5) V_1(\alpha, 4) V_1(\alpha, 3) V_2(\alpha, 1)  	\\
			&	\quad	+ V_1(\alpha, 5) V_2(\alpha, 3) V_2(\alpha, 1) + V_2(\alpha, 4) V_1(\alpha, 3) V_2(\alpha, 1)\big) C(\alpha, 1)		\,	,
\end{split}		
\end{align}
\begin{align}
\begin{split}
\hspace{-6mm}C(\alpha, 7)	&= 	\big(V_1(\alpha, 6) V_1(\alpha, 5) V_1(\alpha, 4) V_1(\alpha, 3) V_1(\alpha, 2) + V_1(\alpha, 6) V_1(\alpha, 5) V_1(\alpha, 4) V_2(\alpha, 2)	\\
			&	\quad 	+ V_1(\alpha, 6) V_1(\alpha, 5) V_2(\alpha, 3) V_1(\alpha, 2) + V_1(\alpha, 6) V_2(\alpha, 4) V_1(\alpha, 3) V_1(\alpha, 2)	\\
			&	\quad	+ V_1(\alpha, 6) V_2(\alpha, 4) V_2(\alpha, 2) + V_2(\alpha, 5) V_1(\alpha, 4) V_1(\alpha, 3) V_1(\alpha, 2) 	\\
			&	\quad	+ V_2(\alpha, 5) V_1(\alpha, 4) V_2(\alpha, 2) + V_2(\alpha, 5) V_2(\alpha, 3) V_1(\alpha, 2)\big) C(\alpha, 2) 	\\
			&	\quad	+ \big(V_1(\alpha, 6) V_1(\alpha, 5) V_1(\alpha, 4) V_1(\alpha, 3) V_2(\alpha, 1) + V_1(\alpha, 6) V_1(\alpha, 5) V_2(\alpha, 3) V_2(\alpha, 1)	\\
			&	\quad	+ V_1(\alpha, 6) V_2(\alpha, 4) V_1(\alpha, 3) V_2(\alpha, 1) + V_2(\alpha, 5) V_1(\alpha, 4) V_1(\alpha, 3) V_2(\alpha, 1) 	\\
			&	\quad	+ V_2(\alpha, 5) V_2(\alpha, 3) V_2(\alpha, 1)\big) C(\alpha, 1)		\,	.
\end{split}								
\end{align}
Also, $C(\alpha,m)$ with $m \geq 4$ can be calculated from $C(\alpha,1)$ and $C(\alpha,2)$.
By using \eqref{sinh-gen} and \eqref{cosh-gen}, they are given by
\begin{align}
\begin{split}
C(\alpha, 1)	
&= \frac{\cosh \alpha y}{(\sinh y)^2} 	\\
&= \sum_{m=0}^{M} 2m (e^{(\alpha - 2m)y} + e^{-(\alpha - 2m)y}) + \frac{1}{P} \sum_{n=0}^{\infty} n \varphi_{2P}^{(P\alpha-2MP)}(n) e^{-\frac{n}{P}y} -\alpha \sum_{n=0}^{\infty} \psi_{2P}^{(P\alpha-2MP)}(n) e^{-\frac{n}{P}y}	\,	,
\end{split}
\end{align}
\begin{align}
\begin{split}
C(\alpha, 2) 	&= \frac{\cosh \alpha y}{(\sinh y)^{4}} 
= \frac{1}{6} \Big( \frac{d^2}{dy^2} - (2+\alpha)^2 \Big) \frac{\cosh y\alpha}{(\sinh y)^{2}} 	
+ \frac{2\alpha}{3} \frac{\sinh y(1+\alpha)}{(\sinh y)^{3}}	\\
&= \frac{4}{3} \sum_{m=0}^{M} m (m-1) (m+1) (e^{(\alpha - 2m)y} + e^{-(\alpha - 2m)y}) 	\\
& -\alpha \sum_{n=0}^{\infty} \Big( \frac{1}{2} \frac{n^2}{P^2} +\frac{1}{6} (\alpha^2 - 4) \Big) \psi_{2P}^{(P\alpha-2MP)}(n) e^{-\frac{n}{P}y}
+\frac{1}{6P}  \sum_{n=0}^{\infty} \Big( (3\alpha^2 - 4) + \frac{n^2}{P^2} \Big) n \varphi_{2P}^{(P\alpha-2MP)}(n) e^{-\frac{n}{P}y}
\end{split}
\label{cosh-gen2}
\end{align}
where $M=0,1, \cdots$ is such that $2MP < P\alpha < 2(M+1)P$.
Thus, when the number of singular fibers is even, the partition function of the analytically continued $SU(2)$ theory is obtained from
\begin{align}
Z_{SU(2)}(M_3) = \frac{B}{2 \pi i} q^{-\phi/4} \sum_{t=0}^{H-1} \int dy \ e^{-\frac{K}{2 \pi i} \frac{H}{P} y^2 - 2 K t y} \
\frac{1}{2^{F-1}} \sum_{\epsilon_j=\pm1} \frac{\epsilon}{2} \, C\Big(\sum_{j=1}^{F} \frac{\epsilon_j}{P_j}, \frac{F-2}{2}\Big)	\,	. 
\label{su2-a-f6}
\end{align}
Also, by doing similar calculations as before, we have similar results as previous cases but with a different coefficient of $e^{-n y/P}$.
For example, when $F=6$ and $H=1$, by integrating $C(\alpha, 2)$ from \eqref{cosh-gen2}, we obtain
\begin{align}
\begin{split}
&\frac{8}{3} \sum_{m=0}^{M} m(m-1)(m+1) q^{\frac{1}{4P} (P\alpha-2mP)^2} 	\\
&\hspace{5mm}-\frac{\alpha}{2} \bigg( \frac{4}{P} q\frac{\partial}{\partial q} + \frac{1}{3}(\alpha^2-4)\bigg) \widetilde{\Psi}_{P}^{(P\alpha-2MP)}(q)	
+\frac{1}{6P} \bigg( \frac{4}{P} q\frac{\partial}{\partial q} + (3\alpha^2-4)\bigg) \widetilde{\Phi}_{P}^{(P\alpha-2MP)}(q)	\,	.
\end{split}
\end{align}
For general $H$, decomposing $\psi_{2P}^{(P\alpha-2MP)}(n)$ and $\varphi_{2P}^{(P\alpha-2MP)}(n)$ as before, from \eqref{su2-a-f6} we obtain
\begin{align}
\begin{split}
&\hspace{-10mm}Z_{SU(2)}(M_3) = \frac{B}{2i} q^{-\phi/4} \bigg( \frac{2i}{K} \frac{P}{H} \bigg)^{1/2} \sum_{t=0}^{H-1} e^{2\pi i K \frac{P}{H}t^2} 	\\
&\times \sum_{s=0}^{31} (-1)^s
\Bigg[
\frac{4}{3} \sum_{m=0}^{M_s} m(m-1)(m+1) (e^{-2 \pi i \frac{P}{H}(R_s-2mP) t} + e^{2 \pi i \frac{P}{H}(R_s-2mP) t}) q^{\frac{1}{4HP} (R_s-2mP)^2} \\
&\hspace{10mm}+
\Bigg( \sum_{h=0}^{\big\lceil \frac{H}{2}-1 \big\rceil} 
e^{2\pi i \frac{t}{H} (2hP+R_s- 2M_s P) } 
\bigg( 
-\frac{R_s}{2P} \bigg( \frac{4H}{P} q\frac{\partial}{\partial q} + \frac{1}{3}(R_s/P)^2-4\bigg) \widetilde\Psi_{HP}^{(2hP+R_s-2M_sP)}(q)	\\
&\hspace{65mm}+\frac{1}{6P} \bigg( \frac{4H}{P} q\frac{\partial}{\partial q} + 3(R_s/P)^2-4\bigg) \widetilde{\Phi}_{HP}^{(2hP+R_s-2M_sP)}(q) 
\bigg) 	\\
&\hspace{15mm}+\sum_{h=0}^{\big\lfloor \frac{H}{2}-1 \big\rfloor} e^{2\pi i \frac{t}{H} (2(h+1)P-R_s+2M_sP) } 
\bigg( 
\frac{R_s}{2P} \bigg( \frac{4H}{P} q\frac{\partial}{\partial q} + \frac{1}{3}(R_s/P)^2-4\bigg) \widetilde\Psi_{HP}^{(2(h+1)P-R_s+ 2M_s P)}(q)	\\
&\hspace{65mm}+\frac{1}{6P} \bigg( \frac{4H}{P} q\frac{\partial}{\partial q} + 3(R_s/P)^2-4\bigg) \widetilde{\Phi}_{HP}^{(2(h+1)P-R_s+ 2M_s P)}(q) 
\bigg)
\Bigg]	\,	\Bigg|_{q \searrow e^{\frac{2\pi i}{K}}}
\end{split}	\label{su2f6}
\end{align}
where $R_s$ with even (odd) $s$ are $P\sum_{j=1}^{6} \frac{\epsilon_j}{P_j}$ with $\epsilon=+1 \, (-1)$ up to an overall sign.
Also, $M_{R_s}=0,1, \cdots$ is such that $2M_{R_s} P < R_s < 2(M_{R_s}+1)P$.
When $F=6$, the maximum value that $R_s$ can take is less than $2P$, so $M_{R_s}$ is zero. 
Therefore, the second line of \eqref{su2f6} vanishes.
We provide an example for $H=3$ in section \ref{ssec:su2fg-ex}.

\subsubsection*{Higher genus}

For the case of higher genus, the power of the denominator increases by $2g$.
Thus, we can use the formula for $\frac{\sinh \alpha y}{(\sinh y)^{2m+1}}$ or $\frac{\cosh \alpha y}{(\sinh y)^{2m}}$ given an $F$ and $g$ to calculate the homological blocks.
Then we obtain the partition function of the analytically continued theory or the WRT invariant with the same structure as before but with different homological blocks.
For example, when $F=3$ and $g=1$, the WRT invariant is given by
{\small
\begin{align}
\begin{split}
&\hspace{-10mm}Z_{SU(2)}(M_3) = \frac{B}{32i} (-2K ) q^{-\phi_3/4} \bigg( \frac{2i}{K} \frac{P}{H} \bigg)^{1/2} \sum_{t=0}^{H-1} e^{2\pi i K \frac{P}{H}t^2}	\\
&\hspace{10mm}\times \sum_{s=0}^{3} 
\Bigg[
\Bigg( \sum_{h=0}^{\frac{H-1}{2}} 
e^{2\pi i \frac{t}{H} (2hP+R_s-(2M_s+1)P) } \bigg( - \frac{1}{2} \Big( \frac{4H}{P} q\frac{\partial}{\partial q} + \frac{R_s^2}{P^2}  -1 \Big) \widetilde{\Psi}_{HP}^{(2hP+R_s-(2M_s+1)P)}(q) \\
&\hspace{75mm}+ \frac{R_s}{P^2} \widetilde{\Phi}_{HP}^{((2hP+R_s-(2M_s+1)P))}(q) \bigg) 	\\
&\hspace{20mm}+\sum_{h=0}^{\frac{H-1}{2}-1}e^{2\pi i \frac{t}{H} (2(h+1)P-R_s+(2M_s+1)P) } \bigg( \frac{1}{2} \Big( \frac{4H}{P} q \frac{\partial}{\partial q} + \frac{R_s^2}{P^2} -1 \Big) \widetilde{\Psi}_{HP}^{(2(h+1)P-R_s+(2M_s+1)P)}(q) 	\\
&\hspace{80mm}+ \frac{R_s}{P^2} \widetilde{\Phi}_{HP}^{((2(h+1)P-R_s+(2M_s+1)P))}(q) 
\bigg)
\Bigg]	\,	\Bigg|_{q \searrow e^{\frac{2\pi i}{K}}}
\end{split}
\end{align}
}where $R_s$'s are given by $P\sum_{j=1}^{3}\frac{\epsilon_j}{P_j}$ with $\epsilon=1$ and $M_s =-1,0$ is such that $(2M_s+1)P<R_s<(2M_s+3)P$.
Since $M_s =-1,0$, there is no additional term in the expression as in the case of \eqref{su2f5}.
The explicit expression of homological blocks for $H=3$ with a choice $(P_1,P_2,P_3)=(2,5,7)$ is available in section \ref{ssec:su2fg-ex}.\\

We see that the structure of the WRT invariant for the Seifert manifolds that we consider stays the same for arbitrary number of singular fibers and arbitrary genus.
Thus, the WRT invariant for Seifert manifolds with arbitrary number of singular fibers and arbitrary genus are written similarly as in section \ref{sec:su2f3}.

\subsection{Some Examples}
\label{ssec:su2fg-ex}

\subsubsection*{\textbullet \ $F=5$ and $H=3$}
When the number of singular fibers is 5, $H=3$ can be obtained, for example, by choosing $(P_1,P_2,P_3,P_4,P_5)=(2,5,7,11,13)$ and $(Q_1,Q_2,Q_3,Q_4,Q_5)=(-1,4,-2,-1,1)$.
The WRT invariant can be written as
\begin{align}
Z_{SU(2)}(M_3) = \frac{B}{i} q^{-\phi_5/4} \bigg( \frac{10010i}{K} \bigg)^{1/2}  \frac{1}{2^4 \cdot 2 \cdot 70^2}  \sum_{a,b=0}^{1} e^{2 \pi i K CS_a} S_{ab} \widehat{Z}_b(q)	\,	\Big|_{q \searrow e^{\frac{2\pi i}{K}}}
\end{align}
where $(CS_0, CS_1)=(0,\frac{2}{3})$ and $S_{ab}=\frac{1}{\sqrt{3}}\begin{pmatrix} 1 & 1 \\ 2 &-1 \end{pmatrix}$.
The homological blocks $\widehat{Z}_b(q)$ are given by
{\scriptsize
\begin{align}
\begin{split}
&\widehat{Z}_0=
-16594 \widetilde{\Phi }_{30030}^{(1713)}
-12226 \widetilde{\Phi }_{30030}^{(3897)}
+11434 \widetilde{\Phi }_{30030}^{(4293)}
-5506\widetilde{\Phi }_{30030}^{(7257)}
-2866 \widetilde{\Phi }_{30030}^{(8577)}
-214 \widetilde{\Phi }_{30030}^{(9903)}
+6506 \widetilde{\Phi }_{30030}^{(13263)}
+9146 \widetilde{\Phi }_{30030}^{(14583)}	\\	&
+10874 \widetilde{\Phi }_{30030}^{(15447)}
+13514 \widetilde{\Phi }_{30030}^{(16767)}
+20234 \widetilde{\Phi }_{30030}^{(20127)}
-17154 \widetilde{\Phi }_{30030}^{(21453)}
-14514 \widetilde{\Phi }_{30030}^{(22773)}
-7794 \widetilde{\Phi }_{30030}^{(26133)}
-3426 \widetilde{\Phi }_{30030}^{(28317)}
+8586 \widetilde{\Phi }_{30030}^{(34323)}	\\	&
+(31359891-120120 q\partial/\partial q) \widetilde{\Psi }_{30030}^{(1713)}
+(62831331-120120 q\partial/\partial q) \widetilde{\Psi}_{30030}^{(3897)}
+3 (-22505337+40040 q\partial/\partial q) \widetilde{\Psi }_{30030}^{(4293)}	\\	&
+(92621091-120120 q\partial/\partial q) \widetilde{\Psi}_{30030}^{(7257)}
+(98146611-120120 q\partial/\partial q) \widetilde{\Psi }_{30030}^{(8577)}
-3 (-33396217+40040 q\partial/\partial q) \widetilde{\Psi}_{30030}^{(9903)}	\\	&
+(89618091-120120 q\partial/\partial q) \widetilde{\Psi }_{30030}^{(13263)}
+(79287771-120120 q\partial/\partial q) \widetilde{\Psi}_{30030}^{(14583)}
+(70639131-120120 q\partial/\partial q) \widetilde{\Psi }_{30030}^{(15447)}	\\	&
+(54543051-120120 q\partial/\partial q) \widetilde{\Psi}_{30030}^{(16767)}
-3 (717863+40040 q\partial/\partial q) \widetilde{\Psi }_{30030}^{(20127)}
+(26635171-120120 q\partial/\partial q) \widetilde{\Psi}_{30030}^{(21453)}	\\	&
+(47536051-120120 q\partial/\partial q) \widetilde{\Psi }_{30030}^{(22773)}
+(85013491-120120 q\partial/\partial q) \widetilde{\Psi}_{30030}^{(26133)}
+(97265731-120120 q\partial/\partial q) \widetilde{\Psi }_{30030}^{(28317)}	\\	&
+(81770251-120120 q\partial/\partial q) \widetilde{\Psi}_{30030}^{(34323)}		\,	,
\end{split}
\end{align}
\begin{align}
\begin{split}
&\widehat{Z}_1 =
20234 \widetilde{\Phi }_{30030}^{(107)}
-17154 \widetilde{\Phi }_{30030}^{(1433)}
-14514 \widetilde{\Phi }_{30030}^{(2753)}
+13514 \widetilde{\Phi }_{30030}^{(3253)}
+10874 \widetilde{\Phi }_{30030}^{(4573)}
+9146 \widetilde{\Phi }_{30030}^{(5437)}
+8586 \widetilde{\Phi }_{30030}^{(5717)}	\\	&
-7794 \widetilde{\Phi }_{30030}^{(6113)}
+6506 \widetilde{\Phi }_{30030}^{(6757)}
-3426 \widetilde{\Phi }_{30030}^{(8297)}
-214 \widetilde{\Phi }_{30030}^{(10117)}
-2866 \widetilde{\Phi }_{30030}^{(11443)}
-3426 \widetilde{\Phi }_{30030}^{(11723)}
-5506 \widetilde{\Phi }_{30030}^{(12763)}	\\	&
-7794 \widetilde{\Phi }_{30030}^{(13907)}
+8586 \widetilde{\Phi }_{30030}^{(14303)}
+11434 \widetilde{\Phi }_{30030}^{(15727)}
-12226 \widetilde{\Phi }_{30030}^{(16123)}
-14514 \widetilde{\Phi }_{30030}^{(17267)}
-16594 \widetilde{\Phi }_{30030}^{(18307)}
-17154 \widetilde{\Phi }_{30030}^{(18587)}	\\	&
+20234 \widetilde{\Phi }_{30030}^{(19913)}
-16594 \widetilde{\Phi }_{30030}^{(21733)}
-12226 \widetilde{\Phi }_{30030}^{(23917)}
-5506 \widetilde{\Phi }_{30030}^{(27277)}
-2866 \widetilde{\Phi }_{30030}^{(28597)}
-214 \widetilde{\Phi }_{30030}^{(29923)}
+6506 \widetilde{\Phi }_{30030}^{(33283)}	\\	&
+9146 \widetilde{\Phi }_{30030}^{(34603)}
+10874 \widetilde{\Phi }_{30030}^{(35467)}
+11434 \widetilde{\Phi }_{30030}^{(35747)}
+13514 \widetilde{\Phi }_{30030}^{(36787)}	\\	&
-3 (717863+40040 q\partial/\partial q) \widetilde{\Psi}_{30030}^{(107)}
+(26635171-120120 q\partial/\partial q) \widetilde{\Psi }_{30030}^{(1433)}
+(47536051-120120 q\partial/\partial q) \widetilde{\Psi}_{30030}^{(2753)}	\\	&
+3 (-18181017+40040 q\partial/\partial q) \widetilde{\Psi }_{30030}^{(3253)}
+3 (-23546377+40040 q\partial/\partial q) \widetilde{\Psi}_{30030}^{(4573)}
+3 (-26429257+40040 q\partial/\partial q) \widetilde{\Psi }_{30030}^{(5437)}	\\	&
+(-81770251+120120 q\partial/\partial q) \widetilde{\Psi}_{30030}^{(5717)}
+(85013491-120120 q\partial/\partial q) \widetilde{\Psi }_{30030}^{(6113)}
+3 (-29872697+40040 q\partial/\partial q) \widetilde{\Psi}_{30030}^{(6757)}	\\	&
+(97265731-120120 q\partial/\partial q) \widetilde{\Psi }_{30030}^{(8297)}
+3 (-33396217+40040 q\partial/\partial q) \widetilde{\Psi}_{30030}^{(10117)}
+3 (-32715537+40040 q\partial/\partial q) \widetilde{\Psi }_{30030}^{(11443)}	\\	&
+(-97265731+120120 q\partial/\partial q) \widetilde{\Psi}_{30030}^{(11723)}
+3 (-30873697+40040 q\partial/\partial q) \widetilde{\Psi }_{30030}^{(12763)}
+(-85013491+120120 q\partial/\partial q) \widetilde{\Psi}_{30030}^{(13907)}	\\	&
+(81770251-120120 q\partial/\partial q) \widetilde{\Psi }_{30030}^{(14303)}
+(67516011-120120 q\partial/\partial q) \widetilde{\Psi}_{30030}^{(15727)}
+3 (-20943777+40040 q\partial/\partial q) \widetilde{\Psi }_{30030}^{(16123)}	\\	&
+(-47536051+120120 q\partial/\partial q) \widetilde{\Psi}_{30030}^{(17267)}
+3 (-10453297+40040 q\partial/\partial q) \widetilde{\Psi }_{30030}^{(18307)}
+(-26635171+120120 q\partial/\partial q) \widetilde{\Psi}_{30030}^{(18587)}	\\	&
+3 (717863+40040 q\partial/\partial q) \widetilde{\Psi }_{30030}^{(19913)}
+(31359891-120120 q\partial/\partial q) \widetilde{\Psi}_{30030}^{(21733)}
+(62831331-120120 q\partial/\partial q) \widetilde{\Psi }_{30030}^{(23917)}	\\	&
+(92621091-120120 q\partial/\partial q) \widetilde{\Psi}_{30030}^{(27277)}\
+(98146611-120120 q\partial/\partial q) \widetilde{\Psi }_{30030}^{(28597)}
-3 (-33396217+40040 q\partial/\partial q) \widetilde{\Psi}_{30030}^{(29923)}	\\	&
+(89618091-120120 q\partial/\partial q) \widetilde{\Psi }_{30030}^{(33283)}
+(79287771-120120 q\partial/\partial q) \widetilde{\Psi}_{30030}^{(34603)}
+(70639131-120120 q\partial/\partial q) \widetilde{\Psi }_{30030}^{(35467)}	\\	&
+(67516011-120120 q\partial/\partial q) \widetilde{\Psi}_{30030}^{(35747)}
+(54543051-120120 q\partial/\partial q) \widetilde{\Psi }_{30030}^{(36787)}
\end{split}
\end{align}
}
where $\widehat{Z}_0 = q^{\frac{355507}{680680}} \mathbb{Z}[[q]]$ and $\widehat{Z}_1 = q^{\frac{1747201}{2042040}} \widehat{Z}[[q]]$.

\subsubsection*{\textbullet \ $F=6$ and $H=3$}
When $F=6$, we can have $H=3$, for example, by choosing $(P_1,P_2,P_3,P_4,P_5,P_6)=(2,5,7,11,13,17)$ and $(Q_1,Q_2,Q_3,Q_4,Q_5)=(-1,2,4,-2,-3,-1)$.
The WRT invariant can be written as
\begin{align}
Z_{SU(2)}(M_3) = \frac{B}{i} q^{-\phi_6/4} \bigg( \frac{170170i}{K} \bigg)^{1/2}  \frac{1}{2^5 \cdot 6 \cdot 170170^3}  \sum_{a,b=0}^{1} e^{2 \pi i K CS_a} S_{ab} \widehat{Z}_b(q)	\,	\Big|_{q \searrow e^{\frac{2\pi i}{K}}}
\end{align}
where $(CS_0,CS_1)=(0,\frac{1}{3})$ and $S_{ab}=\frac{1}{\sqrt{3}}\begin{pmatrix} 1 & 1 \\ 2 &-1 \end{pmatrix}$.
The homological blocks $\widehat{Z}_b(q)$ are given by
{\tiny
\begin{align}
\begin{split}
&\hspace{-5mm}\widehat{Z}_0=
(115411539877-2042040 q\partial/\partial q) \widetilde{\Phi }_{510510}^{(11829)}
+(112287218677-2042040 q\partial/\partial q) \widetilde{\Phi}_{510510}^{(34371)}
+(109677491557-2042040 q\partial/\partial q) \widetilde{\Phi }_{510510}^{(45291)}	\\	&\hspace{-5mm}
+(106148846437-2042040 q\partial/\partial q) \widetilde{\Phi}_{510510}^{(56811)}
+(102068850517-2042040 q\partial/\partial q) \widetilde{\Phi }_{510510}^{(67731)}
+(95452640917-2042040 q\partial/\partial q) \widetilde{\Phi}_{510510}^{(82419)}	\\	&\hspace{-5mm}
+(82845085957-2042040 q\partial/\partial q) \widetilde{\Phi }_{510510}^{(104859)}
+(-81356438797+2042040 q\partial/\partial q) \widetilde{\Phi}_{510510}^{(107199)}
+(76890497317-2042040 q\partial/\partial q) \widetilde{\Phi }_{510510}^{(113931)}	\\	&\hspace{-5mm}
+(47374851157-2042040 q\partial/\partial q) \widetilde{\Phi}_{510510}^{(151059)}
+(37119726277-2042040 q\partial/\partial q) \widetilde{\Phi }_{510510}^{(161979)}
+(42992633317-2042040 q\partial/\partial q) \widetilde{\Phi}_{510510}^{(184521)}	\\	&\hspace{-5mm}
+(62461442677-2042040 q\partial/\partial q) \widetilde{\Phi }_{510510}^{(206961)}
+(-64317657037+2042040 q\partial/\partial q) \widetilde{\Phi}_{510510}^{(209301)}
+(-89373487837+2042040 q\partial/\partial q) \widetilde{\Phi }_{510510}^{(246429)}	\\	&\hspace{-5mm}
+(93030781477-2042040 q\partial/\partial q) \widetilde{\Phi}_{510510}^{(253161)}
+(-95168797357+2042040 q\partial/\partial q) \widetilde{\Phi }_{510510}^{(257349)}
+(98385010357-2042040 q\partial/\partial q) \widetilde{\Phi}_{510510}^{(264081)}	\\	&\hspace{-5mm}
+(111237610117-2042040 q\partial/\partial q) \widetilde{\Phi }_{510510}^{(301209)}
+(-111770582557+2042040 q\partial/\partial q) \widetilde{\Phi }_{510510}^{(303549)}
+(-115213461997+2042040 q\partial/\partial q) \widetilde{\Phi }_{510510}^{(325989)}	\\	&\hspace{-5mm}
+(103935275077-2042040 q\partial/\partial q) \widetilde{\Phi }_{510510}^{(403311)}
+(-103034735437+2042040 q\partial/\partial q) \widetilde{\Phi}_{510510}^{(405651)}
+(-92730601597+2042040 q\partial/\partial q) \widetilde{\Phi }_{510510}^{(428091)}	\\	&\hspace{-5mm}
+(-84350069437+2042040 q\partial/\partial q) \widetilde{\Phi}_{510510}^{(442779)}
+(-77280526957+2042040 q\partial/\partial q) \widetilde{\Phi }_{510510}^{(453699)}
+(-69047021677+2042040 q\partial/\partial q) \widetilde{\Phi}_{510510}^{(465219)}	\\	&\hspace{-5mm}
+(-60507210397+2042040 q\partial/\partial q) \widetilde{\Phi }_{510510}^{(476139)}
+(-16460407597+2042040 q\partial/\partial q) \widetilde{\Phi}_{510510}^{(522339)}
+(-106342840237+2042040 q\partial/\partial q) \widetilde{\Phi }_{510510}^{(624441)}	\\	&\hspace{-5mm}
+(-114735624637+2042040 q\partial/\partial q) \widetilde{\Phi }_{510510}^{(661569)}
+(-115630038157+2042040 q\partial/\partial q) \widetilde{\Phi }_{510510}^{(672489)}	\\	&\hspace{-5mm}
+35487 (-115784673853+2042040 q\partial/\partial q) \widetilde{\Psi }_{510510}^{(11829)}
+34371 (-346312581159+6126120 q\partial/\partial q) \widetilde{\Psi}_{510510}^{(34371)}
+45291 (-345442672119+6126120 q\partial/\partial q) \widetilde{\Psi }_{510510}^{(45291)}	\\	&\hspace{-5mm}
+56811 (-344266457079+6126120 q\partial/\partial q) \widetilde{\Psi }_{510510}^{(56811)}
+67731 (-342906458439+6126120 q\partial/\partial q) \widetilde{\Psi }_{510510}^{(67731)}
+82419 (-340701055239+6126120 q\partial/\partial q) \widetilde{\Psi }_{510510}^{(82419)}	\\	&\hspace{-5mm}
+314577 (-112166178973+2042040 q\partial/\partial q) \widetilde{\Psi}_{510510}^{(104859)}
-321597 (-112000773733+2042040 q\partial/\partial q) \widetilde{\Psi }_{510510}^{(107199)}
+341793 (-111504558013+2042040 q\partial/\partial q) \widetilde{\Psi }_{510510}^{(113931)}	\\	&\hspace{-5mm}
+453177 (-108225041773+2042040 q\partial/\partial q) \widetilde{\Psi}_{510510}^{(151059)}
+485937 (-107085583453+2042040 q\partial/\partial q) \widetilde{\Psi }_{510510}^{(161979)}
-155819 (-323214386039+6126120 q\partial/\partial q) \widetilde{\Psi }_{510510}^{(184521)}	\\	&\hspace{-5mm}
-133379 (-329703989159+6126120 q\partial/\partial q) \widetilde{\Psi}_{510510}^{(206961)}
+131039 (-330322727279+6126120 q\partial/\partial q) \widetilde{\Psi }_{510510}^{(209301)}
+93911 (-338674670879+6126120 q\partial/\partial q) \widetilde{\Psi }_{510510}^{(246429)}	\\	&\hspace{-5mm}
-87179 (-339893768759+6126120 q\partial/\partial q) \widetilde{\Psi }_{510510}^{(253161)}
+82991 (-340606440719+6126120 q\partial/\partial q) \widetilde{\Psi }_{510510}^{(257349)}
-76259 (-341678511719+6126120 q\partial/\partial q) \widetilde{\Psi}_{510510}^{(264081)}	\\	&\hspace{-5mm}
-39131 (-345962711639+6126120 q\partial/\partial q) \widetilde{\Psi }_{510510}^{(301209)}
+36791 (-346140369119+6126120 q\partial/\partial q) \widetilde{\Psi }_{510510}^{(303549)}
+14351 (-347287995599+6126120 q\partial/\partial q) \widetilde{\Psi }_{510510}^{(325989)}	\\	&\hspace{-5mm}
+62971 (-343528599959+6126120 q\partial/\partial q) \widetilde{\Psi }_{510510}^{(403311)}
-65311 (-343228420079+6126120 q\partial/\partial q) \widetilde{\Psi}_{510510}^{(405651)}
-87751 (-339793708799+6126120 q\partial/\partial q) \widetilde{\Psi }_{510510}^{(428091)}	\\	&\hspace{-5mm}
-102439 (-337000198079+6126120 q\partial/\partial q) \widetilde{\Psi }_{510510}^{(442779)}
-113359 (-334643683919+6126120 q\partial/\partial q) \widetilde{\Psi }_{510510}^{(453699)}
-124879 (-331899182159+6126120 q\partial/\partial q) \widetilde{\Psi }_{510510}^{(465219)}	\\	&\hspace{-5mm}
-135799 (-329052578399+6126120 q\partial/\partial q) \widetilde{\Psi}_{510510}^{(476139)}
-181999 (-314370310799+6126120 q\partial/\partial q) \widetilde{\Psi }_{510510}^{(522339)}
+56239 (-344331121679+6126120 q\partial/\partial q) \widetilde{\Psi }_{510510}^{(624441)}	\\	&\hspace{-5mm}
+19111 (-347128716479+6126120 q\partial/\partial q) \widetilde{\Psi }_{510510}^{(661569)}
+8191 (-347426854319+6126120 q\partial/\partial q) \widetilde{\Psi }_{510510}^{(672489)}	\,	,
\end{split}
\end{align}
}
{\tiny
\begin{align}
\begin{split}
&\hspace{-5mm}\widehat{Z}_1=
(-115630038157+2042040 q\partial/\partial q) \widetilde{\Phi }_{510510}^{(8191)}
+(-115213461997+2042040 q\partial/\partial q) \widetilde{\Phi}_{510510}^{(14351)}
+(-114735624637+2042040 q\partial/\partial q) \widetilde{\Phi }_{510510}^{(19111)}	\\	&\hspace{-5mm}
+(-111770582557+2042040 q\partial/\partial q) \widetilde{\Phi}_{510510}^{(36791)}
+(111237610117-2042040 q\partial/\partial q) \widetilde{\Phi }_{510510}^{(39131)}
+(-106342840237+2042040 q\partial/\partial q) \widetilde{\Phi}_{510510}^{(56239)}	\\	&\hspace{-5mm}
+(103935275077-2042040 q\partial/\partial q) \widetilde{\Phi }_{510510}^{(62971)}
+(-103034735437+2042040 q\partial/\partial q) \widetilde{\Phi}_{510510}^{(65311)}
+(98385010357-2042040 q\partial/\partial q) \widetilde{\Phi }_{510510}^{(76259)}	\\	&\hspace{-5mm}
+(-95168797357+2042040 q\partial/\partial q) \widetilde{\Phi}_{510510}^{(82991)}
+(93030781477-2042040 q\partial/\partial q) \widetilde{\Phi }_{510510}^{(87179)}
+(-92730601597+2042040 q\partial/\partial q) \widetilde{\Phi}_{510510}^{(87751)}	\\	&\hspace{-5mm}
+(-89373487837+2042040 q\partial/\partial q) \widetilde{\Phi }_{510510}^{(93911)}
+(-84350069437+2042040 q\partial/\partial q) \widetilde{\Phi}_{510510}^{(102439)}
+(-77280526957+2042040 q\partial/\partial q) \widetilde{\Phi }_{510510}^{(113359)}	\\	&\hspace{-5mm}
+(-69047021677+2042040 q\partial/\partial q) \widetilde{\Phi}_{510510}^{(124879)}
+(-64317657037+2042040 q\partial/\partial q) \widetilde{\Phi }_{510510}^{(131039)}
+(62461442677-2042040 q\partial/\partial q) \widetilde{\Phi}_{510510}^{(133379)}	\\	&\hspace{-5mm}
+(-60507210397+2042040 q\partial/\partial q) \widetilde{\Phi }_{510510}^{(135799)}
+(42992633317-2042040 q\partial/\partial q) \widetilde{\Phi}_{510510}^{(155819)}
+(-16460407597+2042040 q\partial/\partial q) \widetilde{\Phi }_{510510}^{(158341)}	\\	&\hspace{-5mm}
+(37119726277-2042040 q\partial/\partial q) \widetilde{\Phi}_{510510}^{(178361)}
+(-16460407597+2042040 q\partial/\partial q) \widetilde{\Phi }_{510510}^{(181999)}
+(47374851157-2042040 q\partial/\partial q) \widetilde{\Phi}_{510510}^{(189281)}	\\	&\hspace{-5mm}
+(-60507210397+2042040 q\partial/\partial q) \widetilde{\Phi }_{510510}^{(204541)}
+(-69047021677+2042040 q\partial/\partial q) \widetilde{\Phi}_{510510}^{(215461)}
+(76890497317-2042040 q\partial/\partial q) \widetilde{\Phi }_{510510}^{(226409)}	\\	&\hspace{-5mm}
+(-77280526957+2042040 q\partial/\partial q) \widetilde{\Phi}_{510510}^{(226981)}
+(-81356438797+2042040 q\partial/\partial q) \widetilde{\Phi }_{510510}^{(233141)}	
+(82845085957-2042040 q\partial/\partial q) \widetilde{\Phi}_{510510}^{(235481)}	\\	&\hspace{-5mm}
+(-84350069437+2042040 q\partial/\partial q) \widetilde{\Phi }_{510510}^{(237901)}
+(-92730601597+2042040 q\partial/\partial q) \widetilde{\Phi}_{510510}^{(252589)}
+(95452640917-2042040 q\partial/\partial q) \widetilde{\Phi }_{510510}^{(257921)}	\\	&\hspace{-5mm}
+(102068850517-2042040 q\partial/\partial q) \widetilde{\Phi}_{510510}^{(272609)}
+(-103034735437+2042040 q\partial/\partial q) \widetilde{\Phi }_{510510}^{(275029)}
+(103935275077-2042040 q\partial/\partial q) \widetilde{\Phi }_{510510}^{(277369)}	\\	&\hspace{-5mm}
+(106148846437-2042040 q\partial/\partial q) \widetilde{\Phi }_{510510}^{(283529)}
+(-106342840237+2042040 q\partial/\partial q) \widetilde{\Phi }_{510510}^{(284101)}
+(109677491557-2042040 q\partial/\partial q) \widetilde{\Phi }_{510510}^{(295049)}	\\	&\hspace{-5mm}
+(112287218677-2042040 q\partial/\partial q) \widetilde{\Phi }_{510510}^{(305969)}
+(-114735624637+2042040 q\partial/\partial q) \widetilde{\Phi}_{510510}^{(321229)}
+(115411539877-2042040 q\partial/\partial q) \widetilde{\Phi }_{510510}^{(328511)}	\\	&\hspace{-5mm}
+(-115630038157+2042040 q\partial/\partial q) \widetilde{\Phi }_{510510}^{(332149)}
+(115411539877-2042040 q\partial/\partial q) \widetilde{\Phi }_{510510}^{(352169)}
+(109677491557-2042040 q\partial/\partial q) \widetilde{\Phi }_{510510}^{(385631)}	\\	&\hspace{-5mm}
+(102068850517-2042040 q\partial/\partial q) \widetilde{\Phi }_{510510}^{(408071)}
+(95452640917-2042040 q\partial/\partial q) \widetilde{\Phi }_{510510}^{(422759)}
+(-95168797357+2042040 q\partial/\partial q) \widetilde{\Phi }_{510510}^{(423331)}	\\	&\hspace{-5mm}
+(93030781477-2042040 q\partial/\partial q) \widetilde{\Phi }_{510510}^{(427519)}
+(-89373487837+2042040 q\partial/\partial q) \widetilde{\Phi }_{510510}^{(434251)}
+(82845085957-2042040 q\partial/\partial q) \widetilde{\Phi }_{510510}^{(445199)}	\\	&\hspace{-5mm}
+(-81356438797+2042040 q\partial/\partial q) \widetilde{\Phi }_{510510}^{(447539)}
+(76890497317-2042040 q\partial/\partial q) \widetilde{\Phi }_{510510}^{(454271)}
+(-64317657037+2042040 q\partial/\partial q) \widetilde{\Phi }_{510510}^{(471379)}	\\	&\hspace{-5mm}
+(62461442677-2042040 q\partial/\partial q) \widetilde{\Phi }_{510510}^{(473719)}
+(47374851157-2042040 q\partial/\partial q) \widetilde{\Phi }_{510510}^{(491399)}
+(42992633317-2042040 q\partial/\partial q) \widetilde{\Phi }_{510510}^{(496159)}	\\	&\hspace{-5mm}
+(37119726277-2042040 q\partial/\partial q) \widetilde{\Phi }_{510510}^{(502319)}
+(98385010357-2042040 q\partial/\partial q) \widetilde{\Phi }_{510510}^{(604421)}
+(106148846437-2042040 q\partial/\partial q) \widetilde{\Phi }_{510510}^{(623869)}	\\	&\hspace{-5mm}
+(111237610117-2042040 q\partial/\partial q) \widetilde{\Phi }_{510510}^{(641549)}
+(-111770582557+2042040 q\partial/\partial q) \widetilde{\Phi }_{510510}^{(643889)}
+(112287218677-2042040 q\partial/\partial q) \widetilde{\Phi }_{510510}^{(646309)}	\\	&\hspace{-5mm}
+(-115213461997+2042040 q\partial/\partial q) \widetilde{\Phi }_{510510}^{(666329)}	\\	&\hspace{-5mm}
-8191 (-347426854319+6126120 q\partial/\partial q) \widetilde{\Psi }_{510510}^{(8191)}
-14351 (-347287995599+6126120 q\partial/\partial q) \widetilde{\Psi}_{510510}^{(14351)}
-19111 (-347128716479+6126120 q\partial/\partial q) \widetilde{\Psi }_{510510}^{(19111)}	\\	&\hspace{-5mm}
-36791 (-346140369119+6126120 q\partial/\partial q) \widetilde{\Psi }_{510510}^{(36791)}
+39131 (-345962711639+6126120 q\partial/\partial q) \widetilde{\Psi }_{510510}^{(39131)}
-56239 (-344331121679+6126120 q\partial/\partial q) \widetilde{\Psi }_{510510}^{(56239)}	\\	&\hspace{-5mm}
+62971 (-343528599959+6126120 q\partial/\partial q) \widetilde{\Psi}_{510510}^{(62971)}
-65311 (-343228420079+6126120 q\partial/\partial q) \widetilde{\Psi }_{510510}^{(65311)}
+76259 (-341678511719+6126120 q\partial/\partial q) \widetilde{\Psi }_{510510}^{(76259)}	\\	&\hspace{-5mm}
-82991 (-340606440719+6126120 q\partial/\partial q) \widetilde{\Psi }_{510510}^{(82991)}
+87179 (-339893768759+6126120 q\partial/\partial q) \widetilde{\Psi }_{510510}^{(87179)}
-87751 (-339793708799+6126120 q\partial/\partial q) \widetilde{\Psi}_{510510}^{(87751)}	\\	&\hspace{-5mm}
-93911 (-338674670879+6126120 q\partial/\partial q) \widetilde{\Psi }_{510510}^{(93911)}
-102439 (-337000198079+6126120 q\partial/\partial q) \widetilde{\Psi }_{510510}^{(102439)}
-113359 (-334643683919+6126120 q\partial/\partial q) \widetilde{\Psi }_{510510}^{(113359)}	\\	&\hspace{-5mm}
-124879 (-331899182159+6126120 q\partial/\partial q) \widetilde{\Psi }_{510510}^{(124879)}
-131039 (-330322727279+6126120 q\partial/\partial q) \widetilde{\Psi}_{510510}^{(131039)}
+133379 (-329703989159+6126120 q\partial/\partial q) \widetilde{\Psi }_{510510}^{(133379)}	\\	&\hspace{-5mm}
-135799 (-329052578399+6126120 q\partial/\partial q) \widetilde{\Psi }_{510510}^{(135799)}
+155819 (-323214386039+6126120 q\partial/\partial q) \widetilde{\Psi}_{510510}^{(155819)}
+181999 (-314370310799+6126120 q\partial/\partial q) \widetilde{\Psi }_{510510}^{(158341)}	\\	&\hspace{-5mm}
-485937 (-107085583453+2042040 q\partial/\partial q) \widetilde{\Psi }_{510510}^{(178361)}
-181999 (-314370310799+6126120 q\partial/\partial q) \widetilde{\Psi}_{510510}^{(181999)}
-453177 (-108225041773+2042040 q\partial/\partial q) \widetilde{\Psi }_{510510}^{(189281)}	\\	&\hspace{-5mm}
+135799 (-329052578399+6126120 q\partial/\partial q) \widetilde{\Psi }_{510510}^{(204541)}
+124879 (-331899182159+6126120 q\partial/\partial q) \widetilde{\Psi}_{510510}^{(215461)}
-341793 (-111504558013+2042040 q\partial/\partial q) \widetilde{\Psi }_{510510}^{(226409)}	\\	&\hspace{-5mm}
+113359 (-334643683919+6126120 q\partial/\partial q) \widetilde{\Psi }_{510510}^{(226981)}
+321597 (-112000773733+2042040 q\partial/\partial q) \widetilde{\Psi}_{510510}^{(233141)}
-314577 (-112166178973+2042040 q\partial/\partial q) \widetilde{\Psi }_{510510}^{(235481)}	\\	&\hspace{-5mm}
+102439 (-337000198079+6126120 q\partial/\partial q) \widetilde{\Psi }_{510510}^{(237901)}
+87751 (-339793708799+6126120 q\partial/\partial q) \widetilde{\Psi}_{510510}^{(252589)}
-82419 (-340701055239+6126120 q\partial/\partial q) \widetilde{\Psi }_{510510}^{(257921)}	\\	&\hspace{-5mm}
-67731 (-342906458439+6126120 q\partial/\partial q) \widetilde{\Psi }_{510510}^{(272609)}
+65311 (-343228420079+6126120 q\partial/\partial q) \widetilde{\Psi }_{510510}^{(275029)}
-62971 (-343528599959+6126120 q\partial/\partial q) \widetilde{\Psi }_{510510}^{(277369)}	\\	&\hspace{-5mm}
-56811 (-344266457079+6126120 q\partial/\partial q) \widetilde{\Psi}_{510510}^{(283529)}
+56239 (-344331121679+6126120 q\partial/\partial q) \widetilde{\Psi }_{510510}^{(284101)}
-45291 (-345442672119+6126120 q\partial/\partial q) \widetilde{\Psi }_{510510}^{(295049)}	\\	&\hspace{-5mm}
-34371 (-346312581159+6126120 q\partial/\partial q) \widetilde{\Psi }_{510510}^{(305969)}
+19111 (-347128716479+6126120 q\partial/\partial q) \widetilde{\Psi }_{510510}^{(321229)}
-35487 (-115784673853+2042040 q\partial/\partial q) \widetilde{\Psi}_{510510}^{(328511)}	\\	&\hspace{-5mm}
+8191 (-347426854319+6126120 q\partial/\partial q) \widetilde{\Psi }_{510510}^{(332149)}
+35487 (-115784673853+2042040 q\partial/\partial q) \widetilde{\Psi }_{510510}^{(352169)}
+45291 (-345442672119+6126120 q\partial/\partial q) \widetilde{\Psi }_{510510}^{(385631)}	\\	&\hspace{-5mm}
+67731 (-342906458439+6126120 q\partial/\partial q) \widetilde{\Psi }_{510510}^{(408071)}
+82419 (-340701055239+6126120 q\partial/\partial q) \widetilde{\Psi}_{510510}^{(422759)}
-82991 (-340606440719+6126120 q\partial/\partial q) \widetilde{\Psi }_{510510}^{(423331)}	\\	&\hspace{-5mm}
+87179 (-339893768759+6126120 q\partial/\partial q) \widetilde{\Psi }_{510510}^{(427519)}
-93911 (-338674670879+6126120 q\partial/\partial q) \widetilde{\Psi }_{510510}^{(434251)}
+314577 (-112166178973+2042040 q\partial/\partial q) \widetilde{\Psi }_{510510}^{(445199)}	\\	&\hspace{-5mm}
-321597 (-112000773733+2042040 q\partial/\partial q) \widetilde{\Psi}_{510510}^{(447539)}
+341793 (-111504558013+2042040 q\partial/\partial q) \widetilde{\Psi }_{510510}^{(454271)}
-131039 (-330322727279+6126120 q\partial/\partial q) \widetilde{\Psi }_{510510}^{(471379)}	\\	&\hspace{-5mm}
+133379 (-329703989159+6126120 q\partial/\partial q) \widetilde{\Psi}_{510510}^{(473719)}
+453177 (-108225041773+2042040 q\partial/\partial q) \widetilde{\Psi }_{510510}^{(491399)}\
+155819 (-323214386039+6126120 q\partial/\partial q) \widetilde{\Psi }_{510510}^{(496159)}	\\	&\hspace{-5mm}
+485937 (-107085583453+2042040 q\partial/\partial q) \widetilde{\Psi}_{510510}^{(502319)}
-76259 (-341678511719+6126120 q\partial/\partial q) \widetilde{\Psi }_{510510}^{(604421)}
-56811 (-344266457079+6126120 q\partial/\partial q) \widetilde{\Psi }_{510510}^{(623869)}	\\	&\hspace{-5mm}
-39131 (-345962711639+6126120 q\partial/\partial q) \widetilde{\Psi }_{510510}^{(641549)}
+36791 (-346140369119+6126120 q\partial/\partial q) \widetilde{\Psi }_{510510}^{(643889)}
-34371 (-346312581159+6126120 q\partial/\partial q) \widetilde{\Psi}_{510510}^{(646309)}	\\	&\hspace{-5mm}
+14351 (-347287995599+6126120 q\partial/\partial q) \widetilde{\Psi }_{510510}^{(666329)}
\end{split}
\end{align}
}
where $\widehat{Z}_0 = q^{\frac{355507}{680680}}\mathbb{Z} [[ q ]]$ and $\widehat{Z}_1=q^{\frac{1747201}{2042040}} \mathbb{Z} [[q]]$.

\subsubsection*{\textbullet \ $F=3$, $H=3$, and $g=1$}
When $F=3$ and $g=1$, we can have $H=3$, for example, from $(P_1,P_2,P_3)=(2,5,7)$ and $(Q_1,Q_2,Q_3)=(1,2,-6)$.
The WRT invariant is given by
\begin{align}
Z_{SU(2)}(M_3) = \frac{B}{i}  (-2K) q^{-\phi_3/4} \bigg( \frac{70i}{K} \bigg)^{1/2}  \frac{1}{2^4 \cdot 2 \cdot 70^2}  \sum_{a,b=0}^{1} e^{\pi i K S_a} S_{ab} \widehat{Z}_b(q)	\,	\Big|_{q \searrow e^{\frac{2\pi i}{K}}}
\end{align}
where $(CS_0,CS_1)=(0,\frac{1}{3})$ and $S_{ab}=\frac{1}{\sqrt{3}}\begin{pmatrix} 1 & 1 \\ 2 &-1 \end{pmatrix}$.
The homological blocks $\widehat{Z}_b(q)$ are given by
\begin{align}
\begin{split}
\widehat{Z}_0 =& \,
-62 \widetilde{\Phi }_{210}^{(39)}
+22 \widetilde{\Phi }_{210}^{(81)}
+118 \widetilde{\Phi }_{210}^{(129)}
-78 \widetilde{\Phi}_{210}^{(171)}	\\
&+(3939-840 q\partial/\partial q) \widetilde{\Psi }_{210}^{(39)}
+(4779-840 q\partial/\partial q) \widetilde{\Psi }_{210}^{(81)}	\\
&+(1419-840 q\partial/\partial q) \widetilde{\Psi }_{210}^{(129)}
+(3379-840 q\partial/\partial q) \widetilde{\Psi }_{210}^{(171)}	\,	,
\end{split}	\\	
\begin{split}
\widehat{Z}_1 = & \,
118 \widetilde{\Phi }_{210}^{(11)}
-78 \widetilde{\Phi }_{210}^{(31)}
+22 \widetilde{\Phi }_{210}^{(59)}
-62 \widetilde{\Phi}_{210}^{(101)}
-78 \widetilde{\Phi }_{210}^{(109)}
-62 \widetilde{\Phi }_{210}^{(179)}
+22 \widetilde{\Phi }_{210}^{(221)}
+118 \widetilde{\Phi }_{210}^{(269)}	\\	&
+3 (-473+280 q\partial/\partial q) \widetilde{\Psi }_{210}^{(11)}
+(3379-840 q\partial/\partial q) \widetilde{\Psi}_{210}^{(31)}
+(-4779+840 q\partial/\partial q) \widetilde{\Psi }_{210}^{(59)}	\\	&
+(-3939+840 q\partial/\partial q) \widetilde{\Psi }_{210}^{(101)}
+(-3379+840 q\partial/\partial q) \widetilde{\Psi }_{210}^{(109)}
+(3939-840 q\partial/\partial q) \widetilde{\Psi }_{210}^{(179)}	\\	&
+(4779-840 q\partial/\partial q) \widetilde{\Psi}_{210}^{(221)}
+(1419-840 q\partial/\partial q) \widetilde{\Psi }_{210}^{(269)}	\,	.
\end{split}
\end{align}

\end{appendices}

\bibliographystyle{JHEP}
\bibliography{ref}

\end{document}